\documentclass[twocolumn,preprintnumbers,amsmath,amssymb,superscriptaddress,nofootinbib,english]{revtex4-1}
\pdfoutput=1
\usepackage{times,amsmath,amsfonts,amssymb,epstopdf}
\usepackage{graphicx}
\usepackage{dcolumn}
\usepackage{bm}
\usepackage{enumerate}
\usepackage{epsfig}
\usepackage{graphicx}
\usepackage{hyperref}
\usepackage[usenames]{color}
\usepackage{url}
\usepackage[normalem]{ulem}
\usepackage[T1]{fontenc}
\usepackage{amsmath}

\usepackage[usenames]{color}

\def\be{\begin{equation}}
\def\ee{\end{equation}}
\def\ben{\begin{eqnarray}}
\def\een{\end{eqnarray}}
\def\ba{\begin{array}}
\def\ea{\end{array}}

\def\ba#1\ea{\begin{align}#1\end{align}}

\newcommand{\bq}{\begin{eqnarray}}
\newcommand{\eq}{\end{eqnarray}}
\newcommand{\bes}{\begin{subequations}}
\newcommand{\ees}{\end{subequations}}

\begin{document}
\newcommand{\half}{{\textstyle\frac{1}{2}}}
\allowdisplaybreaks[3]
\def\triangledown{\nabla}
\def\grad3{\hat{\nabla}}
\def\a{\alpha}
\def\b{\beta}
\def\g{\gamma}\def\G{\Gamma}
\def\d{\delta}\def\D{\Delta}
\def\ep{\epsilon}
\def\et{\eta}
\def\z{\zeta}
\def\t{\theta}\def\T{\Theta}
\def\l{\lambda}\def\L{\Lambda}
\def\m{\mu}
\def\f{\phi}\def\F{\Phi}
\def\n{\nu}
\def\r{\rho}
\def\s{\sigma}\def\S{\Sigma}
\def\ta{\tau}
\def\x{\chi}
\def\o{\omega}\def\O{\Omega}
\def\k{\kappa}
\def\pa {\partial}
\def\ov{\over}
\def\br{\\}
\def\ud{\underline}

\def\lcdm{\Lambda{\rm CDM}}
\def\msun{M_{\odot}/h}
\def\ndgp{\rm nDGP}
\def\lens{{\rm nDGP}_{lens}}
\def\gtw{G_{20}}
\def\gei{G_{80}}
\def\gfi{G_{5}}
\def\gni{G_{95}}
\def\gfo{G_{40}}
\def\gsi{G_{60}}

\def\mnras{MNRAS}

\newcommand{\comment}[1]{}

\title{Weak lensing by galaxy troughs with modified gravity}

\author{Alexandre Barreira}
\email[Electronic address: ]{barreira@mpa-garching.mpg.de}
\affiliation{Max-Planck-Institut f{\"u}r Astrophysik, Karl-Schwarzschild-Str. 1, 85741 Garching, Germany}

\author{Sownak Bose}
\email[Electronic address: ]{sownak.bose@durham.ac.uk}
\affiliation{Institute for Computational Cosmology, Durham University, South Road DH1 3LE, Durham, U.K.}

\author{Baojiu Li}
\email[Electronic address: ]{baojiu.li@durham.ac.uk}
\affiliation{Institute for Computational Cosmology, Durham University, South Road DH1 3LE, Durham, U.K.}

\author{Claudio Llinares}
\email[Electronic address: ]{claudio.llinares@durham.ac.uk}
\affiliation{Institute for Computational Cosmology, Durham University, South Road DH1 3LE, Durham, U.K.}

\begin{abstract}

We study the imprints that theories of gravity beyond GR can leave on the lensing signal around line of sight directions that are predominantly halo-underdense (called troughs) and halo-overdense. To carry out our investigations, we consider the normal branch of DGP gravity, as well as a phenomenological variant thereof that directly modifies the lensing potential. The predictions of these models are obtained with N-body simulation and ray-tracing methods using the {\tt ECOSMOG} and {\tt Ray-Ramses} codes. We analyse the stacked lensing convergence profiles around the underdense and overdense lines of sight, which exhibit, respectively, a suppression and a boost w.r.t.~the mean in the field of view. The modifications to gravity in these models strengthen the signal w.r.t.~$\lcdm$ in a scale-independent way. We find that the size of this effect is the same for both underdense and overdense lines of sight, which implies that the density field along the overdense directions on the sky is not sufficiently evolved to trigger the suppression effects of the screening mechanism. These results are robust to variations in the minimum halo mass and redshift ranges used to identify the lines of sight, as well as to different line of sight aperture sizes and criteria for their underdensity and overdensity thresholds.

\end{abstract}

\maketitle

\section{Introduction}\label{sec:intr}

There is currently a number of ongoing (e.g.~CFHTLenS \cite{2012MNRAS.427..146H}, BOSS \cite{2014MNRAS.441...24A}, DES \cite{Abbott:2015swa}) and planned (e.g.~Euclid \cite{2011arXiv1110.3193L}, DESI \cite{2013arXiv1308.0847L}, LSST \cite{2012arXiv1211.0310L}) large scale structure surveys that are aiming to constrain deviations from General Relativity (GR) using cosmological data. The types of models characterized by such deviations are generically referred to as {\it modified gravity} models and they have been the target of growing interest in recent years for mostly two reasons. First, there is the aforementioned desire to extend tests of gravity onto cosmological scales. This requires extensive and rigorous investigations of the imprints of modified gravity on cosmological observables in order determine "what to look for" in current and future data. Another major motivation for modified gravity studies comes from the possibility to explain cosmic acceleration. The premise here is that the accelerated expansion of the Universe may not be due to a new and exotic form of dark energy or the cosmological constant $\Lambda$, and instead it is simply a manifestation of departures from GR on sufficiently large scales. The body of work on modified theories of gravity has notably increased over the past few years, with this field being now a well developed branch of theoretical and observational cosmology (see e.g.~Refs.~\cite{Jain:2007yk, 2012PhR...513....1C, Joyce:2014kja, 2015arXiv150404623K, 2016arXiv160106133J} for reviews).

In modified gravity, the deviations from the GR force law typically arise in the form of a {\it fifth force} that is sourced by a new scalar degree of freedom. However, the existence of such an additional force quite naturally leads to the concern of how these theories can be made compatible with Solar System tests of gravity \cite{Will:2014xja}, whilst retaining potentially detectable features on cosmological scales. The standard way to ensure this is via an effect that has been dubbed {\it screening} \cite{Brax:2013ida}. In short, screening arises from nonlinear terms that exist in the model equations and work to suppress the relative size of the fifth force in regions where the gravitational potential or its derivatives reach a sufficiently large value. Normally, this tends to happen on small length scales ($\lesssim 1-5\ {\rm Mpc}$), thereby allowing the fifth force to be small locally, but sufficiently large on larger scales. The most popular examples of types of screening include the Chameleon \cite{khoury:2003aq, physrevd.69.044026}, Symmetron \cite{Olive:2007aj, hinterbichler:2010es, hinterbichler:2011ca}, Dilaton \cite{brax:2010gi, brax:2011ja}, Vainshtein \cite{Vainshtein1972393, Babichev:2013usa, Koyama:2013paa} and K-mouflage \cite{Babichev:2009ee, brax:2014gra} mechanisms. The additional scale-dependence introduced by the screening mechanisms enriches the phenomenology of these theories and represents a unique feature w.r.t.~standard GR. This motivates research to determine which cosmological observations stand the best chances to unveil the presence of screening.

Here, our goal is to determine the types of observational signatures of modified gravity (including eventual scale-dependent screening effects) on the lensing signal along lines of sight (LOS) that are predominantly devoid of or have an excess of haloes/galaxies. A recent observational effort that reported the detection of this lensing signal was carried out in Ref.~\cite{2016MNRAS.455.3367G} using data from DES. There, this was dubbed {\it trough lensing}, where the word {\it trough} represents a LOS along which the galaxy number count is sufficiently below the mean of all LOS (we shall define this more rigorously in Sec.~\ref{sec:defi}). Reference \cite{2016MNRAS.455.3367G} detected the suppression of the lensing signal that one would expect if photons had travelled through mostly underdense regions. This is similar to the lensing associated with cosmic voids \cite{melchior:2013gxd, clampitt:2014gpa, 2016arXiv160503982S}, except that troughs are much more extended along the LOS, which improves the signal to noise. The interest in searching for modified gravity effects in trough lensing arises because this signal is sensitive to low density regions, where the screening is normally not at play, and hence the effects of the fifth force can be manifest at their full strength \cite{2015JCAP...08..028B}. Conversely, by looking at the lensing signal associated with overdense LOS (the {\it opposite} of a trough, and that we also define in Sec.~\ref{sec:defi}), one might hope to find evidence for the suppresion effects of the screening.

One way to organize the theory space of modified gravity is by splitting it into (i) models {whose extra terms source both the lensing potential $\Phi_{\rm len}$ and the dynamical potential $\Psi$, and (ii) models whose extra terms source only $\Psi$, but leave the equation that governs $\Phi_{\rm len}$ unchanged (one can also consider models whose extra terms only contribute to $\Phi_{\rm len}$, but there are fewer concrete examples of such cases).} Naturally, models that modify the way $\Phi_{\rm len}$ reacts to the density distribution are more likely to leave stronger imprints on lensing observations such as trough lensing. There are, however, some difficulties that arise from applying conventional numerical ray-tracing methods using N-body simulations (which are needed to model the trough lensing signal) to cases where $\Phi_{\rm len}$ is governed by a nonlinear (to have screening) Poisson equation (cf.~Eqs.~(\ref{eq:modpoisson}) and (\ref{eq:eomvarphi}) below). In these conventional methods (see e.g.~Refs.~\cite{2000ApJ...530..547J, 2003ApJ...592..699V, Hilbert:2008kb, 2015arXiv151108211G} and references therein), the lensing signal along the LOS is evaluated at a finite number of planes, onto which the three-dimensional density field has been projected. If $\Phi_{\rm len}$ is sourced also by an additional scalar field, then this must be carefully taken into account in the construction of the several planes, which adds complication to the numerical procedures. Partly due to this difficulty, lensing studies on nonlinear scales of models with modified $\Phi_{\rm len}$ have relied on simplifying assumptions such as spherical symmetry \cite{2011PhRvL.106t1102W, 2015PhRvD..91f4012P, 2015JCAP...08..028B, 2015MNRAS.454.4085B}, and studies based on N-body simulations have mostly focused on models that do not modify the lensing potential \cite{2015MNRAS.451.1036C, 2015JCAP...10..036T, 2016MNRAS.459.2762H}.

In this paper, we make use of the ray-tracing methods of the {\tt Ray-Ramses} code \cite{2016arXiv160102012B, whitehu2000, li2001}, which provide a straightforward  way out of the complications mentioned above. This code, which consists of a series of {\it add-on} modules to the publicly available adaptive mesh refinement (AMR) {\tt Ramses} code \cite{2002A&A...385..337T}, computes the lensing signal on-the-fly with the N-body simulation by making use of the full three-dimensional $\Phi_{\rm len}$ that is evaluated at every cell of the AMR structure. This feature is particularly useful for modified gravity because it allows to calculate the lensing signal without {resorting} to methods based on plane projections: once the three-dimensional distribution of the additional scalar field is determined, it is used to construct $\Phi_{\rm len}$, which is what is directly used by {\tt Ray-Ramses} \cite{2016arXiv160102012B}. We carry out our simulations of modified gravity with the {\tt ECOSMOG} code \cite{2012JCAP...01..051L, baojiudgp}, which is itself an extension of {\tt Ramses} that solves for the additional scalar degree of freedom. We study the influence of modified gravity on the trough lensing signal by taking the normal branch of the Dvali-Gabadadze-Porrati (DGP) model \cite{Dvali:2000hr} (with a $\lcdm$ background) as our working case. {This is an example of a model in which the modifications to gravity only contribute to the dynamical potential $\Psi$, leaving the equation that governs the lensing potential $\Phi_{\rm len} = (\Phi + \Psi)/2$ as in GR.} Hence, to enrich our (mostly phenomenological) analysis, we also consider a variant of the DGP model {that contains terms that also directly source the lensing potential.} The comparison of the results of these two variants of DGP gravity allows one to disentangle the contribution to lensing that arises from the modified matter distribution and that which arises from the explicit modifications to the lensing potential.

This paper is organized as follows. In Sec.~\ref{sec:work}, we present the main aspects of the DGP gravity models we study and in Sec.~\ref{sec:rayt} we describe the N-body simulation and lensing numerical setups that we use to obtain our results. In Sec.~\ref{sec:defi}, we specify the procedure to identify the halo-underdense and halo-overdense LOS around which we measure the lensing signal. Section \ref{sec:prof} contains our main results for the stacked lensing convergence profiles around these LOS. We discuss, in particular, the impact of the modifications to gravity on the lensing profiles, as well as the impact of minimum mass and redshift range of the haloes used to construct the halo lightcones that are used to identify the LOS. We summarize and conclude in Sec.~\ref{sec:conc}. Appendix \ref{sec:app1} presents a few validation checks of the numerical setup adopted in the main body of the text. {In Appendix \ref{sec:app2}, we display a few tests that ensure that our main conclusions are not affected by sample variance effects on the simulated lensing maps.}
\begin{figure*}
	\centering
	\includegraphics[scale=0.40]{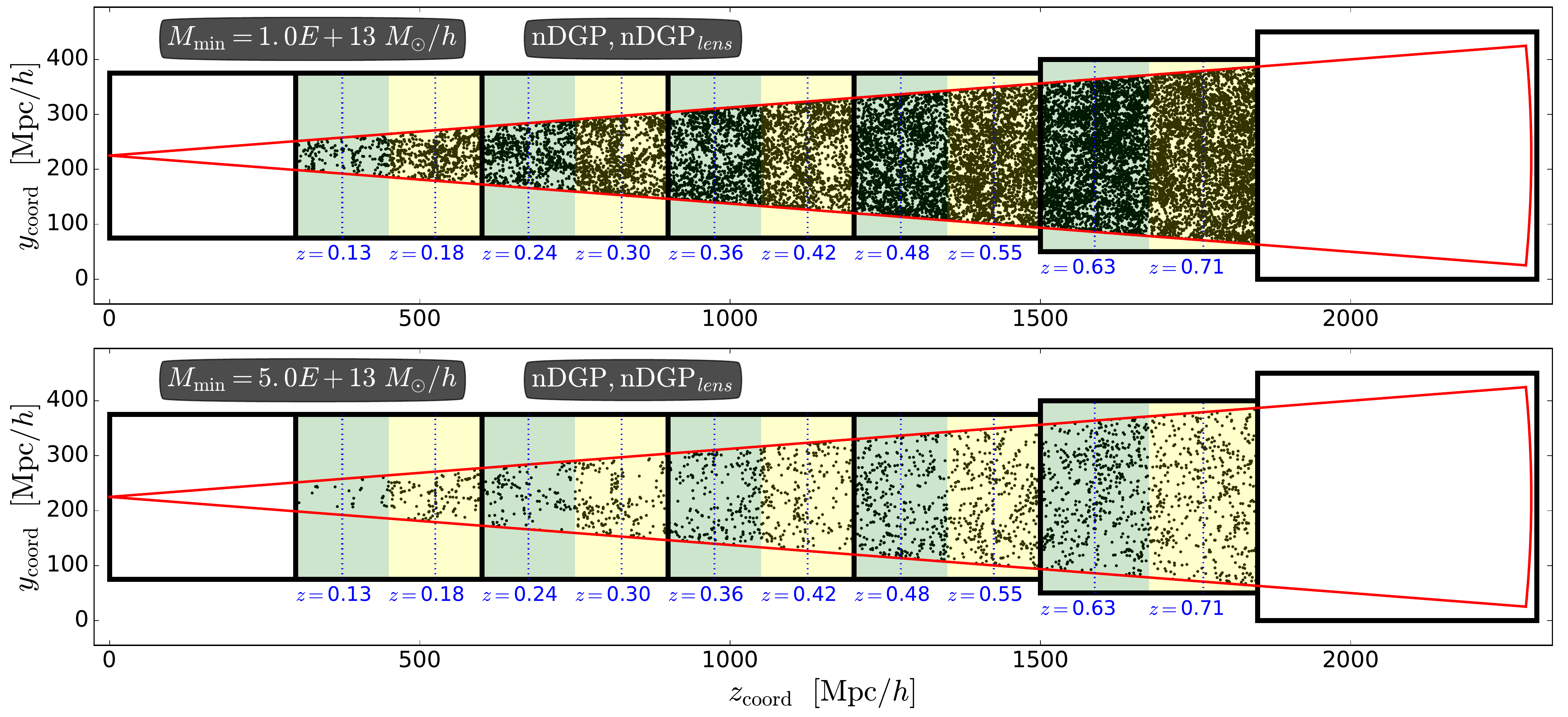}
	\caption{Lensing tiling setup and "pseudo" halo lightcones used in this paper. In both panels, the solid red lines show the lensing lightcone geometry, which spans a FOV of $10\times10\ {\rm deg}^2$. The solid black lines depict the simulation boxes that make up the tile that encompasses the lensing lightcone. From left to right, the boxes have sizes $L=300{\rm Mpc}/h$ (first five), $L=350{\rm Mpc}/h$ and $L=450{\rm Mpc}/h$, respectively. The dots show the positions of dark matter haloes found in the simulations. Haloes located on top of a given colored background correspond to the simulation snapshot associated with the redshift value indicated below. The vertical dotted lines indicate the approximate $z_{\rm coord}$ of the rays at the listed redshift values (see Sec.~\ref{sec:tili} for more details about the construction of the halo lightcone). The two panels show the halo positions in the $\ndgp$ and $\lens$ models for different minimum halo mass cutoffs, as labelled (recall that the halo distribution is the same in these two models).}
\label{fig:tile}
\end{figure*}

\section{Working case gravity models}\label{sec:work}

In this section, we describe the cosmological models that we use as working cases.  We shall be brief and limit ourselves to only laying out the relevant aspects for the analysis in this paper, and refer the interested reader to the cited literature (and references therein) for more details.

\subsection{Normal branch of DGP gravity}\label{sec:ndgp}

We consider the normal branch of the DGP model \cite{Dvali:2000hr} (henceforth referred to as $\ndgp$), which together with its self-accelerating branch counterpart, is one of the most well studied toy-models of gravity in cosmology. This is both at the theoretical and observational levels \cite{DeffayetEtal02, 2003JHEP...09..029L, SahniShtanov, 2004JHEP...06..059N, 2007CQGra..24R.231K, 2006PhLB..642..432F, 2006PhRvD..74b3004M, 2008PhRvD..78j3509F, 2009PhRvD..80f3536L, 2010PhRvD..82d4032W, 2013MNRAS.436...89R, 2014JCAP...02..048X}, including also several studies of nonlinear structure formation with N-body simulations \cite{2009PhRvD..80d3001S, 2009PhRvD..80l3003S, baojiudgp, 2012PhRvL.109e1301L, 2013MNRAS.436...89R, 2014MNRAS.445.1885Z, 2014JCAP...07..058F, 2015JCAP...07..049F, codecomp, 2015JCAP...12..059B, 2016arXiv160503965B}. {The normal branch, while not very theoretically appealing (in the sense that it lacks self-accelerating solutions), is nevertheless a useful toy-model to constrain deviations from GR using large scale structure.}

In a perturbed Friedmann-Robertson-Walker (FRW) four dimensional flat spacetime
\bq\label{eq:metric}
{\rm d}s^2 = \left(1 + 2\Psi\right){\rm d}t^2 - a(t)^2\left(1 - 2\Phi\right){\rm d}x^2,
\eq
structure formation in the $\ndgp$ model on sub-horizon scales is governed by the equations (see e.g.~\cite{2007PhRvD..75h4040K, 2009PhRvD..80d3001S, 2015arXiv150503539W})
\bq
\label{eq:modpoisson}&&\nabla^2\Psi = 4\pi G a^2 \delta\rho_m + \frac{1}{2}\nabla^2\varphi, \\
\label{eq:eomvarphi}&&\nabla^2\varphi + \frac{r_c^2}{3\beta(a) a^2} \left[\left(\nabla^2\varphi\right)^2 - \left(\nabla_i\nabla_j\varphi\right)^2\right] = \frac{8\pi G}{3\beta(a)}a^2 \delta\rho_m, \nonumber \\
\eq
where $r_c$ is a model parameter, $\delta\rho_m = \rho_m - \bar{\rho}_{m}$ is the matter density perturbation ($\rho_{m}$ is the total matter density and an overbar indicates background averaged quantities) and $\varphi$ is a scalar field. {The lensing potential $\Phi_{\rm len} = (\Phi + \Psi)/2$ is governed by the same equation as in GR, $\nabla^2\Phi_{\rm len} = 4\pi G a^2 \delta\rho_m$.} The function $\beta(a)$ is given by
\bq\label{eq:beta}
\beta(a) = 1 + 2Hr_c\left(1 + \frac{\dot{H}(a)}{3H^2(a)}\right),
\eq
where a dot indicates a derivative w.r.t.~physical time $t$. The expansion rate $H(a)$ in the $\ndgp$ model is given by
\bq\label{eq:ndgpH}
H(a) = H_0\sqrt{\Omega_{m0}a^{-3} + \Omega_{\rm de}(a) + \Omega_{rc}} + \sqrt{\Omega_{rc}},
\eq
where $\Omega_{m0} = 8\pi G \bar{\rho}_{m0}/(3H_0^2)$ is the fractional nonrelativistic matter density at the present time (we ignore the contribution from radiation, which is negligible at the late times we are interested in), $\Omega_{rc} = 1/(4H_0^2r_c^2)$ and $H_0 = 100h\ {\rm km/s/Mpc}$ is the present-day value of the Hubble rate. The term $\Omega_{\rm de}(a)$ represents the fractional energy density of some dark energy component, which we tune such that the expansion rate in the $\ndgp$ model becomes the same as in a flat $\lcdm$ cosmology with the same $\Omega_m$ and $H_0$ (see e.g.~Ref.~\cite{2009PhRvD..80l3003S}):
\bq\label{eq:lcdmH}
H(a) = H_0\sqrt{\Omega_{m0}a^{-3} + \left(1 - \Omega_{m0}\right)}.
\eq
This allows one to single out the effects of the modified gravitational potentials from those of modified background dynamics on any observed differences in our $\ndgp$ and $\lcdm$ results.

\subsection{A phenomenological variant of the DGP model with modified lensing}\label{sec:vari}

An important aspect of the DGP model that is very relevant to the analysis in this paper is that, in this theory of gravity, the lensing potential {is governed by the same equation as in GR, but the dynamical potential gets a contribution also from the scalar field $\varphi$.} This means that differences in the lensing signal in $\ndgp$ and $\lcdm$ cosmologies are induced by the different matter distribution (which reacts to the dynamical potential), and not because photons themselves react to a modified lensing potential. In other words, for a fixed matter source, photon geodesics are the same in the $\ndgp$ and $\lcdm$ models.

Models of gravity {with modified Poisson equations for $\Phi_{\rm len}$} provide us with a richer phenomenology to be tested by observational data. Known examples of such models include the Covariant Galileon model \cite{PhysRevD.79.064036, PhysRevD.79.084003, Deffayet:2009mn, 2012PhRvD..86l4016B, Barreira:2014jha, 2016arXiv160403487R} (and variants/generalizations thereof \cite{2011PhRvL.106t1102W, 2015PhRvD..91f4012P, 2015JCAP...10..064T}), Nonlocal gravity \cite{Deser:2007jk, Deser:2013uya, Maggiore:2014sia, Barreira:2014kra, 2016arXiv160203558D}, K-mouflage \cite{Babichev:2009ee, Brax:2014wla, Brax:2014yla, 2015PhRvD..91f3528B}, and other corners of Horndeski's general theory \cite{Horndeski:1974wa, Deffayet:2009mn} (or theories beyond it \cite{2014PhRvD..89f4046Z, 2015PhRvL.114u1101G, 2015arXiv151005964S, 2016arXiv160306368S}). Here, instead of taking one of the models listed above, we follow a more phenomenological approach and consider a toy model we call $\lens$, which has the same equations as the standard $\ndgp$ model, but with the important difference {that $\nabla^2\Phi_{\rm len} = 4\pi G a^2 \delta\rho_m + \frac{1}{2}\nabla^2\varphi$.} 

In this paper, we therefore show results for three cosmological scenarios: $\lcdm$, $\ndgp$ and $\lens$. The comparison of the results of $\lcdm$ and $\ndgp$ measures the impact of the modified matter distribution on the lensing signal. On the other hand, the matter distribution is the same in $\ndgp$ and $\lens$, and as a result, the differences in the lensing signal arise because photons also "feel" the fifth force in $\lens$.

\subsection{Screening mechanism}\label{sec:screening}

In the $\ndgp$ model (as well as in our $\lens$ variant), there is a dynamical mechanism called {\it Vainshtein screening} \cite{Vainshtein1972393, Babichev:2013usa, Koyama:2013paa} that suppresses the modifications to gravity in regions of high matter density. The existence of such a screening mechanism is what makes nDGP gravity a viable alternative to GR, as it allows the physics of the model to be within the tight bounds set by Solar System experiments \cite{Will:2014xja}. The implementation of the Vainshtein screening is triggered by the nonlinear derivative terms in the equation of the scalar field $\varphi$, Eq.~(\ref{eq:eomvarphi}). To get a quick feeling for the physics of the screening, we can work assuming spherical symmetry, in which case the fifth force sourced by the scalar field is given by $F_{5\rm th} = \varphi,_r/2$ (a comma denotes a partial derivative, in this case w.r.t.~the physical radius $r$). The gradient of $\varphi$ can be obtained by solving the quadratic algebraic equation that follows from integrating Eq.~(\ref{eq:eomvarphi}) over $r$, after which one can write
\bq\label{eq:eomvarphi_sph2}
F_{5\rm th} = \frac{\varphi,_r}{2} = \frac{2}{3\beta}\left(\frac{r}{r_V}\right)^3\left[-1 + \sqrt{1 + \left(\frac{r_V}{r}\right)^3}\right]\frac{GM(r)}{r^2}, \nonumber \\
\eq
where $M(r)$ is the mass enclosed by radius $r$ and we have defined a distance scale called the {\it Vainshtein radius}, which is given by
\bq\label{eq:rv}
r_V(r) = \left(\frac{16r_c^2GM(r)}{9\beta^2}\right)^{1/3}.
\eq
At a given $r$, $r_V$ determines the distance from the center of some matter source below which the size of the fifth force becomes small, in comparison to the standard GR force. To illustrate this, consider for simplicity a top-hat profile with size $R_{\rm TH}$ and mass $M_{\rm TH}$. In this case, if $r \gg r_V(r) > R_{\rm TH}$, then
\bq\label{eq:forcelarge}
F_{5\rm th} = \frac{\varphi,_r}{2} \approx \frac{1}{3\beta(a)}\frac{G M_{\rm TH}}{r^2} = \frac{1}{3\beta(a)}F_{\rm GR}.
\eq
The value of $\beta$ grows with redshift, which suppresses the modifications to gravity at {these early epochs}. On the other hand, at late times $\beta \sim \mathcal{O}(1)$, which yields a sizeable positive fifth force (note that $\beta > 0$). However, if $R_{\rm TH} < r \ll r_V(r)$, then 
\bq\label{eq:forcesmall}
\frac{F_{5\rm th}}{F_{GR}} \rightarrow 0, \ \ \ \ \ \ {\rm as} \ \ \ \ \ \frac{r}{r_V} \rightarrow 0,
\eq
and the effects of the fifth force become negligible, regardless of the cosmological epoch.

The detection of the additional scale dependence introduced by the screening mechanism would constitute "smoking gun-like" evidence for theories beyond GR, which strongly motivates research on these types of observational signatures. One of our goals in this paper is precisely to determine whether or not such scale dependent behaviour is noticeable in the trough lensing signal.

\section{Ray-tracing N-body simulations}\label{sec:rayt}

In this section, we describe our N-body setup to calculate the lensing signal on the fly with simulations of modified gravity.

\subsection{Numerical methods}\label{sec:nume}

\begin{figure}
	\centering
	\includegraphics[scale=0.37]{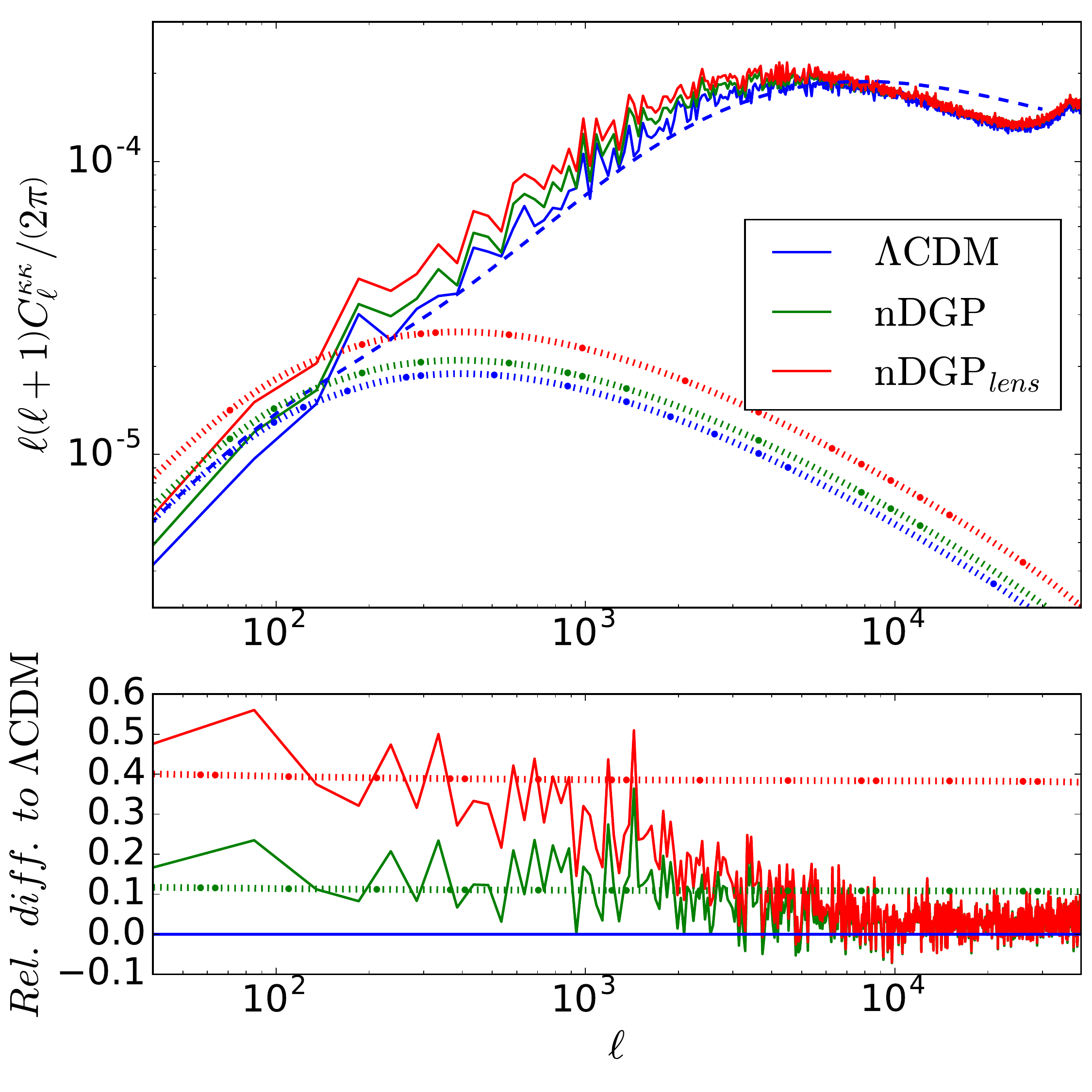}
	\caption{Lensing convergence power spectra for the $\lcdm$, $\ndgp$ and $\lens$ models, as labelled. {The dotted lines display the linear theory result (with the same color coding as the solid lines in the legend). The dashed blue line shows the $\lcdm$ result obtained with the ${\rm Halofit}$ prescription for the three-dimensional matter power spectrum.} The amplitude mismatch between our measured $\lcdm$ spectrum and the ${\rm Halofit}$ curve is due to sample variance.}
\label{fig:pk}
\end{figure}

\begin{figure*}
	\centering
	\includegraphics[scale=0.40]{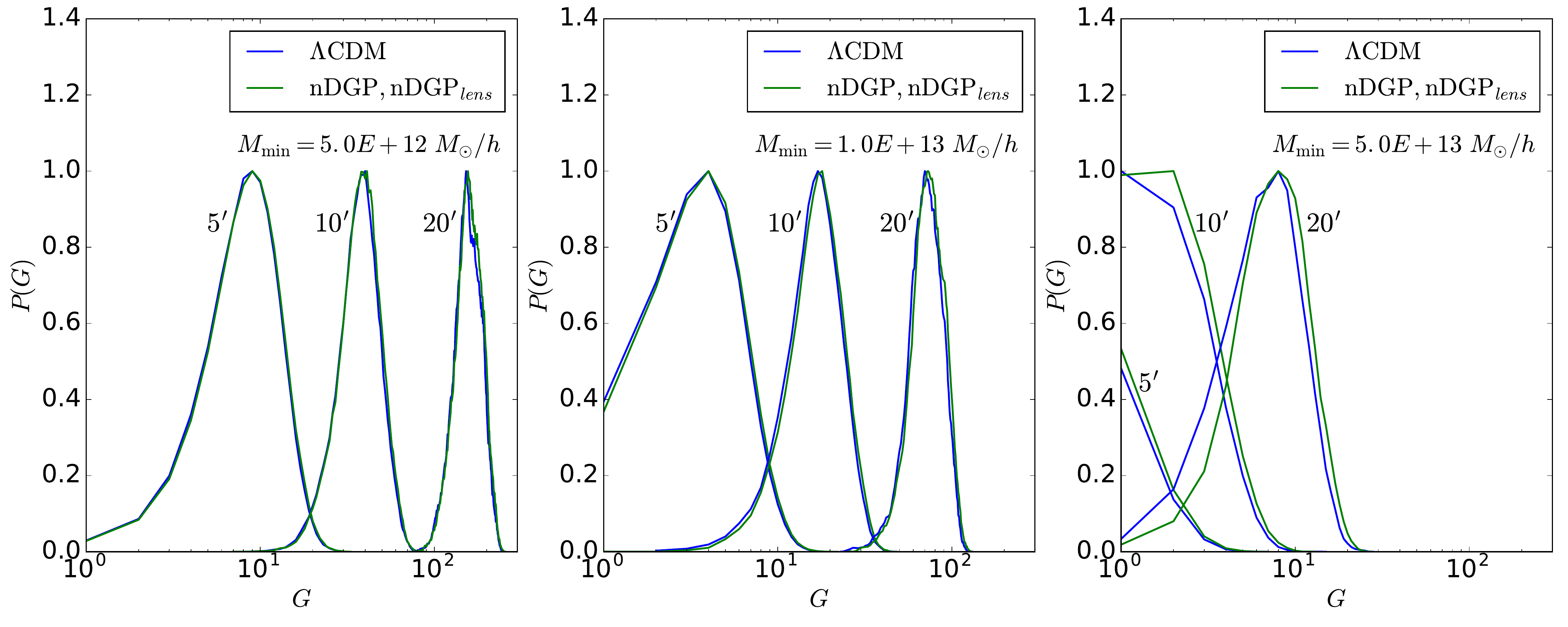}
	\caption{$G$ field distributions (cf.~Eq.~(\ref{eq:gfield})) for three $\theta_T$ and $M_{\rm min}$ values, and for the $\lcdm$ and $\ndgp$/$\lens$ models, as labelled. The halo-underdense ($\gtw$) and halo-overdense ($\gei$) LOS are associated with the lower and upper $20\%$ percentiles of these distributions. For cases when more than $20\%$ of the $G$ field pixels have $G=0$, the $\gtw$ LOS are selected randomly out of all these pixels. The distributions correspond to $z_{\rm halo} \in \left[0.1, 0.76\right]$ and they are normalized such that their maximum is unity. The $x$-axis is ${\rm log}$-scaled, which is why $G=0$ does not appear.} 
\label{fig:hist}
\end{figure*}

Our numerical results are obtained by combining the ray-tracing modules of the {\tt Ray-Ramses} code \cite{2016arXiv160102012B} with the modified gravity N-body code {\tt ECOSMOG} \cite{2012JCAP...01..051L, baojiudgp}, both being extensions of the publicly available AMR {\tt Ramses} code \cite{2002A&A...385..337T}. The {\tt ECOSMOG} code differs from {\tt Ramses} by including a number of additional routines that calculate the fifth force. These routines solve a discretized version of Eq.~(\ref{eq:eomvarphi}) via a Gauss-Seidel iterative procedure to find the value of $\varphi$ at the center of every cell of the AMR structure. The fifth force, which is proportional to $\vec{\nabla}\varphi$, is computed at the center of each cell by finite-differencing the value of $\varphi$ on neighbouring cells. The fifth force at particle positions is obtained via interpolation using a cloud-in-cell (CIC) scheme. To ensure consistency and momentum-conservation, the same CIC interpolation scheme is used to construct the density field on the grid from the particle distribution. For the lensing calculations we perform in this paper, one also needs to evaluate $\Phi_{\rm len}$ at the cell centers. For $\lcdm$ and $\ndgp$, this is given by the gravitational potential\footnote{The superscript $^{GR}$ indicates that the potential is governed by the GR Poisson equation, but note that the density field that sources it is in general different in between these two models.} $\Phi_{\rm len}^{\rm GR}$ computed by default {\tt Ramses}, whereas for the $\lens$ case we have $\Phi_{\rm len} = \Phi_{\rm len}^{\rm GR} + \varphi/2$. We refer the interested reader to Ref.~\cite{baojiudgp} for more details about the application of {\tt ECOSMOG} in simulations of DGP gravity, and to Ref.~\cite{codecomp} for a comparison project of modified gravity N-body codes.

The {\tt Ray-Ramses} code consists of an {\it add-on} extension of {\tt Ramses} that computes projected cosmological observables by integrating some relevant quantity (potential, density, etc.) along ray trajectories in the simulation. {For instance, to calculate the lensing convergence $\kappa$}, the quantity that is integrated is the two-dimensional Laplacian of the lensing potential, ${\nabla}^2_{2D}\Phi_{\rm len} = \nabla_1\nabla^1\Phi_{\rm len} + \nabla_2\nabla^2\Phi_{\rm len}$ (where $1,2$ denote the two directions on the sky perpendicular to the LOS). More specifically, in this paper we have {\tt Ray-Ramses} evaluating the integral
\bq\label{eq:kappa}
\kappa = \frac{1}{c^2}\int_0^{\chi_s}\frac{\chi\left(\chi_s - \chi\right)}{\chi_s} {\nabla}^2_{2D}\Phi_{\rm len} {\rm d}\chi,
\eq
where $c$ is the speed of light, $\chi$ is the comoving distance and $\chi_s$ is the comoving distance to the lensing sources. {The calculation of the two components $\gamma_1$ and $\gamma_2$ of the lensing shear, $\gamma = \gamma_1 + i\gamma_2$, is analogous to that of $\kappa$, but with ${\nabla}^2_{2D}\Phi_{\rm len}$ replaced by $\nabla_1\nabla^1\Phi_{\rm len} - \nabla_2\nabla^2\Phi_{\rm len}$ for $\gamma_1$ and by $2\nabla_1\nabla^2\Phi_{\rm len}$ for $\gamma_2$ in Eq.~(\ref{eq:kappa}). The value of these two-dimensional derivatives of $\Phi_{\rm len}$} can be obtained from the values of $\Phi_{\rm len}$ at the center of the AMR cells via finite-differencing and some geometrical considerations (see Refs.~\cite{li2001, 2016arXiv160102012B}). The above integral is split into the contribution from each AMR cell that is crossed by a ray, which ensures that the ray integrations take full advantage of the (time and spatial) resolution attained by the N-body run. For the weak lensing signal we wish to study in this paper, we can employ the Born approximation, in which the lensing signal is accumulated along unperturbed ray trajectories. Moreover, since {\tt Ray-Ramses} can run on-the-fly with the simulation, the lensing maps are readily available once the N-body run is done. This spares the user from having to output the density distribution several times during the N-body run to later compute the lensing signal at post-processing.

For the case of lensing studies of the $\lens$ model, there is one interesting advantage of {\tt Ray-Ramses} over the more conventional lensing ray-tracing methods that rely on the multiple plane lens approximation. In the latter methods, the lensing signal is calculated on a series of two-dimensional planes onto which the three-dimensional particle distribution is projected (see e.g.~Refs.~\cite{2000ApJ...530..547J, 2003ApJ...592..699V, Hilbert:2008kb, 2015arXiv151108211G}). A central assumption of this method is that the superposition principle holds, i.e., the lensing signal associated with the "chunk" of the density field used to construct the plane is the same as the lensing signal computed at the plane location. The superposition principle holds for GR and $\ndgp$ because the Poisson equation that governs $\Phi_{\rm len}$ is linear in $\delta\rho_m$. This is, however, not the case in $\lens$ because, in this model, $\nabla^2\Phi_{\rm len}$ depends nonlinearly on the density via $\nabla^2\varphi$ (cf.~Eqs.~(\ref{eq:modpoisson}) and (\ref{eq:eomvarphi})). This is a subtle (but important) point that is not very often highlighted in the literature \cite{2015MNRAS.454.4085B}. This may be due to the fact that the most thoroughly modified gravity models in the nonlinear regime (such as the likes of $f(R)$ \cite{2010RvMP...82..451S} and DGP) do not modify $\Phi_{\rm len}$ directly, and hence, conventional multiple plane lens methods are straightforwardly applicable (see e.g.~Ref.~\cite{2015JCAP...10..036T, 2016MNRAS.459.2762H}). {For models like $\lens$, on the other hand, these conventional methods need forcibly to be generalized to account for the nonlinearity of the equations. In the particular case of the model studied here, one possible first step in that direction could be to design ways to project the term $\nabla^2\varphi$, which enters linearly in Eq.~(\ref{eq:modpoisson}). This would imply storing also the scalar field distribution during the N-body run to be used at post-processing, which would only aggravate the rather large data storage requirements of these methods. These issues are not reason for concern with the {\tt Ray-Ramses} code, since its calculations do not rely on plane projections or the need to store any given quantity for post-processing.}

Except for the additional modified gravity solver, the structure of {\tt ECOSMOG} remains otherwise the same as standard {\tt Ramses}. The details of the operation and implementation of the {\tt Ray-Ramses} routines in {\tt ECOSMOG} are therefore in all similar to its implementation in {\tt Ramses}, which is explained with detail in Ref.~\cite{2016arXiv160102012B}.

\subsection{Lensing setup}\label{sec:tili}

\begin{figure}
	\centering
	\includegraphics[scale=0.76]{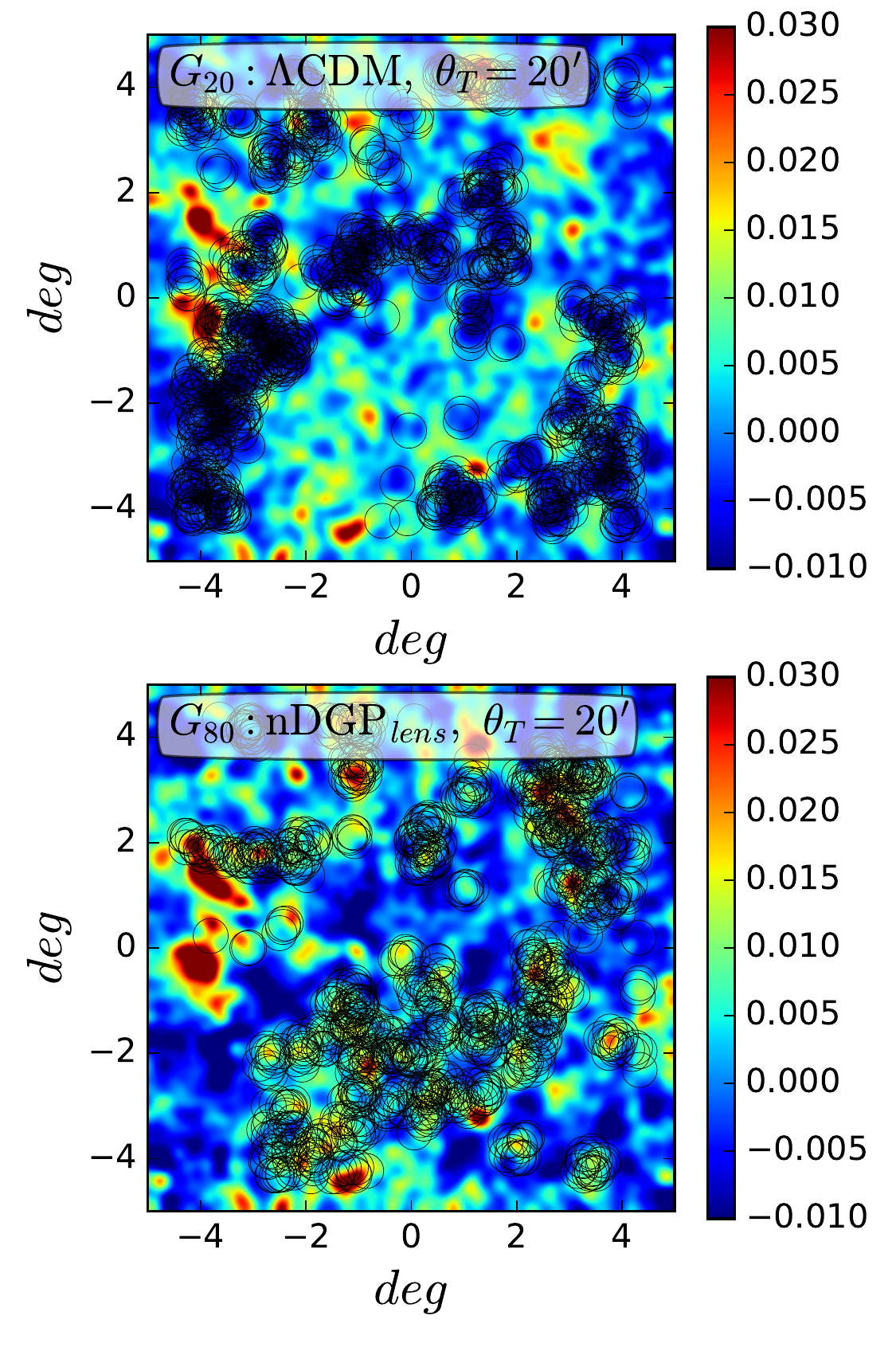}
	\caption{Location of the $\gtw$ and $\gei$ LOS with $\theta_T = 20'$, $M_{\rm min} = 5.0\times10^{12}\msun$ and $z_{\rm halo} \in \left[0.1, 0.76\right]$ on the lensing convergence maps. The upper panel corresponds to $\lcdm$ and $\gtw$, while the lower panel corresponds to the $\lens$ model and $\gei$. Note that $\gtw$ ($\gei$) LOS are predominantly on top of negative (positive) $\kappa$ regions. The $\kappa$ maps have been smoothed with a Gaussian filter with size $7'$. To facilitate visualization, we only show one every fifty $\gtw$ and $\gei$ LOS.}
\label{fig:onmap}
\end{figure}

The setup used in our analysis corresponds to a lensing lightcone that extends out to a source redshift of $z_s = 1$ with a field of view (FOV) of $10\times10\ {\rm deg}^2$. The lightcone geometry is illustrated by the solid red line in Fig.~\ref{fig:tile} and we consider $2048\times2048$ rays. The solid black lines depict the simulation boxes that we tiled to encompass the lensing lightcone. From the observer to the sources, the tile is made up of five $L = 300{\rm Mpc}/h$ boxes (called boxes 1 to 5, respectively), one $L = 350{\rm Mpc}/h$ box (box 6) and one $L = 450{\rm Mpc}/h$ box (box 7) that contains the source plane. For all simulation boxes, the N-body tracer particle number is $N_p = 512^3$ and the AMR grid refinement criterion is taken to be 8 for all AMR levels. The Gauss-Seidel iterations in the $\ndgp$ and $\lens$ simulations are only performed on the domain level of the AMR structure, with the value of $\varphi$ on finer levels being obtained via interpolation from the solution on the domain level. Reference \cite{2015JCAP...12..059B} has shown that for similar N-body resolution setups this approach leads to a substantial boost in the code performance with negligible sacrifice in accuracy. 

The initial conditions were generated at $z = 49$ for the following cosmological parameters
\bq\label{eq:cosmoparams}
&&\left\{\Omega_{b0}, \Omega_{c0}, h, n_s, \sigma_8^{\Lambda{\rm CDM}} \right\} = \\ \nonumber
&&\left\{0.049, 0.267 , 0.6711, 0.9624, 0.8344\right\},
\eq
where $\Omega_{b0}, \Omega_{c0}, h, n_s, \sigma_8^{\Lambda{\rm CDM}}$ are, respectively, the present-day fractional baryon density, the present-day fractional dark matter density, the dimensionless Hubble rate today, the primordial scalar spectral index and the root mean squared fluctation of the density field on $8\ {\rm Mpc}/h$ scales in $\lcdm$. At $z = 49$, all our cosmologies are indistinguishable, and hence the initial conditions can be generated assuming $\lcdm$ (we have used the {\tt MPGRAFIC} code \cite{2008ApJS..178..179P}). {In Ref.~\cite{2016arXiv160503965B}, the authors constrained $r_cH_0 \lesssim 1\ (2\sigma)$ using measurements of the growth rate of structure. Here, for the simulations of the $\ndgp$ variants we take $r_cH_0 = 1$, which results from a compromise between having sizeable fifth force effects, while remaining in acceptable regions of the parameter space.} Note also that by virtue of the boosted growth of structure in the $\ndgp$ and $\lens$ models, their values of $\sigma_8$ at $z = 0$ are larger than the corresponding one in $\lcdm$. {The primary goal of this paper is to determine the types (and estimate the size) of the signatures that modified gravity effects can inprint on the lensing signal around over- and underdense LOS. For this, it is sufficient to have the simulations of all models starting from the same initial conditions and cosmological parameters. This allows to better single out the effects that are intrinsic to the presence of the fifth force, from those that would be induced by changes in the cosmological parameters. Naturally, a formal comparison between theory and observations should involve an exploration of other regions of parameter space, but this is beyond the scope of the present work.} In the {\tt Ray-Ramses} code, every box in the tile takes as input its relative position w.r.t.~the observer, which is used to determine the redshift interval during which each box should initialize the rays on one face of the box and integrate them until they reach the other face. Once each box finishes its integration, then the simulations can be stopped. The initial conditions for different boxes were generated with different random seeds to avoid repetition of structures along the LOS. The simulations of $\lcdm$, $\ndgp$ and $\lens$ for the same box evolve from the exact same initial conditions though. At the end, the total lensing signal of the tile is obtained by simply adding the contribution from each of the boxes. 

{Figure \ref{fig:pk} shows the power spectrum of the $\kappa$ maps obtained from the $\lcdm$, $\ndgp$ and $\lens$ tiles (solid lines, as labelled). The figure shows also the corresponding result from linear theory (dotted lines), which is given by
\bq\label{eq:analy}
C_{\ell}^{\kappa\kappa} = \frac{9\Omega_{m0}^2H_0^4}{4c^4\chi_s^2}\int_0^{\chi_s} \left(\chi_s - \chi\right)^2 \mathcal{G}_{\rm eff}^2(\chi) \frac{P_{\rm lin}(k=\frac{\ell}{\chi}, \chi)}{a(\chi)^2} {\rm d}\chi, \nonumber \\
\eq
where $\Omega_{m0} = \Omega_{c0} + \Omega_{b0}$ and $\mathcal{G}_{\rm eff}$ is an effective gravitational strength for lensing (to be distinguished from the same quantity for dynamics $G_{\rm eff}$). For $\lcdm$, $\mathcal{G}_{\rm eff} = G$ and the linear matter power spectrum is that associated with the parameters of Eq.~(\ref{eq:cosmoparams}), $P_{\rm lin} = P_{\rm lin}^{\lcdm}$. For $\ndgp$, $\mathcal{G}_{\rm eff} = G$, and $P_{\rm lin}^{\ndgp} = \left(D^{\ndgp}/D^{\lcdm}\right)^2P_{\rm lin}^{\lcdm}$, where $D^{\rm model} \equiv D^{\rm model}(\chi)$ is the growth factor of linear density perturbations for a given model\footnote{{In DGP gravity cosmologies, the linear growth factor is governed by the equation $\ddot{D} + 2H\dot{D} - 4\pi G_{\rm eff}\bar{\rho}_m D = 0$, where $G_{\rm eff}$ is the effective gravitational strength for dynamics (not lensing) in the linear regime.}}. Finally, in the case of the $\lens$ model, the linear matter power spectrum is the same as in the $\ndgp$ case, but 
\bq\label{eq:geff}
\mathcal{G}_{\rm eff}(\chi) = G_{\rm eff}(\chi) = \left(1 + \frac{1}{3\beta(a(\chi))}\right)G,
\eq
which captures the additional modifications to the lensing potential in this model ($G_{\rm eff}$ can be derived from Eq.~(\ref{eq:forcelarge})). Shown also in Fig.~\ref{fig:pk} is the $\lcdm$ result\footnote{{Computed with the {\tt CAMB-Sources} software (http://camb.info/sources/).}} obtained by using the semi-analytical ${\rm Halofit}$ formula for the nonlinear matter power spectrum \cite{2003MNRAS.341.1311S, 2012ApJ...761..152T}, instead of $P_{\rm lin}^{\lcdm}$ in Eq.~(\ref{eq:analy}). } 

\begin{figure}
	\centering
	\includegraphics[scale=0.76]{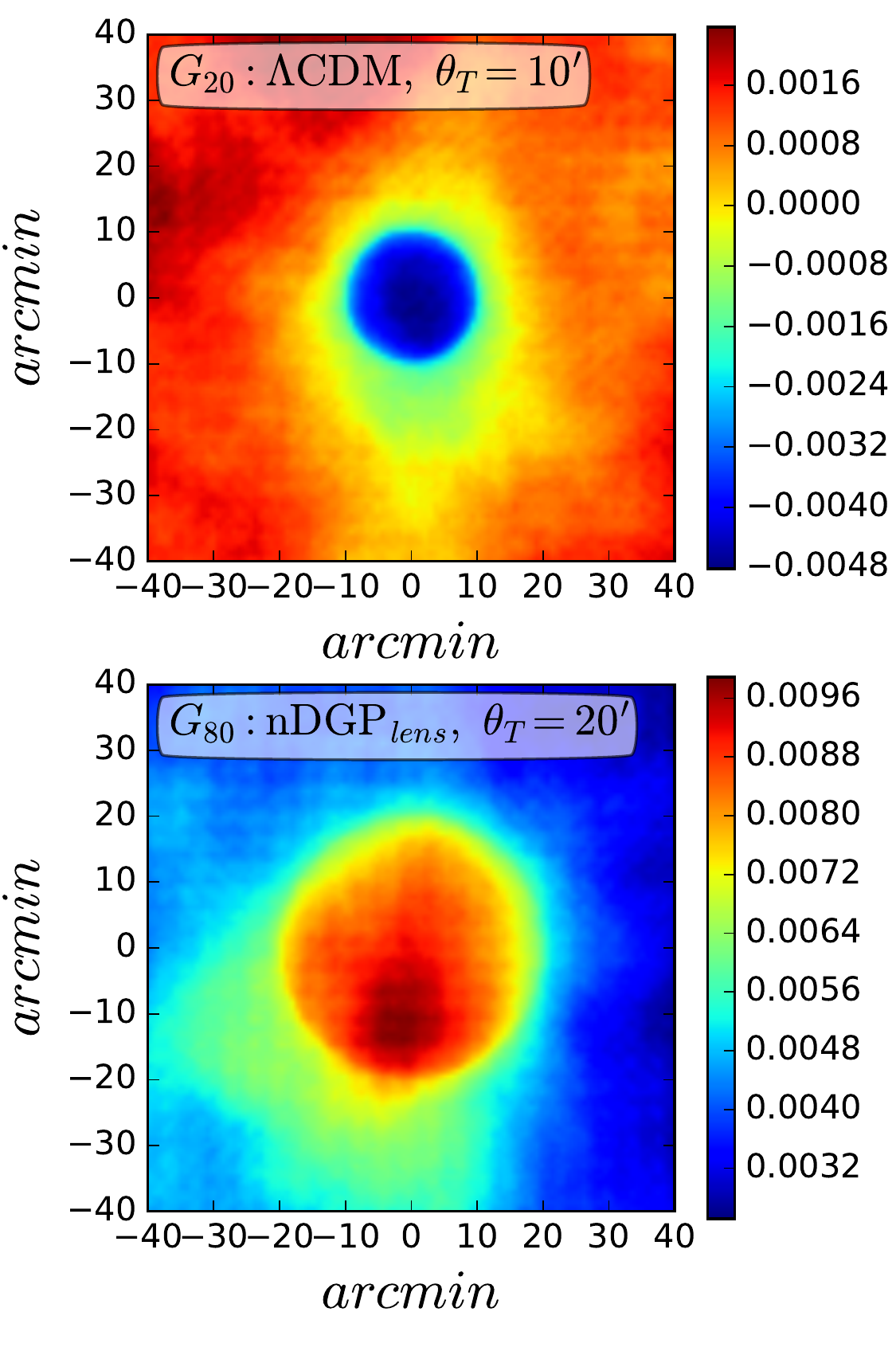}
	\caption{Lensing $\kappa$ maps stacked on $\gtw$ (upper panel, for $\lcdm$ and $\theta_T = 10'$) and $\gei$ (lower panel, for $\lens$ and $\theta_T=20'$) LOS. The color scale in the different panels is not the same to facilitate the visualization. These maps correspond to $M_{\rm min} = 10^{13} \msun$ and $z_{\rm halo} \in \left[0.1, 0.76\right]$.}
\label{fig:stacked}
\end{figure}

One notes that our $\lcdm$ spectrum has as higher amplitude ($\approx 10\%$) than ${\rm Halofit}$. This is attributed to sample variance, i.e.,  our realization of the initial conditions is such that our $10\times10\ {\rm deg}^2$ FOV happens to be "pointing" to a region of predominantly higher matter clustering (see e.g.~the discussion in Ref.~\cite{2008MNRAS.391..435F} or Fig.~9 of Ref.~\cite{2016arXiv160102012B} for a measure of the expected spread due to sample variance).  The comparison of the shape of the measured spectra with ${\rm Halofit}$ can, however, be useful in assessing the resolution attained by our weak-lensing simulations. The upper panel of Fig.~\ref{fig:pk} shows that the absolute power spectra measurements start to lose resolution for $\ell \gtrsim 3000$. In terms of the relative difference to $\lcdm$, the curves show the expected behavior that, on large angular scales {($\ell \lesssim 10^3$)}, the amplitude is higher in the modified gravity models. In the case of the $\ndgp$ model, this is because the density field is more evolved, which amplifies the lensing signal. In the case of the $\lens$ model, on top of the boost in structure formation, there is also the fact that photons in this model are also directly affected by the positive fifth force. {On the scales where linear theory holds ($\ell \lesssim 10^2$), there is also good agreement between the ray-tracing simulation results and the linear expectation, which serves as a successful sanity check of our ray-tracing results. {Note that although sample variance may affect the absolute shape of the power spectra, its impact on the size of the differences induced by modified gravity largely cancel out.} On scales $\ell \gtrsim 10^3$, the curves of the modified gravity models approach that of $\lcdm$, which reflects the operation of the screening mechanism. One should bear in mind that the absolute value of the power spectra becomes affected by the lack of resolution for $\ell \gtrsim 3000$, although it is reasonable to expect that this is less critical when analyzing the relative difference to $\lcdm$.}

\section{Selection of halo-underdense and halo-overdense LOS}\label{sec:defi}

\begin{figure*}
	\centering
	\includegraphics[scale=0.385]{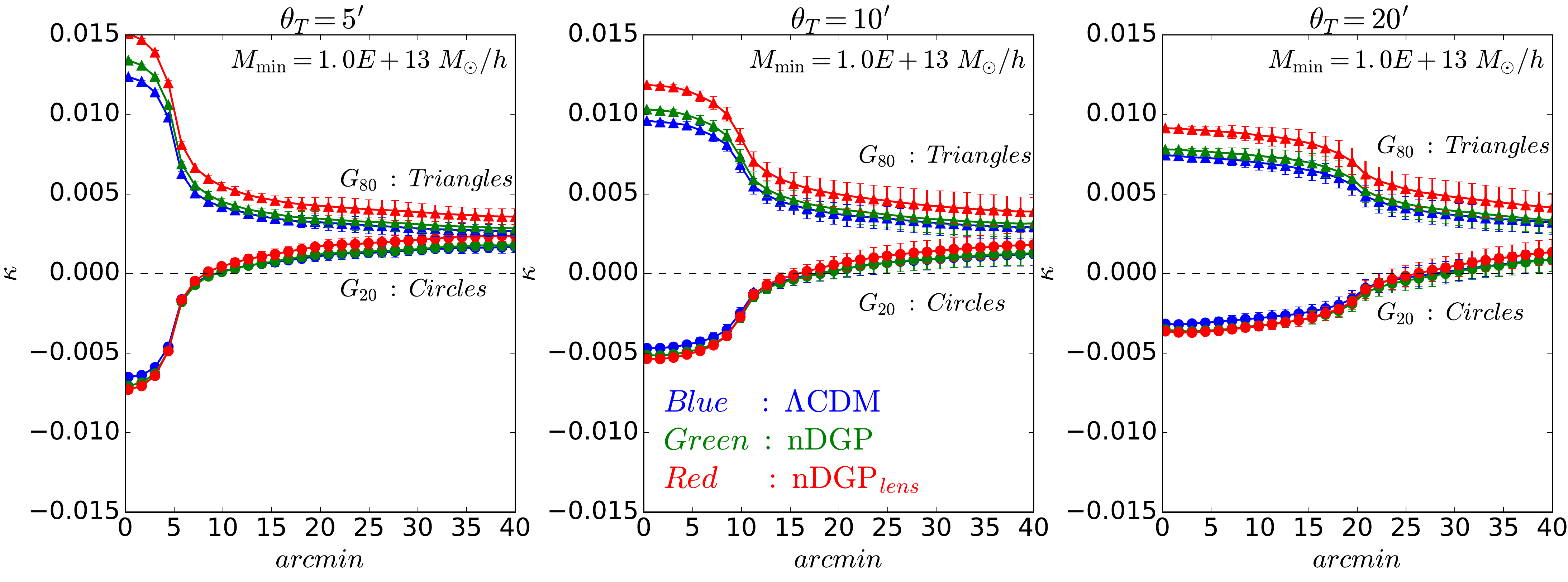}
	\caption{Spherically averaged stacked $\kappa$ profiles around $\gtw$ and $\gei$ LOS, for $\theta_T = 5', 10', 20'$ and for the $\lcdm$, $\ndgp$ and $\lens$ models, as labelled. These profiles correspond to $M_{\rm min} = 10^{13} \msun$ and $z_{\rm halo} \in \left[0.1, 0.76\right]$.}
\label{fig:abs}
\end{figure*}

In this section, we describe the construction of the halo catalogues we use and the procedure to identify the desired underdense and overdense LOS.

\subsection{"Pseudo" halo lightcones}\label{sec:halos}

The black dots in Fig.~\ref{fig:tile} indicate the positions of dark matter haloes found in the simulations of the $\ndgp$ and $\lens$ models. These "pseudo" halo lightcones were constructed as follows. The time at which the rays have travelled a quarter and three quarters of their total trajectory inside each box (vertical dotted lines) is marked by the redshift values displayed below the boxes\footnote{To be precise, this statement holds exactly only for the central ray of the light bundle because a surface of constant redshift in the lightcone does not have the same $z_{\rm coord}$.}. At these times, the box outputs a snapshot of the density field, which we use to identify dark matter haloes using the {\tt Rockstar} code \cite{2013ApJ...762..109B}. We then "split" the boxes in two halves along the LOS direction (green and yellow colors in Fig.~\ref{fig:tile}), and for each of the two snapshots, we consider only those haloes that lie in the half of the box that is closer in redshift to the output redshift of the snapshot. For concreteness, take box 2 as an example. From the snapshot at $z=0.13$, the lightcone contains only those haloes whose $z$-coordinate w.r.t.~the observer is smaller than $450{\rm Mpc}/h$ (green region); from the snapshot at $z = 0.18$ we consider the haloes that lie in the other half of the box (yellow region). Naturally, for all boxes we only consider the haloes that lie within the FOV. We use the word "pseudo" to highlight that, although the haloes are continously distributed along the LOS, their dynamical state and position was recorded only at a finite number of redshifts. In Appendix \ref{sec:app1}, we show that this is an approximation that has little impact on our results and conclusions.

Recall also that the dynamical potential is the same in the two $\ndgp$ variants, which is why they share the same halo catalogues. The two panels in Fig.~\ref{fig:tile} also show the catalogues for two different halo mass cutoffs to visualize how much sparser the halo distribution becomes when increasing minimum halo mass (a point to which we shall return below when we discuss our results).

\subsection{LOS selection: the $G$ field}\label{sec:gfield}

\begin{figure*}
	\centering
	\includegraphics[scale=0.43]{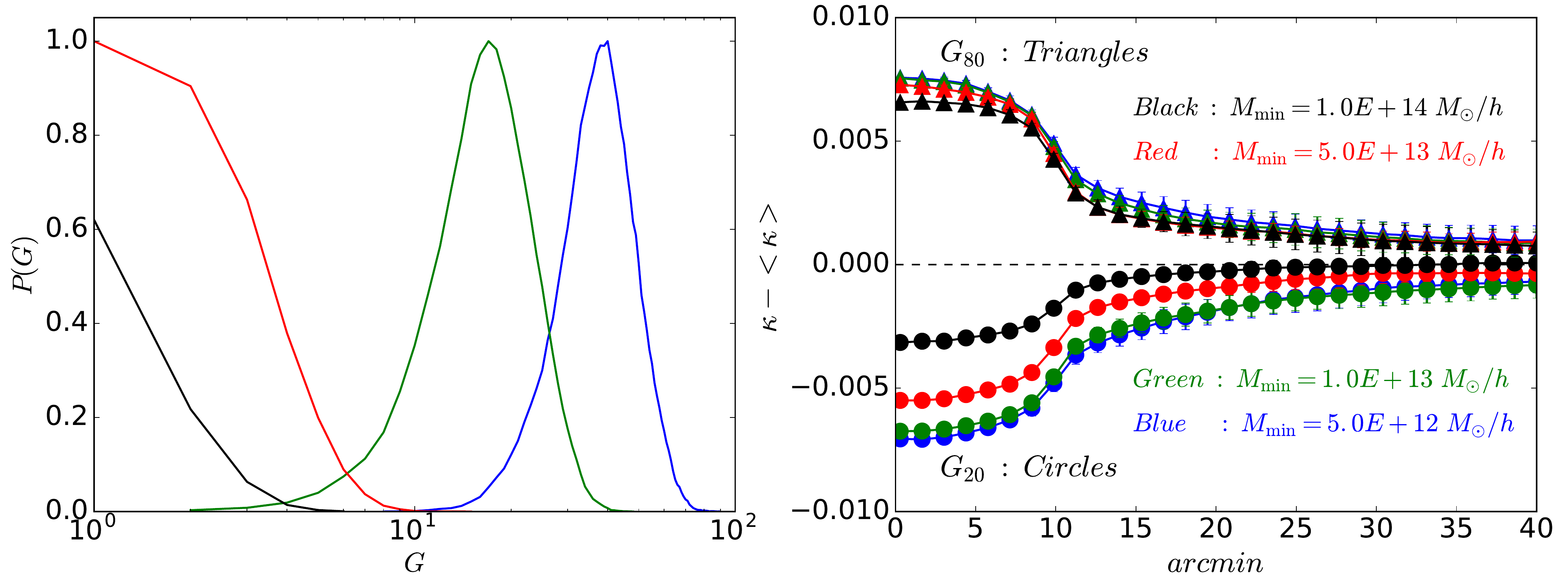}
	\caption{$G$ field distribution (left) and spherically averaged $\kappa - \langle\kappa\rangle$ profiles around $\gtw$ and $\gei$ LOS (right) for varying minimum halo mas cutoffs $M_{\rm min}$, as labelled. The result corresponds to $\theta_T=10'$ and $z_{\rm halo} \in \left[0.1, 0.76\right]$, for the $\lcdm$ model. The distributions are normalized such that their maximum is unity.  The $x$-axis in the left panel is ${\rm log}$-scaled, which is why $G=0$ does not appear.}
\label{fig:mmin1}
\end{figure*}

\begin{figure}
	\centering
	\includegraphics[scale=0.76]{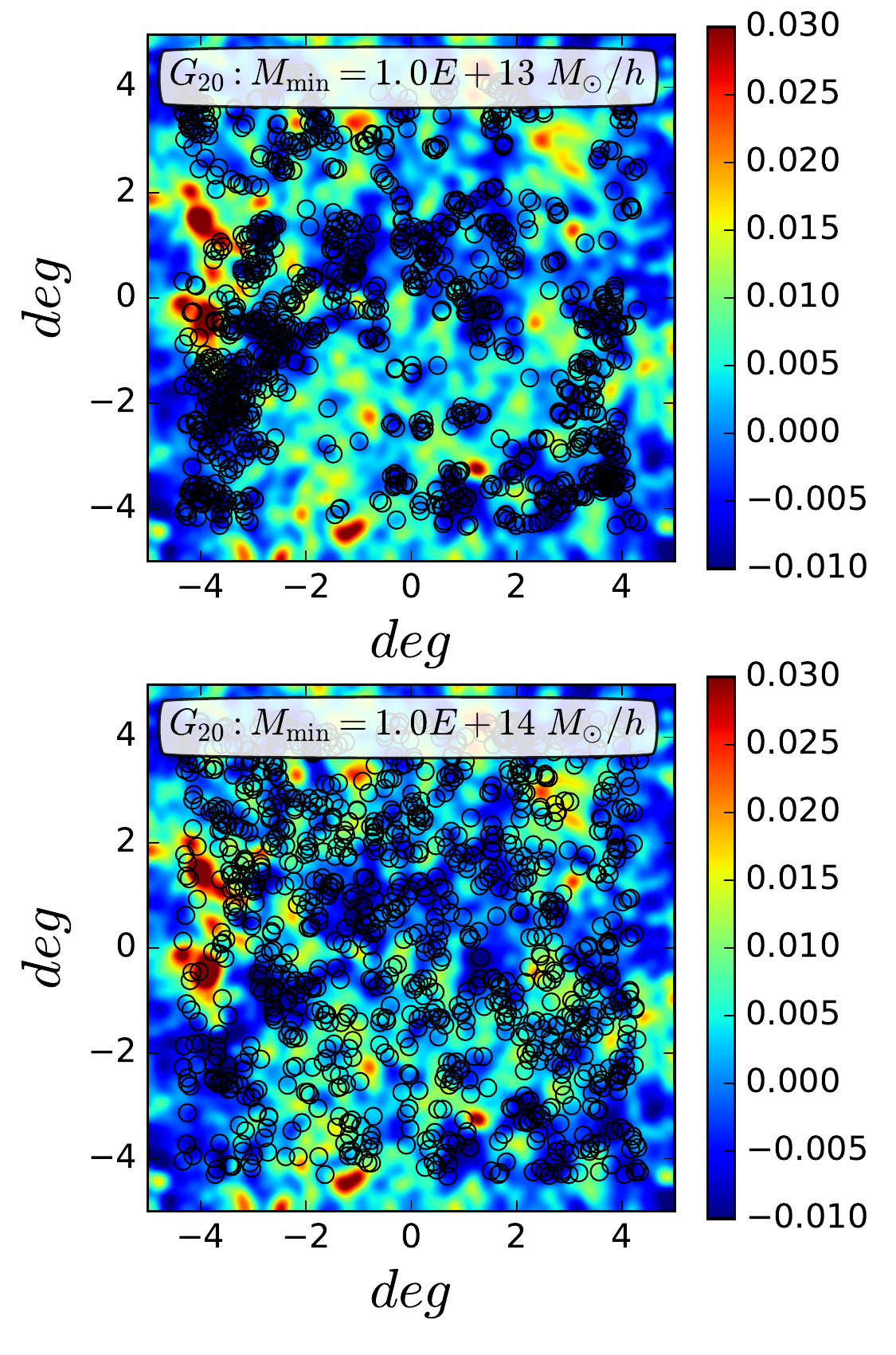}
	\caption{Location of the $\gtw$ LOS on top of the lensing $\kappa$ maps for $M_{\rm min} = 10^{13} \msun$ (top) and $M_{\rm min} = 10^{14}\msun$ (bottom). The result corresponds to $\theta_T=10'$ and $z_{\rm halo} \in \left[0.1, 0.76\right]$, for the $\lcdm$ model as in Fig.~\ref{fig:mmin1}. Note how the $\gtw$ LOS trace better regions of negative $\kappa$ if $M_{\rm min}$ is smaller. To facilitate visualization, we only show one every fifty $\gtw$ LOS.}
\label{fig:mmin2}
\end{figure}

Our trough selection procedure is the same as that employed in the DES observational analysis presented in Ref.~\cite{2016MNRAS.455.3367G}. The first step is to define the projected halo density field $G$, which is constructed from the halo lightcone as
\bq\label{eq:gfield}
G(\vec{\theta}) = \sum_{i=1}^{N_{\rm halo}} W_{\rm sel, i}\left(\theta_T, z_l, z_u, M_{\rm min}\right),
\eq
where the selection function is given by
\[  W_{\rm sel, i} =   \left\{
\begin{array}{ll}
      1, & |\vec{\theta} - \vec{\theta_i}|\leq \theta_T, z_i \in [z_l, z_u], M_i \geq M_{\rm min}\\
      0, & {\rm otherwise} \\
\end{array}.
\right. \]
Here, $i$ runs over all $N_{\rm halo}$ haloes, $\vec{\theta}$ is the two-dimensional coordinate on the FOV, $\theta_T$ is an angular radial size, $M_{\rm min}$ is a halo mass cutoff\footnote{In this paper, whenever we refer to halo mass, we shall be referring to $M_{200}$, i.e., the mass enclosed by a sphere of radius $R_{200}$, which is defined as the radial distance from the halo center within which the mean density is $200$ times the critical density of the Universe.} and $z_l$ and $z_u$ are the lower and upper halo redshift values. The redshift of a halo\footnote{Not to be confused with the output times of the simulation boxes.} $z_{\rm halo}$ is defined as $\chi(z_{\rm halo}) = d_{\rm halo-obs}$, where recall $\chi(z)$ is the cosmological comoving distance at $z$ and $d_{\rm halo-obs}$ is the distance between that halo and the observer in the tiling geometry of Fig.~\ref{fig:tile}. We evaluate the $G$ field on a $N^G_{\rm grid}\times N^G_{\rm grid}$ grid that covers the whole FOV. We work with $N^G_{\rm grid} = 512$, but in Appendix \ref{sec:app1}, we show that our results are robust to this choice of the $G$ field resolution. At every point $\vec{\theta}$ on this grid, the value of $G$ therefore corresponds to the number of haloes with mass higher that $M_{\rm min}$ and redshift $z\in\left[z_l, z_u\right]$ whose position on the FOV lies within $\theta_T$ from $\vec{\theta}$. 

The distribution of the values of the $G$ field across the FOV are shown in Fig.~\ref{fig:hist} for three values of $\theta_T$ and $M_{\rm min}$, for the $\lcdm$, $\ndgp$ and $\lens$ cosmologies, as labelled. In this figure, $z_{\rm halo}\in\left[0.1, 0.76\right]$, which corresponds to using the haloes in boxes 2 to 6 in Fig.~\ref{fig:tile}. As one would expect, for fixed $\theta_T$ and decreasing $M_{\rm min}$, the distributions shift to higher halo counts because the halo number density increases. Also as expected, for fixed $M_{\rm min}$, the number of haloes inside an aperture $\theta_T$ increases, if $\theta_T$ increases. The distributions in $\lcdm$ and $\ndgp$ are almost indistinguishable except for cases of higher mass cutoff $M_{\rm min}$ (cf.~right panel of Fig.~\ref{fig:hist}), for which the distributions in the modified gravity models are slightly shifted torwards higher halo counts. This reflects the known fact that in $\ndgp$ cosmologies, the positive fifth force contributes to a boost in the abundance of massive haloes. The interested reader can check the halo mass function shown in the top left panel of Fig.~6 of Ref.~\cite{2015JCAP...12..059B}, which corresponds to the same value of $r_cH_0 = 1$ analysed here.

We follow Ref.~\cite{2016MNRAS.455.3367G} and identify as trough centers the pixels of the $G$ field that correspond to the lower $20\%$ percentile of the distribution. We call this set of points $G_{20}$. In the same way that troughs are identified as predominantly halo-underdense LOS, we can also define predominantly halo-overdense LOS as the set of points of the $G$ field that lie above the $80\%$ percentile of the distribution. We call this set of points $G_{80}$. Figure \ref{fig:onmap} shows the location of the $\gtw$ (upper panel) and $\gei$ (lower panel) LOS overlaid with smoothed $\kappa$ maps of the $\lcdm$ (upper panel) and $\lens$ (lower panel) simulations, for $\theta_T = 20'$, $M_{\rm min} = 5.0\times10^{12}\msun$ and $z_{\rm halo} \in \left[0.1, 0.76\right]$. The two convergence maps shown are not exactly the same because the theories of gravity are different, but high and low $\kappa$ regions are correlated because the simulations evolved from the same initial phases. Note that $\gtw$ LOS tend to trace regions where $\kappa < 0$, and vice-versa for $\gei$ LOS. One of the main motivations for this paper is precisely to determine whether or not the effects of the fifth force on the lensing signal around $G_{20}$ and $G_{80}$ LOS is different. We point out that the $\gei$ LOS do not necessarily overlap with all the $\kappa$ peaks in the FOV. This is because these peaks can be caused by one or a few very massive haloes, and hence, the pixels of the $G$ field associated with them do not necessarily make it to the upper $20\%$ percentile. {Note also that there can be substantial overlap between different $\gtw$ or $\gei$ LOS, meaning that not all the selected LOS are independent \cite{2016MNRAS.455.3367G}.}

There is a subtle point about the identification of the $\gtw$ LOS for those cases where the $G$ distribution is still significant at $G=0$. In Fig.~\ref{fig:hist}, this is very noticeable for $\theta_T = 5', 10'$ when $M_{\rm min} = 5.0\times10^{13} \msun$ (right panel, and note that $G=0$ does not appear on the $x$-axis because of the ${\rm log}$ scale). In these particular situations, it is the case that more than $20\%$ of the pixels have $G=0$, in which case the definition of the $20\%$ most halo-underdense LOS becomes ill-defined. Whenever this is the case, we randomly select from all pixels with $G=0$ a number of them that makes up for $20\%$ of the distribution. This is an important aspect of our $\gtw$ identification procedure that one should bear in mind when interpreting our results. We shall return to this discussion in Sec.~\ref{sec:mmin}.

\section{Lensing profiles around $\gtw$ and $\gei$ LOS}\label{sec:prof}

In this section, we display our main results from the analysis of the lensing signal around $\gtw$ and $\gei$ LOS. We analyse the spherically averaged profiles of the $\kappa$ lensing map stacked on $\gtw$ and $\gei$ LOS. Below, we first outline the construction of the stacked $\kappa$ profiles and then discuss in turn the impact of $M_{\rm min}$, the impact of the modifications to gravity in the $\ndgp$ and $\lens$ models, the impact of $z_u$ and $z_l$, and the impact of the choice of the percentiles of the $G$ field distribution on our results.

\subsection{Stacked lensing maps}\label{sec:stac}

Figure \ref{fig:stacked} shows the resulting maps obtained by stacking the $\kappa$ maps around $\gtw$ and $\gei$ LOS. For brevity, we only show the maps for $\gtw$ in $\lcdm$ and $\gei$ in $\lens$. These maps are obtained as follows. For each identified $\gtw$ or $\gei$ LOS on the sky, we interpolate $\kappa$ onto a grid with size $100\times100$ that spans a $80\times80\ {\rm arcmin}^2$ FOV centred on the $\gtw$ and $\gei$ LOS. The stacked map corresponds to the average signal over all $\gtw$ and $\gei$ LOS. The spherically averaged profiles that are analysed in this paper are obtained from the stacked $\kappa$ maps as follows. For each radius, we evaluate via interpolation the value of the stacked $\kappa$ map on $40$ points {(we have checked that our conclusions are insensitive to the exact choice of this number)} uniformly distributed along a ring with that radius centred at the center of the stack. Our stacked profiles correspond to the mean value of these $40$ sampled points and the errorbars show the standard deviation around this mean.

The spherically averaged profiles of the maps of Fig.~\ref{fig:stacked} are shown in Fig.~\ref{fig:abs} (together with the corresponding $\gei$ result in $\lcdm$, $\gtw$ in $\lens$, as well as for the $\ndgp$ model and other $\theta_T$ values). For all cases shown, on scales smaller than $\theta_T$,  the $\gtw$ ($\gei$) profiles exhibit a suppression (boost) of the signal w.r.t.~the value at larger radii. The $\gtw$ and $\gei$ signal gets more pronounced with decreasing $\theta_T$, which is in accordance with what is found in the observational DES paper \cite{2016MNRAS.455.3367G}. At larger radii, all curves approach a constant value that is larger than zero, which translates the fact that the FOV is pointed towards a region of high projected density. This is the same reason why the $\kappa$ power spectrum of the $\lcdm$ tile has a higher amplitude than the ${\rm Halofit}$ prediction in Fig.~\ref{fig:pk}. In this paper, we shall be more interested in analysing the suppression (boost) of the lensing signal around $\gtw$ ($\gei$) LOS w.r.t.~the mean in the FOV. For this reason, in the rest of the paper we display the profiles of $\kappa - \langle\kappa\rangle$, where $\langle\kappa\rangle$ is the mean value of $\kappa$ across all the $2048\times2048$ pixels that span our FOV. For the $\lcdm$, $\ndgp$ and $\lens$ maps one has $\langle\kappa\rangle = 0.00205$, $\langle\kappa\rangle = 0.00211$ and $\langle\kappa\rangle = 0.00276$, respectively. {As in the case of the convergence power spectrum results of Fig.~\ref{fig:pk}, we expect that the size of the effects caused by modified gravity on $\kappa - \langle\kappa\rangle$ should not be noticeably affected by sample variance, even if the exact shape of the absolute profiles may be specific to our one realization of the FOV (this expectation is addressed further in Appendix \ref{sec:app2}).} 

\subsection{The impact of minimum halo mass, $M_{\rm min}$}\label{sec:mmin}

\begin{figure*}
	\centering
	\includegraphics[scale=0.475]{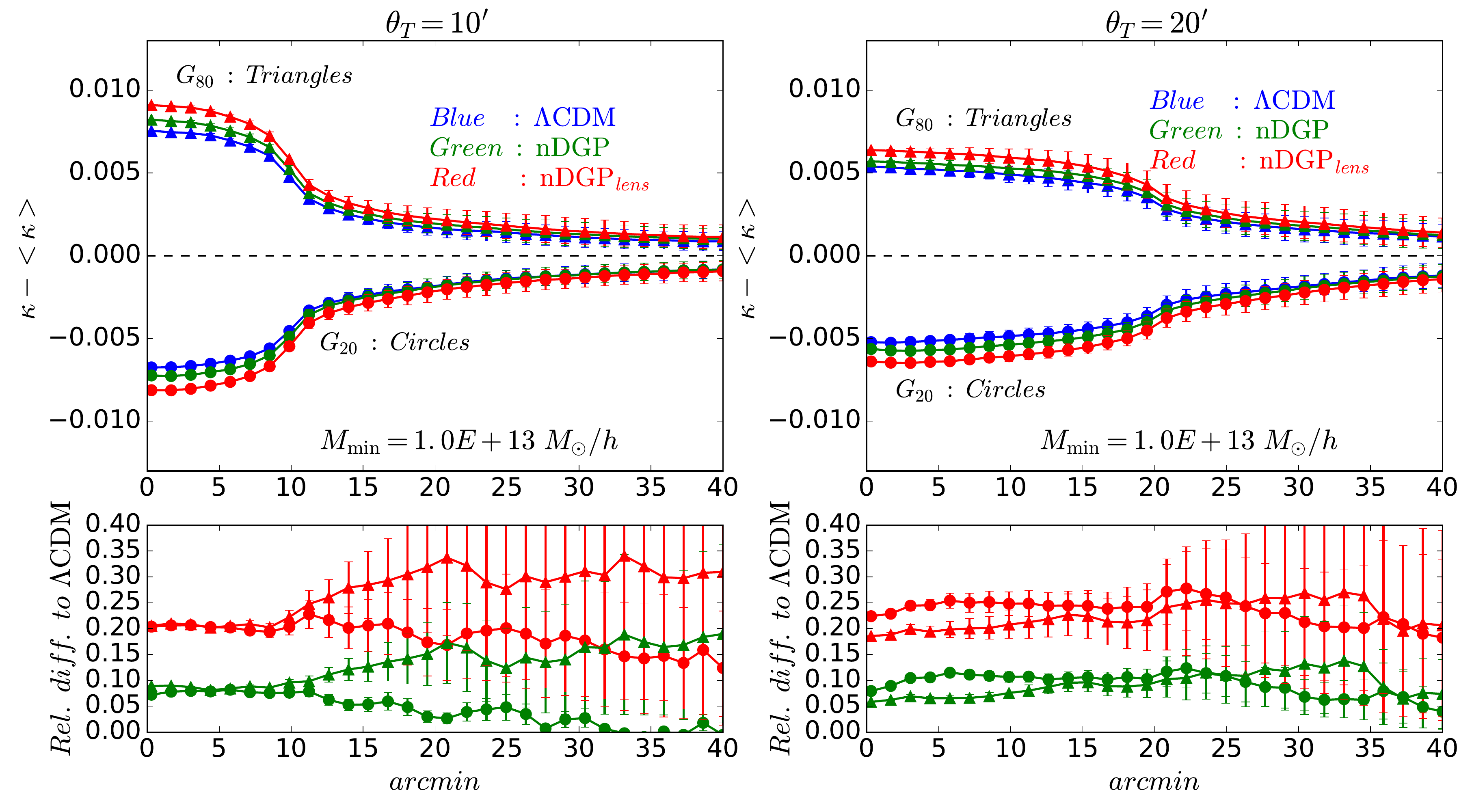}
	\caption{Stacked lensing convergence profiles, plotted as $\kappa - \langle\kappa\rangle$, around $\gtw$ and $\gei$ LOS for $\lcdm$, $\ndgp$ and $\lens$, as labelled. The upper panels show the absolute value, while the lower panels show the relative difference to $\lcdm$. Note that for the $\gtw$ results (circles), a positive relative difference w.r.t.~$\lcdm$ means that the profiles are more negative. The left and right panels show the result for $\theta_T = 10'$ and $\theta_T=20'$, respectively. The result in both panels corresponds to $M_{\rm min} = 10^{13} \msun$ and $z_{\rm halo} \in \left[0.1, 0.76\right]$.}
\label{fig:mgm2}
\end{figure*}

Figure \ref{fig:mmin1} shows the $G$ field distributions (left) and stacked $\kappa$ profiles around $\gtw$ and $\gei$ LOS (right) in $\lcdm$ for a number of halo mass cutoffs $M_{\rm min}$, as labelled. The result corresponds to $\theta_T = 10'$ and $z_{\rm halo} \in \left[0.1, 0.76\right]$. As we have already discussed in Sec.~\ref{sec:gfield}, the distribution of the $G$ field shifts to lower values with increasing $M_{\rm min}$ because the halo distribution becomes sparser. What we wish to analyse here is what happens to the $\gtw$ and $\gei$ profiles. For the two lowest mass cutoff values ($M_{\rm min} = 5\times 10^{12}, 10^{13} \msun$), the profiles are close to one another (perhaps with a slight trend for the signal to be weaker for $M_{\rm min} = 10^{13} \msun$) and the corresponding $\gtw$ and $\gei$ profiles are fairly symmetric around the mean. On the other hand, the $\gtw$ profiles for the two highest mass cutoff values shown ($M_{\rm min} = 5\times 10^{13}, 10^{14} \msun$) are appreciably different from one another, and are also not symmetric to the corresponding $\gei$ profiles around the mean. More specifically, the $\gtw$ signal becomes weaker with increasing $M_{\rm min}$, and in the particular case of the $M_{\rm min} = 10^{14} \msun$ cutoff, the $\gtw$ profiles have the amplitude of the mean convergence in the map on angular scales larger than $20'$.

This behaviour of the $\gtw$ profiles for high $M_{\rm min}$ {holds also for other values of $\theta_T$ and} can be linked to the fact that the $G$ distribution becomes significant at $G=0$. Recall that, whenever the number of pixels with $G=0$ exceeds $20\%$, we randomly select a number of them that adds up to $20\%$ of the total number of pixels. An explanation of the result in Fig.~\ref{fig:mmin1} is therefore as follows. If the halo catalogue is too sparse, then the $\gtw$ LOS are essentially chosen at random across the FOV, and therefore, will not necessarily trace regions with the lowest $\kappa$ values. In other words, the absence of very massive halos along a particular direction in the sky does not guarantee that direction to be devoid of many lower mass haloes. Figure \ref{fig:mmin2} shows the location of $\gtw$ LOS overlaid with the $\kappa$ maps of $\lcdm$ for two mass cutoff values. The figure illustrates that, indeed, for the higher cutoff value $M_{\rm min} = 10^{14} \msun$ (bottom), the location of the $\gtw$ LOS is much more decorrelated with low-$\kappa$ regions, compared to the lower cutoff value $M_{\rm min} = 10^{13} \msun$ (top). One other possible way to get intuition about this result is to consider the rather extreme case of a halo catalogue without any halos. In this case, all pixels of the $G$ field would have $G=0$, and as as result, the lensing signal around randomly chosen pixels would match the mean value of $\kappa$ in the FOV.

Our results for the impact of $M_{\rm min}$ are in apparent constrast with those reported recently in Ref.~\cite{2016MNRAS.459.2762H}. There, the authors find that the lensing signal around $\gtw$ LOS can exhibit an enhancement w.r.t.~the mean for $M_{\rm min} \geq 5\times10^{13} \msun$ (cf.~Fig.~6 of Ref.~\cite{2016MNRAS.459.2762H}), whereas our results show a suppression. Qualitatively, our result is in accordance with our strategy to select the $\gtw$ LOS whenever the distribution of the $G$ field is significant at $G=0$, and so the origin for the difference could lie in different treatments of these special cases.

\subsection{The impact of modified gravity}\label{sec:modi}

\begin{figure*}
	\centering
	\includegraphics[scale=0.40]{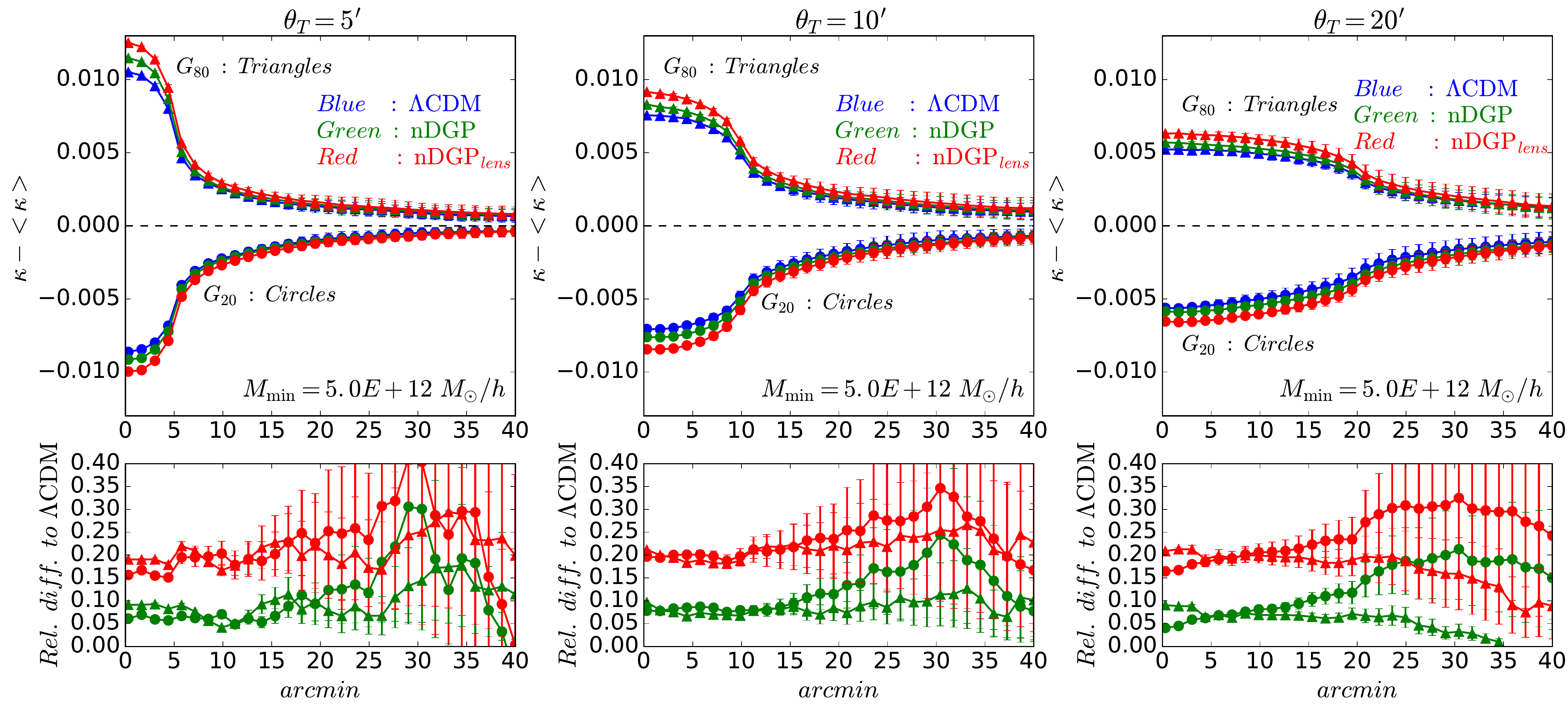}
	\caption{Same as Fig.~\ref{fig:mgm2}, but for $M_{\rm min} = 5\times10^{12} \msun$ and including also the result for $\theta_T = 5'$.}
\label{fig:mgm1}
\end{figure*}

\begin{figure}
	\centering
	\includegraphics[scale=0.375]{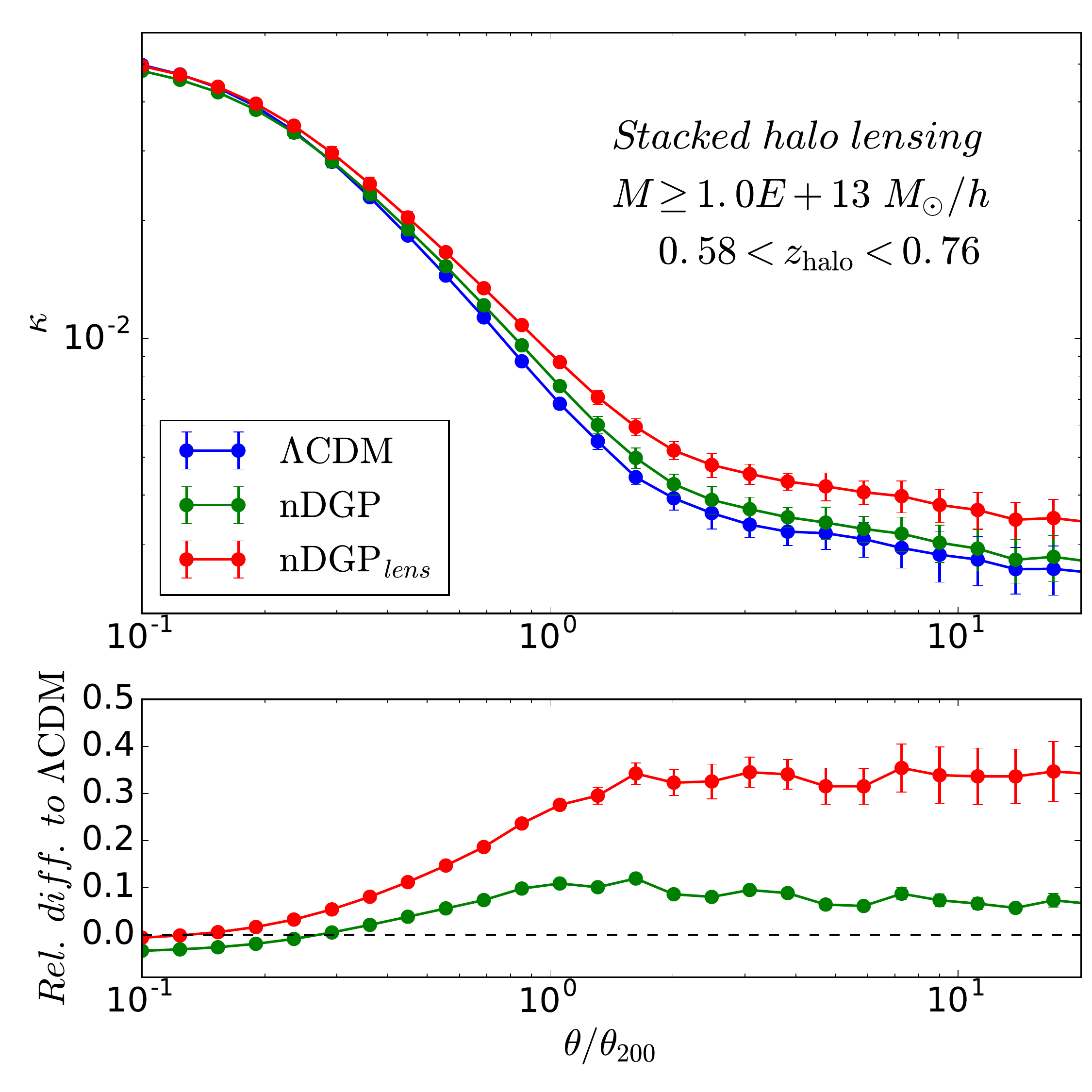}
	\caption{Stacked lensing convergence profiles around the locations of halos with mass $M \geq 10^{13} \msun$ for $\lcdm$, $\ndgp$ and $\lens$, as labelled. The halos were found in box 6 of the lensing tile (cf.~Fig.~\ref{fig:tile}), which corresponds to $z_{\rm halo} \in \left[0.58, 0.76\right]$. The upper panel show the absolute value, while the lower panel shows the relative difference to $\lcdm$.}
\label{fig:halos}
\end{figure}

Figure \ref{fig:mgm2} shows the stacked $\kappa - \langle\kappa\rangle$ profiles around $\gtw$ and $\gei$ LOS for the $\lcdm$, $\ndgp$ and $\lens$ models, and for $\theta_T = 10', 20'$, $M_{\rm min} = 10^{13} \msun$ and $z_{\rm halo} \in \left[0.1, 0.76\right]$. The same is shown in Fig.~\ref{fig:mgm1}, but for $M_{\rm min} = 5\times10^{12} \msun$ and for $\theta_T = 5'$ as well (this smaller aperture size is not shown in Fig.~\ref{fig:mgm2} because the corresponding profiles are noisier). We do not show the results for higher mass cutoffs to ensure that the $G$ field distribution is always negligible at $G=0$ (recall the discussion in the last subsection). The figure shows that the modifications to gravity amplify the lensing signal around $\gtw$ and $\gei$ LOS. In the case of $\ndgp$, this is due to the modified matter distribution caused by the deeper dynamical gravitational potential: there is more matter in regions around haloes and less matter in regions devoid of haloes, compared to $\lcdm$. The boost in the lensing signal in $\ndgp$ is of the order of $5-10\%$, a figure that holds for the different values of $\theta_T$ and $M_{\rm min}$ shown. Naturally, the size of the deviations to $\lcdm$ is larger in the $\lens$ model because of the additional modifications to the lensing potential. Specifically, the boost in the lensing signal becomes of order $15-25\%$ in the $\lens$ model. Note that on larger angular scales, our measurements become noticeably noisier\footnote{We propagate the error estimate of the profiles to their relative difference in quadrature. More specifically, if $f = a/b - 1$, then, $\Delta f = |f|\sqrt{(\Delta a/a)^2 + (\Delta b/b)^2}$, where $\Delta f$, $\Delta a$ and $\Delta b$ are error estimates on the quantities $f$, $a$ and $b$, respectively. In our case, $a$ and $b$ are correlated because the simulations of $\lcdm$, $\ndgp$ and $\lens$ evolved from the same initial conditions. The errors shown on the profiles of the relative difference therefore represent a conservative overestimation.}, which complicates the interpretation of the differences to $\lcdm$. We shall therefore base our analysis on angular scales smaller than $\theta_T$ (on larger scales, the signal becomes small for all models anyway).

One remarkable aspect of the result depicted in Figs.~\ref{fig:mgm2} and \ref{fig:mgm1} is that the boost relative to $\lcdm$ in the lensing signal in $\ndgp$ and $\lens$ is of the same size around $\gtw$ and $\gei$ LOS. The relative difference to $\lcdm$ remains also fairly constant across the radial scales shown (at least where the errorbars are not typically too large). This indicates that the lensing signal around $\gtw$ and $\gei$ LOS shows no evidence for the effects of the screening mechanism. For instance, one could naively expect that, since $\gei$ represent lines of sight that are predominantly halo-overdense, then this could trigger the suppression effects of the Vainshtein screening (cf.~Sec.~\ref{sec:screening}). This nonlinear effect would presumably be less pronounced around $\gtw$ LOS, since these would be lower density regions and so the screening would be less efficient \cite{2015JCAP...08..028B}. If this was the case, then one would expect the size of the deviations from $\lcdm$ to differ between $\gtw$ and $\gei$, but this is not what is shown in Figs.~\ref{fig:mgm2} and \ref{fig:mgm1}. To illustrate the manifestation of the screening mechanism and help understand our results, we show in Fig.~\ref{fig:halos} the stacked lensing signal around halo locations in the FOV. In the case of haloes, indeed, the screening mechanism noticeably suppresses the effects of the modifications to gravity on scales smaller than the angular size of the haloes $\theta_{200} = R_{200}/\chi(z_{\rm halo})$. On the other hand, on scales larger than the typical size of haloes, the screening mechanism is less effective (because the density contrast becomes smaller) and the fifth force effects become larger.

Overall, this suggests that the lensing signal around $\gei$ LOS is not dominated by nonlinear density peaks that exist along these LOS, but instead by the (linear/quasi-linear) density constrast of matter that surrounds these peaks. The lack of evidence for a discriminatory behavior of the fifth force on lensing around $\gtw$ and $\gei$ LOS somewhat dissuades the design of tests of gravity based on a scale- or density-dependent behavior. This leaves the constant boost in the amplitude of the signal as the typical modified gravity signature, at least for models with phenomenology similar to that of DGP (see e.g.~Ref.~\cite{2016MNRAS.459.2762H} for a similar study, but for $f(R)$ gravity).

\subsection{The impact of halo redshift $z_{\rm halo}$}\label{sec:reds}

\begin{figure*}
	\centering
	\includegraphics[scale=0.475]{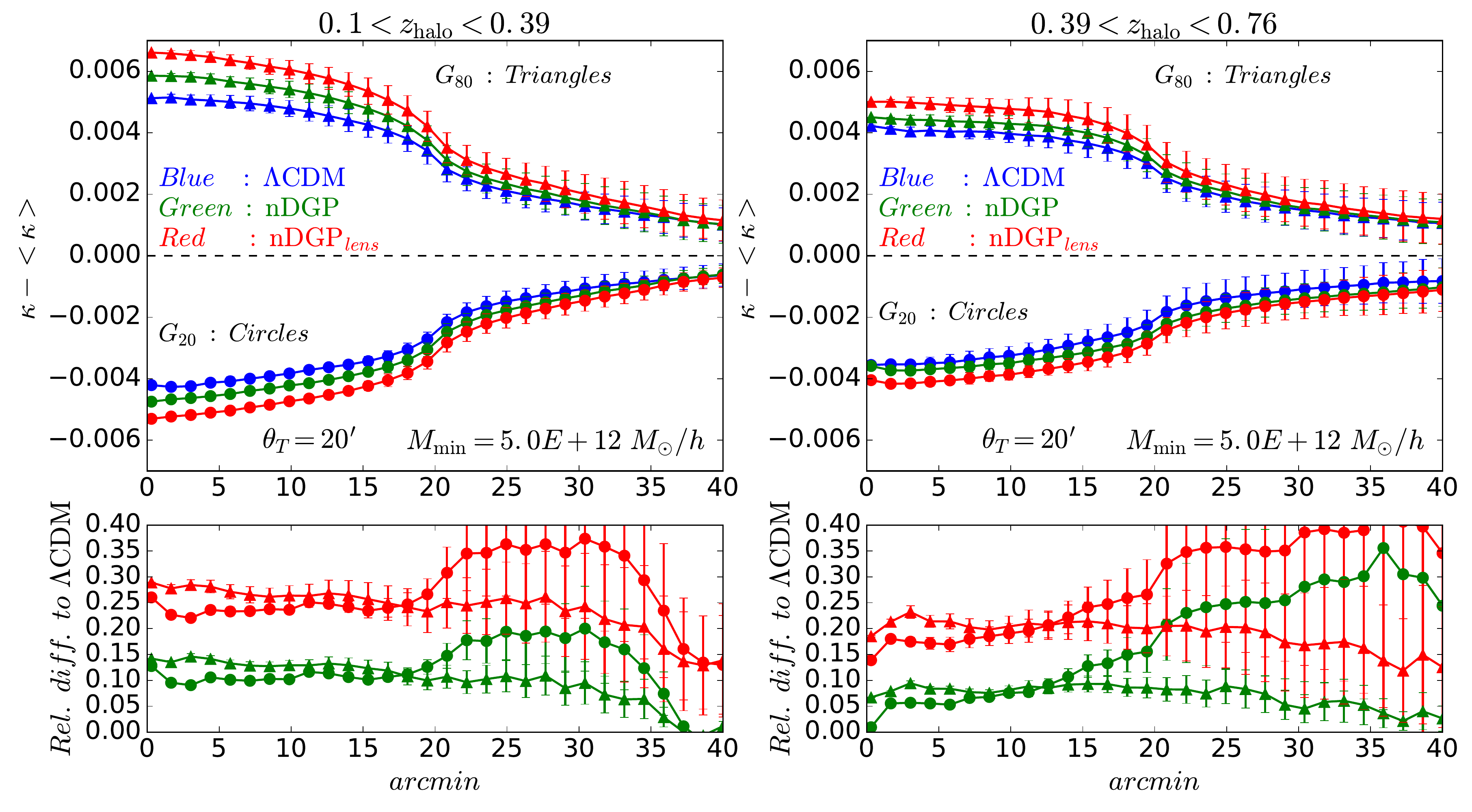}
	\caption{Same as the right panels of Fig.~\ref{fig:mgm1} {($\theta_T = 20'$ and $M_{\rm min} = 5\times10^{12}\msun$)}, but dividing the halo lightcone that is used to construct the $G$ field into two redshift bins: $z_{\rm halo} \in \left[0.1, 0.39\right]$ (left) and $z_{\rm halo} \in \left[0.39, 0.76\right]$ (right).}
\label{fig:tomo}
\end{figure*}

Figure \ref{fig:tomo} shows the impact of $z_l$ and $z_u$ in Eq.~(\ref{eq:gfield}) on the lensing signal around $\gtw$ and $\gei$ LOS for the $\lcdm$, $\ndgp$ and $\lens$ models. The figure shows that the boost (suppression) of the signal w.r.t.~the mean for $\gei$ ($\gtw$) is larger, if the haloes used to construct the $G$ field are at lower redshift. This is in agreement with the results found in the observational analysis of Ref.~\cite{2016MNRAS.455.3367G}. As in the previous subsection, we find that the relative difference between the two modified gravity models and $\lcdm$ remains relatively constant with radius (again, ignoring the scales where the errorbars become too large) and it is of the same order for both $\gtw$ and $\gei$. The amplitude of the difference, however, does seem to depend slightly on $z_l$ and $z_u$: for the lower redshift bin, $z_{\rm halo} \in \left[0.1, 0.39\right]$, the difference to $\lcdm$ in the $\ndgp$ ($\lens$) model is of order $10-15\%$ ($20-30\%$), whereas for the higher redshift bin, $z_{\rm halo} \in \left[0.39, 0.76\right]$, this figure gets reduced to $5-10\%$ ($15-25\%$). This result corresponds to $\theta_T = 20'$ and $M_{\rm min} = 5\times10^{12} \msun$, and we have checked that the same trend exists also for $\theta_T = 10'$ (not shown for brevity). This trend for an increase of the difference to $\lcdm$ with decreasing $z_{\rm halo}$ can be explained by the fact that the effects of the fifth force in the $\ndgp$ and $\lens$ models are larger at later times. We find, however, that the $z_l$, $z_u$ dependence of the relative difference to $\lcdm$ is less clear for $\theta_T = 20'$ and $M_{\rm min} = 10^{13} \msun$ (also not shown).

\subsection{The impact of the $G$ field percentiles}\label{sec:thre}

\begin{figure}
	\centering
	\includegraphics[scale=0.38]{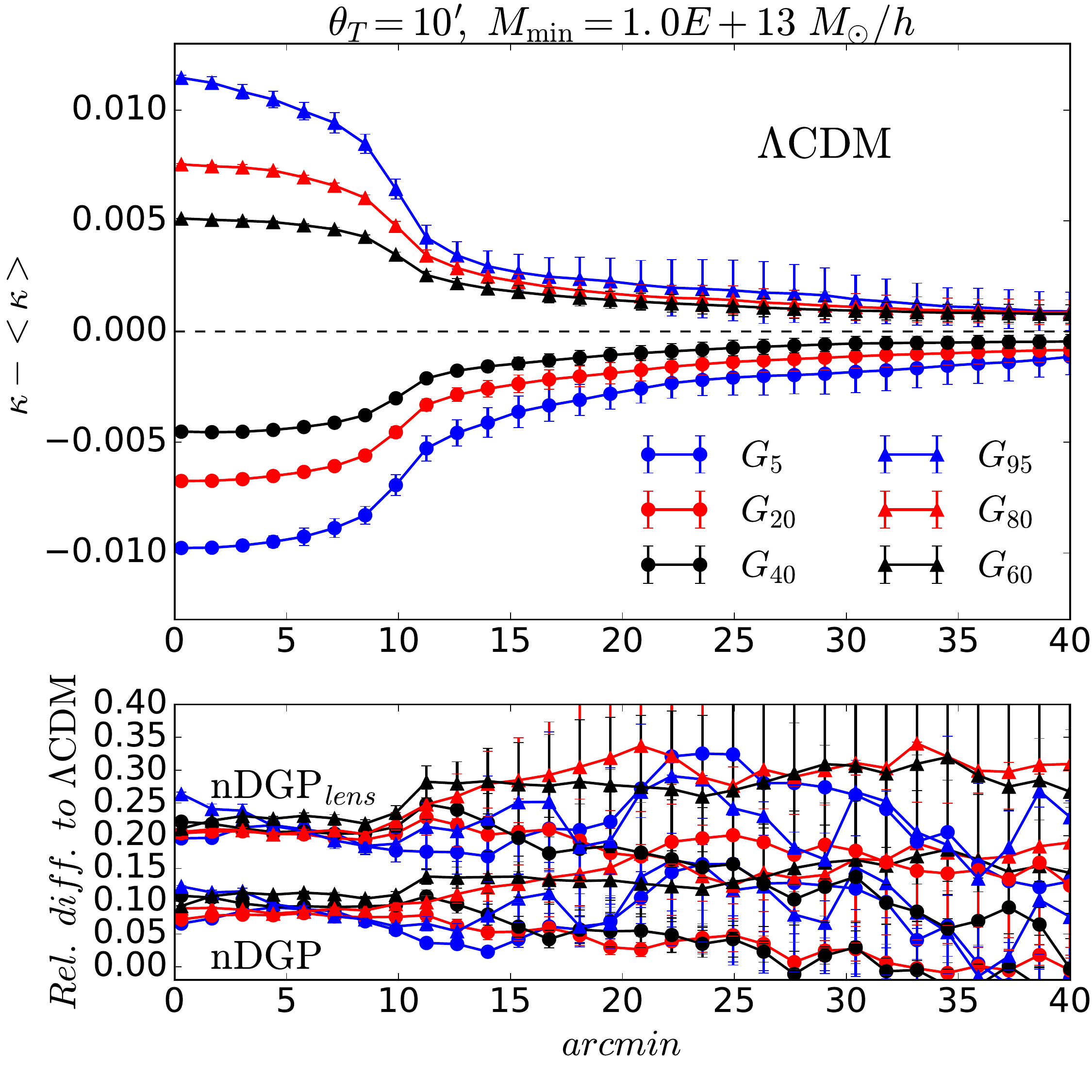}
	\caption{Impact of the choice of the percentile of the $G$ field distribution used to define the halo-underdense and halo-overdense LOS. The upper panel shows the $\kappa - \langle\kappa\rangle$ profiles for $\lcdm$ and for three choices of the percentiles, as labelled. The $\gfi$ and $\gni$ cases correspond, respectively, to taking the lower and upper $5\%$ percentiles (and similarly for the other cases shown). The lower panels display the relative difference to $\lcdm$ in the $\ndgp$ (lower curves at low radii) and $\lens$ (upper curves at low radii) models. The result is for $\theta_T = 10'$, $M_{\rm min} = 10^{13} \msun$ and $z_{\rm halo} \in \left[0.1, 0.76\right]$.}
\label{fig:thre}
\end{figure}

Another test we perform is that of the impact of the choice of the percentile of the $G$ field that is used to define the halo-overdense and halo-underdense LOS. The outcome of the test is shown in Fig.~\ref{fig:thre}.  The $\gfi$ and $\gni$ LOS cases shown are defined analogously to the $\gtw$ and $\gei$ LOS, except that the underdense (overdense) LOS correspond to the lower (upper) $5\%$ percentile of the distribution. The $\gfo$ and $\gsi$ LOS are defined in the same way, but using the lower and upper $40\%$ percentiles, respectively. The result corresponds to $\theta_T = 10'$ and $M_{\rm min} = 10^{13} \msun$, and so the $\gtw$ and $\gei$ results are the same as in the left panel of Fig.~\ref{fig:mgm2}. The figure shows the expected result that the strength of the signal decreases with increasing size of the percentile.  For instance, by increasing the size of the lower percentile used, one considers LOS that are less underdense, thereby effectively reducing the strength of the suppression w.r.t.~the mean (the signal becomes less negative). Similar considerations hold for the upper percentile. As for the impact of the modifications to gravity, the figure displays no evidence that the size of the modifications is dependent on the percentile used. The conclusions we drew before therefore hold also for varying choice of the percentiles.

{\subsection{The tangential shear signal}\label{sec:shear}}

\begin{figure*}
	\centering
	\includegraphics[scale=0.405]{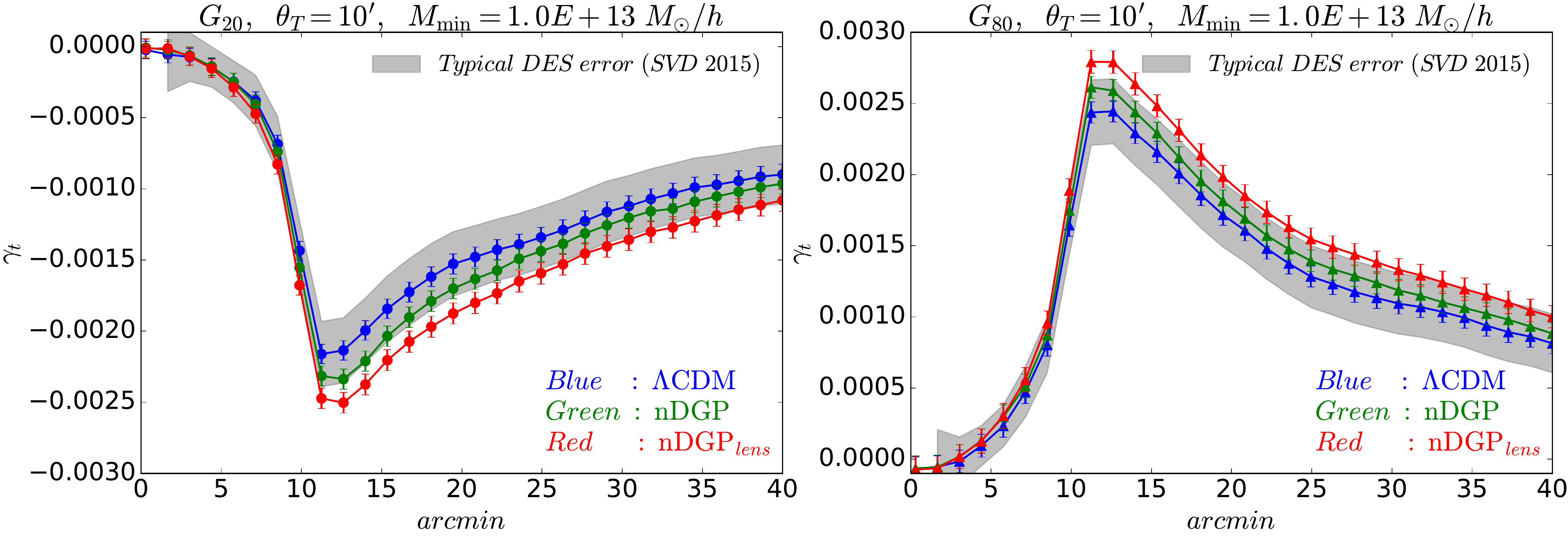}
	\caption{{Tangential shear signal $\gamma_t$ around $\gtw$ (left) and $\gei$ (right) LOS for $\lcdm$, $\ndgp$ and $\lens$, as labelled. The result shown corresponds to $\theta_T = 10'$, $M_{\rm min} = 10^{13}\ M_{\odot}/h$ and $z_{\rm halo} = \left[0.1, 0.76\right]$. The grey area centred around the $\lcdm$ prediction depicts the typical errors attained with the Science Verification Data (SVD) from DES \cite{2016MNRAS.455.3367G}.}}
\label{fig:shear}
\end{figure*}

{In the discussion presented thus far, we were mostly interested in a qualitative comparison of the effects of the modifications to gravity in the lensing signal, for which it sufficed to analyse the corresponding effects on the convergence profiles. The tangential shear signal $\gamma_t$ though, is the quantity that is most commonly reported in observational studies \cite{2016MNRAS.455.3367G}, since it is more directly related to the ellipticity of source galaxy shapes \footnote{{In reality, it is the reduced shear, $g = \gamma_t/(1-\kappa)$ that is most commonly reported in observational studies. In the weak-lensing regime though $\kappa \ll 1$, and hence, $g \approx \gamma_t$.}}. Figure \ref{fig:shear} shows the signal measured around $\gtw$ (left) and $\gei$ (right) LOS in our $\lcdm$, $\ndgp$ and $\lens$ ray-tracing simulations, for $\theta_T = 10'$, $M_{\rm min} = 10^{13}\ M_{\odot}/h$ and $z_{\rm halo} = \left[0.1, 0.76\right]$, as labelled. For a given LOS location $\vec{\theta}$ on the FOV, the tangential shear at an angular distance $\alpha$ from $\vec{\theta}$ is given by
\bq\label{eq:gt}
\gamma_t &=& \langle-{\rm Re}\left[\gamma(\vec{\theta} + \vec{\alpha}){\rm exp}\left(-2i\beta\right)\right]\rangle_{\beta}
\eq
where $\beta$ is the polar angle of the vector $\vec{\alpha} =\alpha\left({\rm cos}\beta, {\rm sin}\beta\right)$ on a coordinate system with origin at $\vec{\theta}$. The notation $\langle\rangle_\beta$ denotes averaging over $\beta$, i.e., averaging over the circunference around $\vec{\theta}$ with radius $\alpha$. The profiles in Fig.~\ref{fig:shear} show the average over all $\gtw$ and $\gei$ locations. As one would expect from the analysis of the convergence profiles in the previous sections, the shear profiles also show no evidence for a discriminatory influence of the fifth force on $\gtw$ and $\gei$ LOS, and hence, that the effects of the screening mechanism remain unnoticed. Note also that since $\gamma_t$ is not sensitive to $\langle\kappa\rangle$, then the result shown in Fig.~\ref{fig:shear} can serve as evidence that our value of $\langle\kappa\rangle > 0$ in the FOV does not have an impact on the conclusions drawn in previous sections. {Moreover, in Appendix \ref{sec:app2} we repeat the analysis shown in Fig.~\ref{fig:shear}, but using only half of the FOV. This test reveals that the size of the modified gravity effects remains the same as that displayed in Fig.~\ref{fig:shear}, which helps to further establish the expectation that our conclusions on the absence of screening effects in the $\gtw$ and $\gei$ lensing signal are not specific to our particular realization of the FOV.}}

{A formal comparison of our lensing results with those presented in the DES analysis of Ref.~\cite{2016MNRAS.455.3367G} requires a few extra modelling steps. For instance, one should use a lens sample (haloes along the LOS) that matches the properties of that used in the observational analyses (e.g., in terms of halo/galaxy mass function and clustering, and their redshift evolution). Also important is the use of matching source redshift distributions. For the time being though, it is still interesting to perform the exercise of comparing the size of the effects from modified gravity with the current precision attained by the data. The grey shaded band in Fig.~\ref{fig:shear} shows the size of the errors from the DES paper analysis for $\theta_T = 10'$, centred around the $\lcdm$ result.  The figure shows that current data can perhaps already be used to place constraints on models that modify directly the lensing potential such as $\lens$. Naturally, constraints on a model like $\ndgp$ require higher precision because of the weaker modifications to the lensing signal. In Ref.~\cite{2016arXiv160503965B}, the authors have used growth rate measurements to constraint $r_cH_0 \lesssim 1\ (2\sigma)$ in the $\ndgp$ model. Hence, recalling that our results correspond to $r_cH_0 = 1$ and noting that the differences to $\lcdm$ in Fig.~\ref{fig:shear} are at least comparable to the current precision of current data, then this suggests that this type of lensing signal may be able to start placing competitive constraints on modified gravity theories in the near future. In fact, the precision from subsequent similar analyses from DES is expected to increase as more area of the sky is used.}

{In Ref.~\cite{2016MNRAS.455.3367G}, an analytical model for the lensing signal associated with $\gtw$ and $\gei$ LOS in $\Lambda{\rm CDM}$ was put forward {(we outline its main equations in Appendix \ref{sec:app2})}. The development of such analytical models and subsequent calibration/validation against N-body simulations can be useful to rapidly span higher-dimensional parameter spaces in cosmological constraint studies. It would therefore be interesting to try to accommodate modified gravity effects onto this analytical framework. Here though, one would benefit from having more realizations of the FOV in order to beat sample variance {(cf.~Appendix \ref{sec:app2})}, and properly validate the model predictions against the measure simulated lensing profiles. These further developments are however beyond the scope of the present paper.}

\section{Summary and Conclusions}\label{sec:conc}

We have carried out a study of the imprints that modifications to GR can leave on the lensing signal around LOS that are predominantly halo-underdense (called trough lensing) and halo-overdense. For a given angular aperture $\theta_T$ centred on the points of a regular grid that covers the FOV (called $G$ field), the underdense LOS (dubbed $\gtw$ LOS) are defined as the grid points with the $20\%$ lowest projected halo count, and the overdense LOS (dubbed $\gei$ LOS) as those with the $20\%$ highest projected halo count (cf.~Sec.~\ref{sec:gfield}). The analysis of the lensing signal around $\gtw$ LOS is particularly interesting for modified gravity studies because it focuses on the signal induced from mostly underdense regions, where the fifth force effects can be large because the screening is weak. Moreover, the comparison with the corresponding result for $\gei$ LOS (which is sensitive to higher density regions) could also offer potentially interesting ways to pinpoint the scale-dependent nature of the screening mechanism. The lensing signal associated with $\gtw$ and $\gei$ LOS has been recently measured by the DES collaboration \cite{2016MNRAS.455.3367G}, which makes our analysis particularly timely.

Our results were obtained from ray-tracing simulations of modified gravity, which were run with a version of the {\tt ECOSMOG} N-body code augmented with the ray-tracing modules of the {\tt Ray-Ramses} algorithm. The {\tt Ray-Ramses} modules compute the lensing signal without {resorting} to projections of the simulation box along the LOS, which facilitates its application to theories of gravity where $\Phi_{\rm len}$ is governed by a nonlinear Poisson equation (cf.~the discussion in Sec.~\ref{sec:nume}). The $\gtw$ and $\gei$ LOS were found using "pseudo" halo lightcones constructed out of snapshots of the simulation boxes that make up the lensing tile, which covers a FOV of $10\times10\ {\rm deg}^2$ from $z=0$ to a source redshift of $z_s = 1$ (cf.~Sec.~\ref{sec:halos} and Fig.~\ref{fig:tile}). We analysed the profiles of the lensing convergence $\kappa$ maps stacked on the locations of the $\gtw$ and $\gei$ LOS. The detailed lensing signal can depend on the minimum mass $M_{\rm min}$ and redshift $z_{\rm halo}$ of the haloes used to construct the lightcone, as well as on the aperture $\theta_T$ and choice of the $G$ field distribution percentiles.

To illustrate possible signatures of modified gravity, we considered the case of the normal branch of the DGP braneworld model with a $\lcdm$ background (cf.~Sec.~\ref{sec:ndgp}), dubbed here by $\ndgp$. This model modifies the lensing signal w.r.t.~$\lcdm$ because the density field evolves differently due to the fifth force, and not because the way photons react to density perturbations is altered. To study the impact of direct modifications to the lensing potential $\Phi_{\rm len}$, we considered also a variant of the DGP model with the same equations, but in which photons react to the same potential as non-relativistic particles. We referred to this model as $\lens$. We adopted the value $r_cH_0 = 1$ for the cross-over scale parameter of the DGP model, which is borderline consistent with current growth rate data.

Our main results can be summarized as follows:

\hspace{0.2 cm} $\bullet$ As a validation check of our ray-tracing calculations, we computed the power spectrum of the lensing convergence maps (cf.~Fig.~\ref{fig:pk}). The $\ndgp$ and $\lens$ spectra on large scales ($\ell \lesssim 10^3$) exhibit an expected enhancement relative to $\lcdm$, with this boost being larger in the case of the $\lens$ model because of the modifications to $\Phi_{\rm len}$. On scales where linear theory is valid, the impact of the modifications to gravity is in good agreement with the linear theory expectation. On scales $\ell \gtrsim 10^3$, both modified gravity spectra approach that of $\lcdm$, which is a manifestation of the Vainshtein screening mechanism.

\hspace{0.2 cm} $\bullet$ The stacked lensing signal around $\gtw$ LOS depends sensitively on $M_{\rm min}$ in cases for which the $G$ field distribution is non-negligible at $G = 0$ (cf.~Fig.~\ref{fig:mmin1}). For instance, if more than $20\%$ of all LOS have no haloes within their aperture size, then there is no good way to identify those with the lower projected halo count. Whenever this was the case, we opted to choose the $\gtw$ LOS at random out of all LOS with $G=0$, which resulted in an expected weakening of the lensing signal (cf.~the discussion in Sec.~\ref{sec:mmin}). The $M_{\rm min}$-dependence of the lensing signal around $\gei$ LOS is less pronounced. This suggests that halo mass (or e.g.~galaxy luminosity in observational studies), or more precisely halo abundance, is an important parameter to take into consideration when studying this type of lensing signal.

\hspace{0.2 cm} $\bullet$ For both the $\ndgp$ and $\lens$ models, the amplitude of the differences w.r.t.~$\lcdm$ are of the same size in $\gtw$ and $\gei$ LOS (cf.~Figs.~\ref{fig:mgm2} and \ref{fig:mgm1}, on radial scales where our measurements are not too noisy). This means that there is no evidence for the suppression effects of the screening mechanism, which one could naively expect to have an impact on the $\gei$ signal that probes overdense regions. In other words, the density field crossed by photons along $\gei$ LOS is overdense, but not overdense enough to trigger the screening effects. To illustrate the implementation of screening, we have stacked the lensing signal at the location of dark matter haloes (cf.~Fig.~\ref{fig:halos}). For haloes, the scale-dependent and suppression effects of the screening become manifest on angular scales smaller than their angular size, where the density is high enough for the Vainshtein mechanism to come into play.

\hspace{0.2 cm} $\bullet$ The conclusion of the bullet point above is robust to various choices of $M_{\rm min}$ and $\theta_T$ (cf.~Figs.~\ref{fig:mgm2} and \ref{fig:mgm1}), as well as to different choices of the redshift range of the haloes used to defined the $\gtw$ and $\gei$ LOS (cf.~Fig.~\ref{fig:tomo}) and to the exact choice of the percentiles of $G$ field distribution used to identify the halo-underdense and halo-overdense LOS (cf.~Fig.~\ref{fig:thre}).

\bigskip
\bigskip

For the case of the $\ndgp$ and $\lens$ models analysed here, which employ the Vainshtein screening mechanism, the main observational signature is an overall shift in the amplitude of the lensing signal, which can in principle be used to place useful constraints on modified gravity {(despite the lack of evident screening effects, cf.~Fig.~\ref{fig:shear})}. This shift is more pronounced in the case of the $\lens$ model because of the modifications to the lensing potential. For the case of $f(R)$ gravity, which employs the Chameleon screening mechanism, the results of Ref.~\cite{2016MNRAS.459.2762H} suggest that trough lensing may be used to place constraints on the model. The determination of the impact of the fifth force on the lensing signal around halo-overdense LOS in $f(R)$ gravity (and corresponding comparison to the lensing signal around underdense LOS) is yet to be performed, and it is beyond the scope of this paper.

In order to allow for a more rigorous assessment of the constraints on a given modified gravity model from trough lensing data, analysis such as ours would benefit from a more elaborate halo lightcone construction (e.g.~to match the properties of catalogues used in observational studies), distribution of lensing source redshifts, estimates of errors in lensing measurements and more realizations of the initial conditions of the simulations {to minimize sample variance (cf.~Appendix \ref{sec:app2})}. It would also be interesting to extend the theoretical model of trough lensing presented in Ref.~\cite{2016MNRAS.455.3367G} to include modified gravity effects. The fact that the screening mechanism does not play a critical role in the signal may help to facilitate the development of such a theoretical framework.

\begin{acknowledgments}

We are in debt to Oliver Friedrich for the very useful help and cross-checks in some of our calculations. We also thank Daniel Gruen for sharing with us the covariance matrices of the DES measurements and Fabian Schmidt for useful comments and suggestions. SB is supported by STFC through grant ST/K501979/1. BL acknowledges support by the UK STFC Consolidated Grant ST/L00075X/1 and RF040335. CLL acknowledges support from STFC consolidated grant ST/L00075X/1. This work used the DiRAC Data Centric system at Durham University, operated by the Institute for Computational Cosmology on behalf of the STFC DiRAC HPC Facility (www.dirac.ac.uk). This equipment was funded by BIS National E-infrastructure capital grant ST/K00042X/1, STFC capital grant ST/H008519/1, and STFC DiRAC Operations grant ST/K003267/1 and Durham University. DiRAC is part of the National E-Infrastructure. 

\end{acknowledgments}

\appendix

\section{The impact of the $G$ field grid size and time resolution of the halo lightcone}\label{sec:app1}

\begin{figure*}
	\centering
	\includegraphics[scale=0.44]{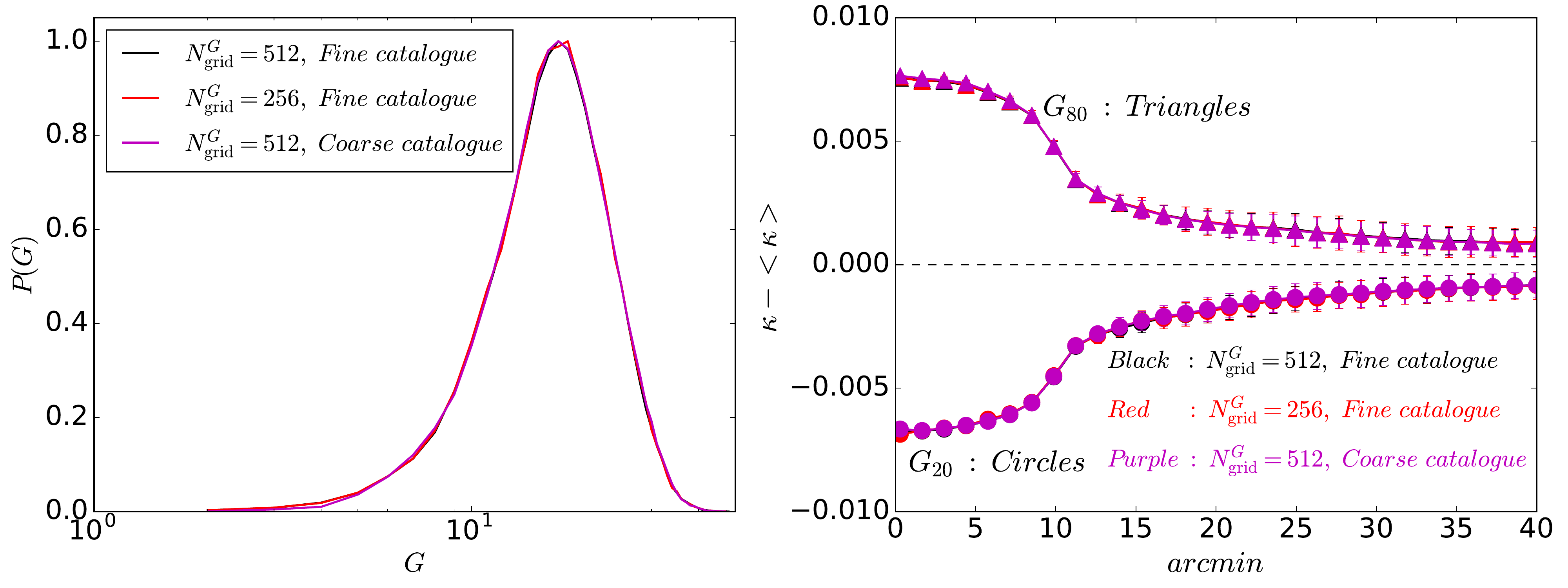}
	\caption{Tests of the impact of the choice of the grid size of the $G$ field and time resolution of the halo lightcone. The result corresponds to $\theta_T=10'$, $M_{\rm min} = 10^{13}\msun$ and $z_{\rm halo} \in \left[0.1, 0.76\right]$ for $\lcdm$. The results displayed with black color correspond to the cases adopted in the main body of the paper. The red colored results correspond to a downgrading of the resolution of the $G$ field grid. The purple colored results correspond to a halo lightcone constructed using only one particle snapshot per box in the tile, as opposed to two snapshot times per box as in the main body of the text. One notes that the $G$ field distributions (left) and $\gtw$ and $\gei$ profiles (right) exhibit no noticeable difference in between the three cases shown.}
\label{fig:tests}
\end{figure*}

Figure \ref{fig:tests} displays a test of the impact of our choice of the $G$ field grid size and time resolution of the "pseudo" halo lightcone. The test corresponds to $\lcdm$ and to $\theta_T=10'$, $M_{\rm min} = 10^{13}\msun$ and $z_{\rm halo} \in \left[0.1, 0.76\right]$. The figure shows that the $G$ field distribution and the corresponding $\kappa$ profiles around $\gtw$ and $\gei$ LOS are barely affected if the grid resolution of the $G$ field is downgraded from $N^G_{\rm grid} = 512$ to $N^G_{\rm grid} = 256$. As explained in Sec.~\ref{sec:halos}, our halo catalogues were constructed by using two particle snapshots per simulation box in the tile. To test whether or not this time resolution is sufficient, we have redone the calculation but using a halo catalogue constructed from only one snapshot per simulation box. The snapshot times chosen for boxes $2$ to $6$ were, respectively, $z = 0.16$, $z = 0.27$, $z = 0.39$, $z = 0.52$ and $z = 0.67$. These redshift values correspond roughly to the epochs when the rays in the bundle are half-way through their "journey" in each box. The results corresponding to this "coarse" halo lightcone are shown in purple, and they are, for all practical purposes, undistinguishable from the case obtained using the "finer" catalogue.

These successful tests show that our results are robust to the exact choices of $N^G_{\rm grid}$ and the number of snapshots used to the build the halo lightcones.

\section{Assessing the impact of sample variance on the trough lensing measurements}\label{sec:app2}

\begin{figure}
	\centering
	\includegraphics[scale=0.39]{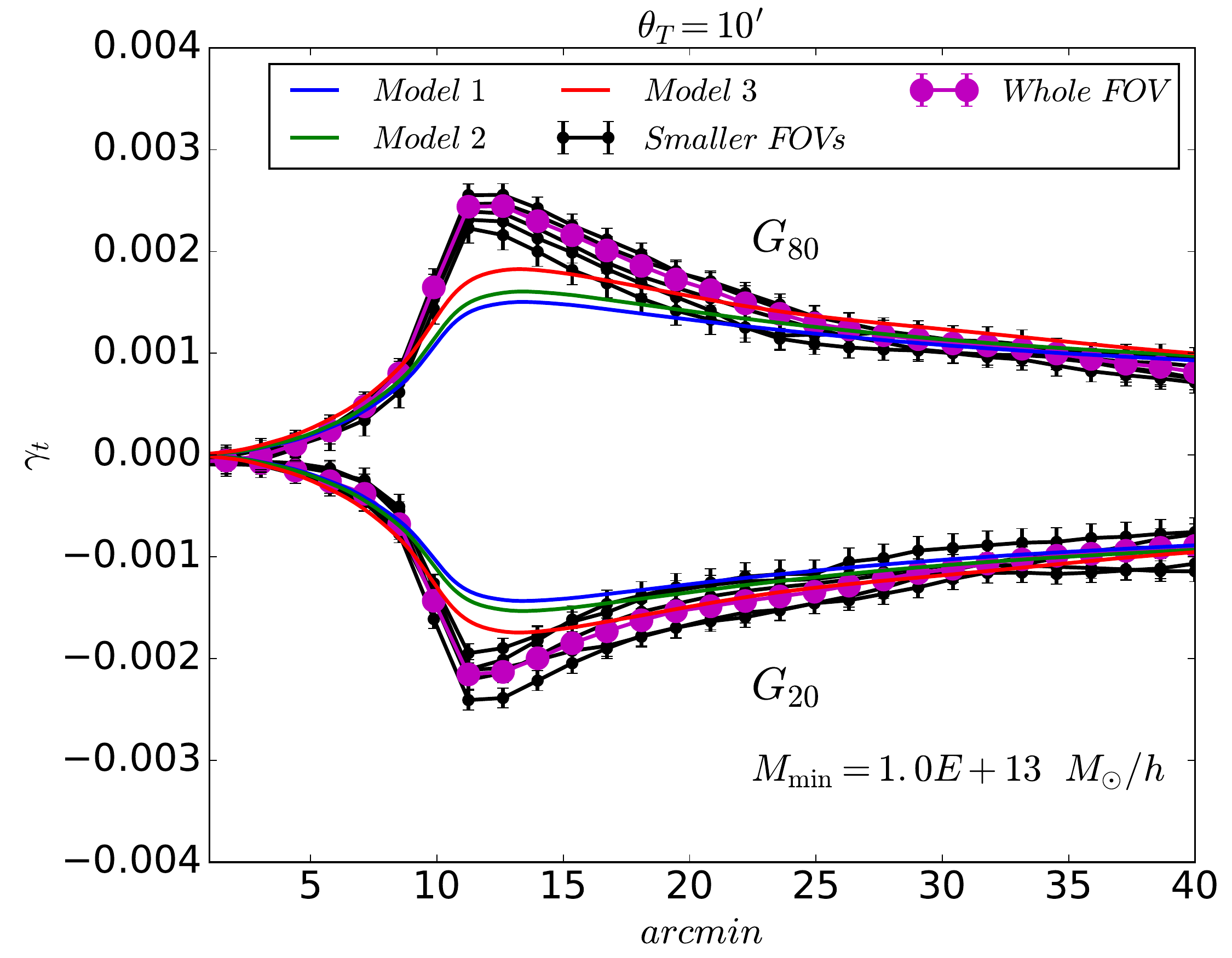}
	\caption{Lensing shear profiles around $\gtw$ and $\gei$ LOS for $\theta_T = 10'$, $M_{\rm min} = 10^{13}M_{\odot}/h$ and $z_{halo} \in \left[0.1, 0.76\right]$, in the $\lcdm$ model. The three solid lines correspond to the three variants (see text for exact meaning) we considered of the analytical model of Ref.~\cite{2016MNRAS.455.3367G}. The purple symbols show the profiles measured across the whole FOV, while the black symbols represent the profiles around LOS in contiguous patches with half the area of the FOV.}
\label{fig:sv1}
\end{figure}

\begin{figure}
	\centering
	\includegraphics[scale=0.39]{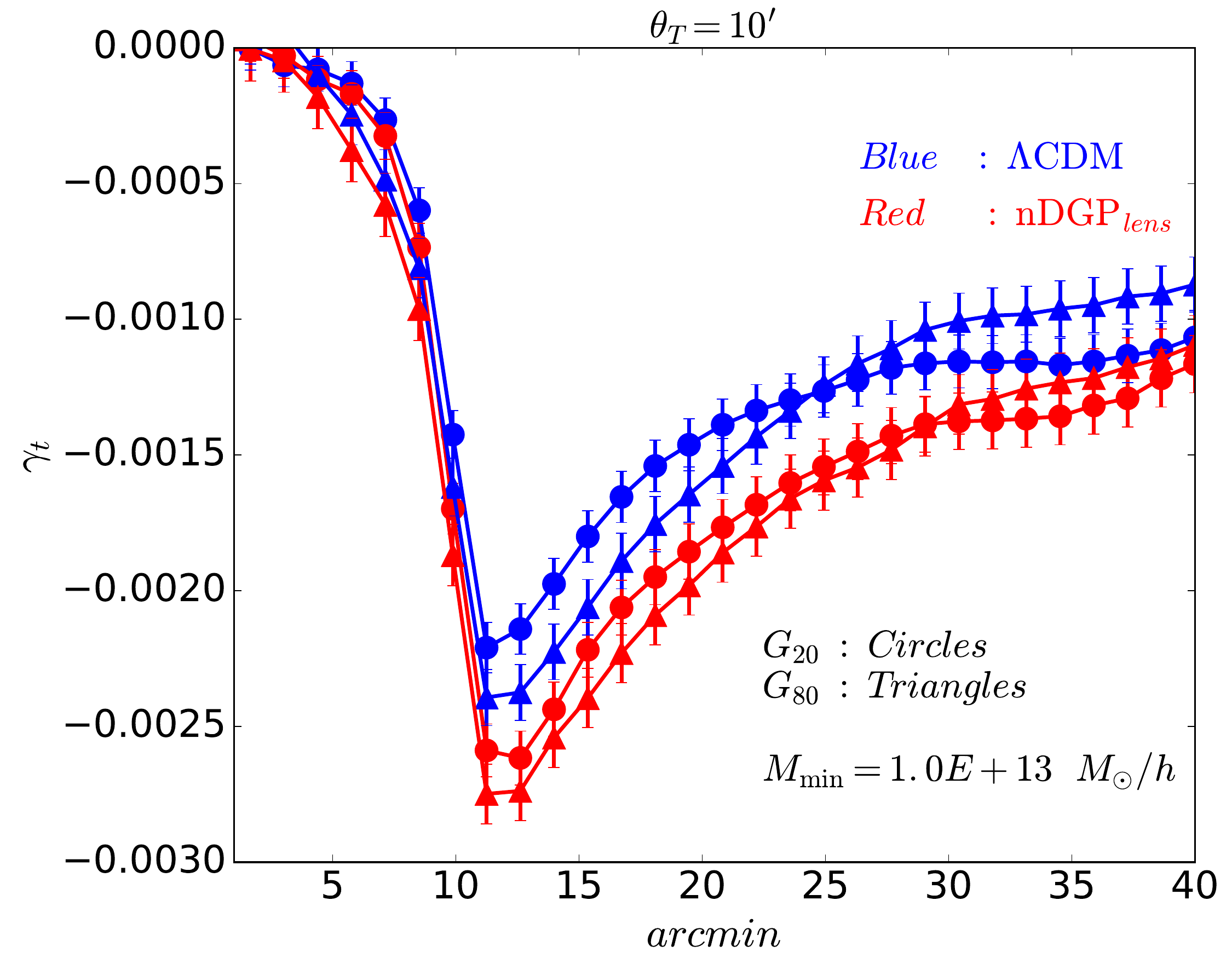}
	\caption{Lensing shear profiles around $\gtw$ (circles) and $\gei$ (triangles) LOS found in one of the smaller patches used in Fig.~\ref{fig:sv1}, in the $\lcdm$ (blue) and $\lens$ (red) models. As in Fig.~\ref{fig:sv1}, the result corresponds to $\theta_T = 10'$, $M_{\rm min} = 10^{13}M_{\odot}/h$ and $z_{halo} \in \left[0.1, 0.76\right]$. Note, the $\gei$ profiles are multiplied by $-1$.}
\label{fig:sv2}
\end{figure}

As noted in Fig.~\ref{fig:pk}, the convergence power spectrum of our $\lcdm$ map is not an exact match to the {\rm Halofit} prediction, which is a signal that the statistics of our FOV may not be representative of the whole sky. In this appendix, we wish the determine whether this sample variance effect may be affecting the conclusions that we drew in the paper. We shall also check how our $\lcdm$ lensing profiles compare to the analytical model put forward in the DES paper \cite{2016MNRAS.455.3367G}.

We start with a brief outline of the analytical model. Here, for brevity, we limit ourselves to displaying the equations that are used explicitly in the calculation. We refer the reader to Ref.~\cite{2016MNRAS.455.3367G} for more details about the derivation. In the model of Ref.~\cite{2016MNRAS.455.3367G}, the lensing convergence around $\gtw$ and $\gei$ LOS is given, respectively, by
\bq\label{eq:anamodel}
\kappa_{\gtw} &=& \frac{{\rm Cov(\delta_T, \mathcal{K}_i)}}{\sigma_T^2}\frac{\sum_{N=0}^{N^{\gtw}}P(N)\big<\delta_T | N\big>}{\sum_{N=0}^{N^{\gtw}}P(N)}, \nonumber \\
\nonumber \\
\kappa_{\gei} &=& \frac{{\rm Cov(\delta_T, \mathcal{K}_i)}}{\sigma_T^2}\frac{\sum_{N=N^{\gei}}^{N^{{\rm max}}}P(N)\big<\delta_T | N\big>}{\sum_{N=N^{\gei}}^{N^{{\rm max}}}P(N)}.
\eq
In the above equations, $\delta_T$ is the projected galaxy count averaged over the aperture size $\theta_T$, and $\mathcal{K}_i$ is the lensing convergence $\kappa$ averaged over an annulus with radius $\theta_i$ and some width $\Delta\theta$. Further, ${\rm Cov(\delta_T, \mathcal{K}_i)}$ denotes the covariance of these two variables and encodes the radial dependence of the profiles, and $\sigma_T^2 = {\rm Cov(\delta_T, \delta_T)}$. The quantity $\big<\delta_T | N\big>$ corresponds to the expectation value of $\delta_T$ for a given galaxy count $N$ in the aperture $\theta_T$ and it is given by
\bq\label{eq:expectation}
\big<\delta_T | N\big> = \int_{-1}^{\infty} \delta_T \frac{P(N|\delta_T)p(\delta_T)}{P(N)} {\rm d}\delta_T,
\eq
where
\bq\label{eq:probs}
P(N|\delta_T) &=& \frac{1}{N!}\left[\bar{N}\left(1+b\delta_T\right)\right]^N {\rm exp}\left[-\bar{N}\left(1+b\delta_T\right)\right] \nonumber \\
p(\delta_T) &=& \frac{1}{\sqrt{2\pi\sigma_T^2}}{\rm exp}\left[-\frac{\delta_T^2}{2\sigma_T^2}\right].
 \eq
The quantity $P(N)$ corresponds to the $G$ field distributions shown in Fig.~\ref{fig:hist} and it is given by
\bq\label{eq:pofN}
&& P(N) = \frac{\bar{N}^N}{N!\sqrt{2\pi\sigma_T^2}}\int_{-1}^{\infty}\left(1 + b\delta_T\right)^N \times \nonumber \\
&&\ \ \ \ \ \ \ \ \ \ \ \ \ \ \ \ \ \ \ \ \ \ \ \ \ \ \ \ \ \ \ \ \ \ \ \ \ \ \ \ \ \ \ \ \ {\rm exp}\Big[-\bar{N}\left(1+b\delta_T\right) - \frac{\delta_T^2}{2\sigma_T^2}\Big] {\rm d}\delta_T. \nonumber \\
\eq
In the sums in Eqs.~(\ref{eq:anamodel}), $N^{\gtw}$ corresponds to the maximum value of $N$ in the lower $20\%$ percentile, $N^{\gei}$ corresponds to the minimum value of $N$ in the upper $80\%$ percentile and $N^{\rm max}$ to the maximum value of $N$ in the upper $80\%$ percentile. The mean value of counts in the aperture $\theta_T$ is $\bar{N} = \int NP(N){\rm d}N$, and $b$ is the linear halo/galaxy bias parameter. What is left to specify are the expressions for the covariances. Effectively, these are given by
\bq\label{eq:covs}
{\rm Cov}(\delta_T, \mathcal{K}_i) &=& \sum_{\ell }C_{\ell}^{\kappa\delta_T}G_{\ell}^{\mathcal{K}_i}G_{\ell}^{\delta_T}, \\
{\rm Cov}(\delta_T, \delta_T) &=& \sum_{\ell }C_{\ell}^{\delta_T\delta_T}\big(G_{\ell}^{\delta_T}\big)^2
\eq
where
\bq\label{eq:thfilters}
G_{\ell}^{\mathcal{K}_i} &=& \frac{2\pi\mathcal{N}_\ell}{A_i} \int_{{\rm cos}(\theta_i + \Delta\theta)}^{{\rm cos}(\theta_i - \Delta\theta)} \mathcal{P}_\ell(x){\rm d}x, \\
G_{\ell}^{\delta_T} &=& \frac{2\pi\mathcal{N}_\ell}{A_t} \int_{{\rm cos}\theta_T}^{1} \mathcal{P}_\ell(x){\rm d}x,
\eq
with $A_i = 2\pi\big[{\rm cos}(\theta_i - \Delta\theta) - {\rm cos}(\theta_i + \Delta\theta)\big]$, $A_t = 2\pi\big[1 - {\rm cos}\theta_t\big]$, $\mathcal{N}_\ell = \sqrt{(2\ell+1)/(4\pi)}$ and $\mathcal{P}_\ell$ is the Legendre polynomial of order $\ell$. Finally, the angular cross-correlations are given by
\bq\label{eq:cls}
C^{\kappa\delta_T}_\ell &=& \int \frac{W_{\kappa}W_{\delta_T}}{\chi^2}P_{\rm nl}\left(k = \ell/\chi, \chi\right) {\rm d}\chi, \\
C^{\delta_T\delta_T}_\ell &=& \int \frac{W_{\delta_T}^2}{\chi^2}P_{\rm nl}\left(k = \ell/\chi, \chi\right) {\rm d}\chi,
\eq
with $W_{\kappa} = \chi(\chi_S-\chi)/(a\chi_S)$ and $W_{\delta_T} = 3\chi^2/(\chi_{\rm max}^3 - \chi_{\rm min}^3)$. The expression for the kernel $W_{\delta_T}$ follows from assuming that the galaxies/halos have constant comoving number density along the LOS ($\chi_{\rm min}$ and $\chi_{\rm max}$ are, respectively, the minimum and maximum comoving distance of tracers in the sample).

The solid curves in Fig.~\ref{fig:sv1} show three variants of the model predictions for $\lcdm$. What is actually shown are the tangential shear profiles, which are obtained from the convergence profiles as
\bq\label{eq:anashear}
\gamma_t(\theta) &=& \bar{\kappa}(\theta) - \kappa(\theta) ; \nonumber \\
\bar{\kappa}(\theta) &=&  \frac{2}{\theta^2}\int_0^\theta \theta'\kappa(\theta'){\rm d}\theta'.
\eq
The curve labeled as $Model\ 1$ (blue) corresponds to the equations displayed above with ${\rm Halofit}$ as the prescription to model the nonlinear matter power spectrum $P_{\rm nl}$. The curve labeled by $Model\ 2$ (green) shows the same, but instead of using the expression for $W_{\delta_T}$ shown above, we use the number density along the LOS measured in the actual halo catalogues. Finally, $Model\ 3$ (red) is obtained in the same manner as $Model\ 2$, but with the angular spectra $C_\ell^{\kappa\delta_T}$ and $C_\ell^{\delta_T\delta_T}$ rescaled by the factor $\mathcal{F} = C_\ell^{\kappa\kappa, {\rm FOV}} / C_\ell^{\kappa\kappa, {\rm Halofit}}$. Here, $C_\ell^{\kappa\kappa, {\rm FOV}}$ corresponds to the measured angular power spectrum of the $\lcdm$ convergence map shown in Fig.~\ref{fig:pk}, and $C_\ell^{\kappa\kappa, {\rm Halofit}}$ is the ${\rm Halofit}$ curve shown there. These two spectra are not the same due to sample variance, and hence, multiplying $C_\ell^{\kappa\delta_T}$ and $C_\ell^{\delta_T\delta_T}$ by $\mathcal{F}$ helps to provide with a rough approximate feeling for the impact of sample variance on the input spectrum, and how this propagates into the model lensing profiles. For all models, $b=1.6$. The purple symbols show the measured profiles using LOS found across the entire FOV (these are the same $\lcdm$ profiles shown in Fig.~\ref{fig:shear}). The five measured profiles shown as black symbols correspond to using LOS across contiguous patches with half the area of the entire FOV. One notes that there is indeed a noticeable amount of variance across these smaller patches. None of these measured profiles is in perfect agreement with the analytical model predictions, but this is expected under the assumption that they are smaller patches of an already peculiar $10\times10\ {\rm deg}^2$ patch of the whole sky. Also as expected, the curve from $Model\ 3$ is that which is closer to the measured profiles, since it takes into account (although only approximately) the effect of sample variance on the input angular power spectrum used in the calculation.

From the exercise performed in Fig.~\ref{fig:sv1} we conclude that the absolute amplitude and shape of our $\gtw$ and $\gei$ profiles are not representative of all LOS in the sky. This poses the question of whether or not our conclusions about the effects of modified gravity on this lensing signal are themselves only applicable to our specific realization of the FOV. To address this question we show in Fig.~\ref{fig:sv2} the lensing shear profiles around $\gtw$ (circles) and $\gei$ (triangles) LOS in one of the patches with half the area of the FOV (we show only one of them to make the figure less busy, but the conclusions we draw hold for all other), for $\lcdm$ and the $\lens$ models. The physical picture displayed in Fig.~\ref{fig:sv2} is the same as the one shown in Fig.~\ref{fig:shear} (which corresponds to the whole FOV), in that there is no evidence that the screening mechanism is at play. From this we conclude that, even though the absolute profiles shown throughout the paper are peculiar to our realization of the FOV, the effects of modified gravity on them (which are the focus of our investigations) are not. 

\bibliography{ndgp_troughs.bib}

%Merlin.mbs v4.21 2009-07-09.
\begin{thebibliography}{10}%
\makeatletter
\providecommand \@ifxundefined [1]{%
 \ifx #1\undefined \expandafter \@firstoftwo
 \else \expandafter \@secondoftwo
\fi
}%
\providecommand \@ifnum [1]{%
 \ifnum #1\expandafter \@firstoftwo
 \else \expandafter \@secondoftwo
\fi
}%
\providecommand \enquote [1]{``#1''}%
\providecommand \bibnamefont  [1]{#1}%
\providecommand \bibfnamefont [1]{#1}%
\providecommand \citenamefont [1]{#1}%
\providecommand\href[0]{\@sanitize\@href}%
\providecommand\@href[1]{\endgroup\@@startlink{#1}\endgroup\@@href}%
\providecommand\@@href[1]{#1\@@endlink}%
\providecommand \@sanitize [0]{\begingroup\catcode`\&12\catcode`\#12\relax}%
\@ifxundefined \pdfoutput {\@firstoftwo}{%
 \@ifnum{\z@=\pdfoutput}{\@firstoftwo}{\@secondoftwo}%
}{%
 \providecommand\@@startlink[1]{\leavevmode\special{html:<a href="#1">}}%
 \providecommand\@@endlink[0]{\special{html:</a>}}%
}{%
 \providecommand\@@startlink[1]{%
  \leavevmode
  \pdfstartlink
   attr{/Border[0 0 1 ]/H/I/C[0 1 1]}%
   user{/Subtype/Link/A<</Type/Action/S/URI/URI(#1)>>}%
  \relax
 }%
 \providecommand\@@endlink[0]{\pdfendlink}%
}%
\providecommand \url  [0]{\begingroup\@sanitize \@url }%
\providecommand \@url [1]{\endgroup\@href {#1}{\urlprefix}}%
\providecommand \urlprefix [0]{URL }%
\providecommand \Eprint[0]{\href }%
\@ifxundefined \urlstyle {%
  \providecommand \doi [1]{doi:\discretionary{}{}{}#1}%
}{%
  \providecommand \doi [0]{doi:\discretionary{}{}{}\begingroup
  \urlstyle{rm}\Url }%
}%
\providecommand \doibase [0]{http://dx.doi.org/}%
\providecommand \Doi[1]{\href{\doibase#1}}%
\providecommand \bibAnnote [3]{%
  \BibitemShut{#1}%
  \begin{quotation}\noindent
    \textsc{Key:}\ #2\\\textsc{Annotation:}\ #3%
  \end{quotation}%
}%
\providecommand \bibAnnoteFile [2]{%
  \IfFileExists{#2}{\bibAnnote {#1} {#2} {\input{#2}}}{}%
}%
\providecommand \typeout [0]{\immediate \write \m@ne }%
\providecommand \selectlanguage [0]{\@gobble}%
\providecommand \bibinfo [0]{\@secondoftwo}%
\providecommand \bibfield [0]{\@secondoftwo}%
\providecommand \translation [1]{[#1]}%
\providecommand \BibitemOpen[0]{}%
\providecommand \bibitemStop [0]{}%
\providecommand \bibitemNoStop [0]{.\EOS\space}%
\providecommand \EOS [0]{\spacefactor3000\relax}%
\providecommand \BibitemShut [1]{\csname bibitem#1\endcsname}%
%</preamble>
\bibitem{2012MNRAS.427..146H}%
  \BibitemOpen
  \bibfield{author}{%
  \bibinfo {author} {\bibfnamefont{C.}~\bibnamefont{{Heymans}}}, \bibinfo
  {author} {\bibfnamefont{L.}~\bibnamefont{{Van Waerbeke}}}, \bibinfo {author}
  {\bibfnamefont{L.}~\bibnamefont{{Miller}}}, \bibinfo {author}
  {\bibfnamefont{T.}~\bibnamefont{{Erben}}}, \bibinfo {author}
  {\bibfnamefont{H.}~\bibnamefont{{Hildebrandt}}}, \bibinfo {author}
  {\bibfnamefont{H.}~\bibnamefont{{Hoekstra}}}, \bibinfo {author}
  {\bibfnamefont{T.~D.}\ \bibnamefont{{Kitching}}}, \bibinfo {author}
  {\bibfnamefont{Y.}~\bibnamefont{{Mellier}}}, \bibinfo {author}
  {\bibfnamefont{P.}~\bibnamefont{{Simon}}}, \bibinfo {author}
  {\bibfnamefont{C.}~\bibnamefont{{Bonnett}}}, \bibinfo {author}
  {\bibfnamefont{J.}~\bibnamefont{{Coupon}}}, \bibinfo {author}
  {\bibfnamefont{L.}~\bibnamefont{{Fu}}}, \bibinfo {author}
  {\bibfnamefont{J.}~\bibnamefont{{Harnois D{\'e}raps}}}, \bibinfo {author}
  {\bibfnamefont{M.~J.}\ \bibnamefont{{Hudson}}}, \bibinfo {author}
  {\bibfnamefont{M.}~\bibnamefont{{Kilbinger}}}, \bibinfo {author}
  {\bibfnamefont{K.}~\bibnamefont{{Kuijken}}}, \bibinfo {author}
  {\bibfnamefont{B.}~\bibnamefont{{Rowe}}}, \bibinfo {author}
  {\bibfnamefont{T.}~\bibnamefont{{Schrabback}}}, \bibinfo {author}
  {\bibfnamefont{E.}~\bibnamefont{{Semboloni}}}, \bibinfo {author}
  {\bibfnamefont{E.}~\bibnamefont{{van Uitert}}}, \bibinfo {author}
  {\bibfnamefont{S.}~\bibnamefont{{Vafaei}}},\ and\ \bibinfo {author}
  {\bibfnamefont{M.}~\bibnamefont{{Velander}}},\ }%
  \bibfield{journal}{%
  \Doi{10.1111/j.1365-2966.2012.21952.x}{\bibinfo {journal} {MNRAS}}\ }%
  \textbf{\bibinfo {volume} {427}},\ \bibinfo {pages} {146} (\bibinfo {month}
  {Nov.}\ \bibinfo {year} {2012}),\
  \Eprint{http://arxiv.org/abs/1210.0032}{arXiv:1210.0032}%
  \bibAnnoteFile{NoStop}{2012MNRAS.427..146H}%
\bibitem{2014MNRAS.441...24A}%
  \BibitemOpen
  \bibfield{author}{%
  \bibinfo {author} {\bibfnamefont{L.}~\bibnamefont{{Anderson}}}\ and\ \bibinfo
  {author} {\bibnamefont{et~al.}},\ }%
  \bibfield{journal}{%
  \Doi{10.1093/mnras/stu523}{\bibinfo {journal} {MNRAS}}\ }%
  \textbf{\bibinfo {volume} {441}},\ \bibinfo {pages} {24} (\bibinfo {month}
  {Jun.}\ \bibinfo {year} {2014}),\
  \Eprint{http://arxiv.org/abs/1312.4877}{arXiv:1312.4877}%
  \bibAnnoteFile{NoStop}{2014MNRAS.441...24A}%
\bibitem{Abbott:2015swa}%
  \BibitemOpen
  \bibfield{author}{%
  \bibinfo {author} {\bibfnamefont{T.}~\bibnamefont{Abbott}} \emph{et~al.}
  (\bibinfo {collaboration} {DES})}%
   (\bibinfo {year} {2015}),\
  \Eprint{http://arxiv.org/abs/1507.05552}{arXiv:1507.05552 [astro-ph.CO]}%
  \bibAnnoteFile{NoStop}{Abbott:2015swa}%
%%CITATION = ARXIV:1507.05552;%%
\bibitem{2011arXiv1110.3193L}%
  \BibitemOpen
  \bibfield{author}{%
  \bibinfo {author} {\bibfnamefont{R.}~\bibnamefont{{Laureijs}}}, \bibinfo
  {author} {\bibfnamefont{J.}~\bibnamefont{{Amiaux}}}, \bibinfo {author}
  {\bibfnamefont{S.}~\bibnamefont{{Arduini}}}, \bibinfo {author}
  {\bibfnamefont{J.~.}\ \bibnamefont{{Augu{\`e}res}}}, \bibinfo {author}
  {\bibfnamefont{J.}~\bibnamefont{{Brinchmann}}}, \bibinfo {author}
  {\bibfnamefont{R.}~\bibnamefont{{Cole}}}, \bibinfo {author}
  {\bibfnamefont{M.}~\bibnamefont{{Cropper}}}, \bibinfo {author}
  {\bibfnamefont{C.}~\bibnamefont{{Dabin}}}, \bibinfo {author}
  {\bibfnamefont{L.}~\bibnamefont{{Duvet}}}, \bibinfo {author}
  {\bibfnamefont{A.}~\bibnamefont{{Ealet}}},\ and\ \bibinfo {author}
  {\bibnamefont{et~al.}},\ }%
  \bibfield{journal}{%
  \bibinfo {journal} {ArXiv e-prints}}%
   (\bibinfo {month} {Oct.}\ \bibinfo {year} {2011}),\
  \Eprint{http://arxiv.org/abs/1110.3193}{arXiv:1110.3193 [astro-ph.CO]}%
  \bibAnnoteFile{NoStop}{2011arXiv1110.3193L}%
\bibitem{2013arXiv1308.0847L}%
  \BibitemOpen
  \bibfield{author}{%
  \bibinfo {author} {\bibfnamefont{M.}~\bibnamefont{{Levi}}}, \bibinfo {author}
  {\bibfnamefont{C.}~\bibnamefont{{Bebek}}}, \bibinfo {author}
  {\bibfnamefont{T.}~\bibnamefont{{Beers}}}, \bibinfo {author}
  {\bibfnamefont{R.}~\bibnamefont{{Blum}}}, \bibinfo {author}
  {\bibfnamefont{R.}~\bibnamefont{{Cahn}}}, \bibinfo {author}
  {\bibfnamefont{D.}~\bibnamefont{{Eisenstein}}}, \bibinfo {author}
  {\bibfnamefont{B.}~\bibnamefont{{Flaugher}}}, \bibinfo {author}
  {\bibfnamefont{K.}~\bibnamefont{{Honscheid}}}, \bibinfo {author}
  {\bibfnamefont{R.}~\bibnamefont{{Kron}}}, \bibinfo {author}
  {\bibfnamefont{O.}~\bibnamefont{{Lahav}}}, \bibinfo {author}
  {\bibfnamefont{P.}~\bibnamefont{{McDonald}}}, \bibinfo {author}
  {\bibfnamefont{N.}~\bibnamefont{{Roe}}}, \bibinfo {author}
  {\bibfnamefont{D.}~\bibnamefont{{Schlegel}}},\ and\ \bibinfo {author}
  {\bibnamefont{{representing the DESI collaboration}}},\ }%
  \bibfield{journal}{%
  \bibinfo {journal} {ArXiv e-prints}}%
   (\bibinfo {month} {Aug.}\ \bibinfo {year} {2013}),\
  \Eprint{http://arxiv.org/abs/1308.0847}{arXiv:1308.0847 [astro-ph.CO]}%
  \bibAnnoteFile{NoStop}{2013arXiv1308.0847L}%
\bibitem{2012arXiv1211.0310L}%
  \BibitemOpen
  \bibfield{author}{%
  \bibinfo {author} {\bibnamefont{{LSST Dark Energy Science Collaboration}}},\
  }%
  \bibfield{journal}{%
  \bibinfo {journal} {ArXiv e-prints}}%
   (\bibinfo {month} {Nov.}\ \bibinfo {year} {2012}),\
  \Eprint{http://arxiv.org/abs/1211.0310}{arXiv:1211.0310 [astro-ph.CO]}%
  \bibAnnoteFile{NoStop}{2012arXiv1211.0310L}%
\bibitem{Jain:2007yk}%
  \BibitemOpen
  \bibfield{author}{%
  \bibinfo {author} {\bibfnamefont{B.}~\bibnamefont{Jain}}\ and\ \bibinfo
  {author} {\bibfnamefont{P.}~\bibnamefont{Zhang}},\ }%
  \bibfield{journal}{%
  \Doi{10.1103/PhysRevD.78.063503}{\bibinfo {journal} {Phys.Rev.}}\ }%
  \textbf{\bibinfo {volume} {D78}},\ \bibinfo {pages} {063503} (\bibinfo {year}
  {2008}),\ \Eprint{http://arxiv.org/abs/0709.2375}{arXiv:0709.2375
  [astro-ph]}%
  \bibAnnoteFile{NoStop}{Jain:2007yk}%
%%CITATION = ARXIV:0709.2375;%%
\bibitem{2012PhR...513....1C}%
  \BibitemOpen
  \bibfield{author}{%
  \bibinfo {author} {\bibfnamefont{T.}~\bibnamefont{{Clifton}}}, \bibinfo
  {author} {\bibfnamefont{P.~G.}\ \bibnamefont{{Ferreira}}}, \bibinfo {author}
  {\bibfnamefont{A.}~\bibnamefont{{Padilla}}},\ and\ \bibinfo {author}
  {\bibfnamefont{C.}~\bibnamefont{{Skordis}}},\ }%
  \bibfield{journal}{%
  \Doi{10.1016/j.physrep.2012.01.001}{\bibinfo {journal} {PHYSREP}}\ }%
  \textbf{\bibinfo {volume} {513}},\ \bibinfo {pages} {1} (\bibinfo {month}
  {Mar.}\ \bibinfo {year} {2012}),\
  \Eprint{http://arxiv.org/abs/1106.2476}{arXiv:1106.2476 [astro-ph.CO]}%
  \bibAnnoteFile{NoStop}{2012PhR...513....1C}%
\bibitem{Joyce:2014kja}%
  \BibitemOpen
  \bibfield{author}{%
  \bibinfo {author} {\bibfnamefont{A.}~\bibnamefont{Joyce}}, \bibinfo {author}
  {\bibfnamefont{B.}~\bibnamefont{Jain}}, \bibinfo {author}
  {\bibfnamefont{J.}~\bibnamefont{Khoury}},\ and\ \bibinfo {author}
  {\bibfnamefont{M.}~\bibnamefont{Trodden}}}%
   (\bibinfo {year} {2014}),\
  \Eprint{http://arxiv.org/abs/1407.0059}{arXiv:1407.0059 [astro-ph.CO]}%
  \bibAnnoteFile{NoStop}{Joyce:2014kja}%
%%CITATION = ARXIV:1407.0059;%%
\bibitem{2015arXiv150404623K}%
  \BibitemOpen
  \bibfield{author}{%
  \bibinfo {author} {\bibfnamefont{K.}~\bibnamefont{{Koyama}}},\ }%
  \bibfield{journal}{%
  \bibinfo {journal} {ArXiv e-prints}}%
   (\bibinfo {month} {Apr.}\ \bibinfo {year} {2015}),\
  \Eprint{http://arxiv.org/abs/1504.04623}{arXiv:1504.04623}%
  \bibAnnoteFile{NoStop}{2015arXiv150404623K}%
\bibitem{2016arXiv160106133J}%
  \BibitemOpen
  \bibfield{author}{%
  \bibinfo {author} {\bibfnamefont{A.}~\bibnamefont{{Joyce}}}, \bibinfo
  {author} {\bibfnamefont{L.}~\bibnamefont{{Lombriser}}},\ and\ \bibinfo
  {author} {\bibfnamefont{F.}~\bibnamefont{{Schmidt}}},\ }%
  \bibfield{journal}{%
  \bibinfo {journal} {ArXiv e-prints}}%
   (\bibinfo {month} {Jan.}\ \bibinfo {year} {2016}),\
  \Eprint{http://arxiv.org/abs/1601.06133}{arXiv:1601.06133}%
  \bibAnnoteFile{NoStop}{2016arXiv160106133J}%
\bibitem{Will:2014xja}%
  \BibitemOpen
  \bibfield{author}{%
  \bibinfo {author} {\bibfnamefont{C.~M.}\ \bibnamefont{Will}},\ }%
  \bibfield{journal}{%
  \Doi{10.12942/lrr-2014-4}{\bibinfo {journal} {Living Rev.Rel.}}\ }%
  \textbf{\bibinfo {volume} {17}},\ \bibinfo {pages} {4} (\bibinfo {year}
  {2014}),\ \Eprint{http://arxiv.org/abs/1403.7377}{arXiv:1403.7377 [gr-qc]}%
  \bibAnnoteFile{NoStop}{Will:2014xja}%
%%CITATION = ARXIV:1403.7377;%%
\bibitem{Brax:2013ida}%
  \BibitemOpen
  \bibfield{author}{%
  \bibinfo {author} {\bibfnamefont{P.}~\bibnamefont{Brax}},\ }%
  \bibfield{journal}{%
  \Doi{10.1088/0264-9381/30/21/214005}{\bibinfo {journal} {Class. Quant.
  Grav.}}\ }%
  \textbf{\bibinfo {volume} {30}},\ \bibinfo {pages} {214005} (\bibinfo {year}
  {2013})%
  \bibAnnoteFile{NoStop}{Brax:2013ida}%
%%CITATION = CQGRD,30,214005;%%
\bibitem{khoury:2003aq}%
  \BibitemOpen
  \bibfield{author}{%
  \bibinfo {author} {\bibfnamefont{J.}~\bibnamefont{Khoury}}\ and\ \bibinfo
  {author} {\bibfnamefont{A.}~\bibnamefont{Weltman}},\ }%
  \bibfield{journal}{%
  \Doi{10.1103/PhysRevLett.93.171104}{\bibinfo {journal} {Phys.Rev.Lett.}}\ }%
  \textbf{\bibinfo {volume} {93}},\ \bibinfo {pages} {171104} (\bibinfo {year}
  {2004}),\
  \Eprint{http://arxiv.org/abs/astro-ph/0309300}{arXiv:astro-ph/0309300
  [astro-ph]}%
  \bibAnnoteFile{NoStop}{khoury:2003aq}%
%%CITATION = ASTRO-PH/0309300;%%
\bibitem{physrevd.69.044026}%
  \BibitemOpen
  \bibfield{author}{%
  \bibinfo {author} {\bibfnamefont{J.}~\bibnamefont{Khoury}}\ and\ \bibinfo
  {author} {\bibfnamefont{A.}~\bibnamefont{Weltman}},\ }%
  \bibfield{journal}{%
  \Doi{10.1103/PhysRevD.69.044026}{\bibinfo {journal} {Phys. Rev. D}}\ }%
  \textbf{\bibinfo {volume} {69}},\ \bibinfo {pages} {044026} (\bibinfo {year}
  {2004})%
  \bibAnnoteFile{NoStop}{physrevd.69.044026}%
\bibitem{Olive:2007aj}%
  \BibitemOpen
  \bibfield{author}{%
  \bibinfo {author} {\bibfnamefont{K.~A.}\ \bibnamefont{Olive}}\ and\ \bibinfo
  {author} {\bibfnamefont{M.}~\bibnamefont{Pospelov}},\ }%
  \bibfield{journal}{%
  \Doi{10.1103/PhysRevD.77.043524}{\bibinfo {journal} {Phys.Rev.}}\ }%
  \textbf{\bibinfo {volume} {D77}},\ \bibinfo {pages} {043524} (\bibinfo {year}
  {2008}),\ \Eprint{http://arxiv.org/abs/0709.3825}{arXiv:0709.3825 [hep-ph]}%
  \bibAnnoteFile{NoStop}{Olive:2007aj}%
%%CITATION = ARXIV:0709.3825;%%
\bibitem{hinterbichler:2010es}%
  \BibitemOpen
  \bibfield{author}{%
  \bibinfo {author} {\bibfnamefont{K.}~\bibnamefont{Hinterbichler}}\ and\
  \bibinfo {author} {\bibfnamefont{J.}~\bibnamefont{Khoury}},\ }%
  \bibfield{journal}{%
  \Doi{10.1103/PhysRevLett.104.231301}{\bibinfo {journal} {Phys.Rev.Lett.}}\ }%
  \textbf{\bibinfo {volume} {104}},\ \bibinfo {pages} {231301} (\bibinfo {year}
  {2010}),\ \Eprint{http://arxiv.org/abs/1001.4525}{arXiv:1001.4525 [hep-th]}%
  \bibAnnoteFile{NoStop}{hinterbichler:2010es}%
%%CITATION = ARXIV:1001.4525;%%
\bibitem{hinterbichler:2011ca}%
  \BibitemOpen
  \bibfield{author}{%
  \bibinfo {author} {\bibfnamefont{K.}~\bibnamefont{Hinterbichler}}, \bibinfo
  {author} {\bibfnamefont{J.}~\bibnamefont{Khoury}}, \bibinfo {author}
  {\bibfnamefont{A.}~\bibnamefont{Levy}},\ and\ \bibinfo {author}
  {\bibfnamefont{A.}~\bibnamefont{Matas}},\ }%
  \bibfield{journal}{%
  \Doi{10.1103/PhysRevD.84.103521}{\bibinfo {journal} {Phys.Rev.}}\ }%
  \textbf{\bibinfo {volume} {D84}},\ \bibinfo {pages} {103521} (\bibinfo {year}
  {2011}),\ \Eprint{http://arxiv.org/abs/1107.2112}{arXiv:1107.2112
  [astro-ph.CO]}%
  \bibAnnoteFile{NoStop}{hinterbichler:2011ca}%
%%CITATION = ARXIV:1107.2112;%%
\bibitem{brax:2010gi}%
  \BibitemOpen
  \bibfield{author}{%
  \bibinfo {author} {\bibfnamefont{P.}~\bibnamefont{Brax}}, \bibinfo {author}
  {\bibfnamefont{C.}~\bibnamefont{van~de Bruck}}, \bibinfo {author}
  {\bibfnamefont{A.-C.}\ \bibnamefont{Davis}},\ and\ \bibinfo {author}
  {\bibfnamefont{D.}~\bibnamefont{Shaw}},\ }%
  \bibfield{journal}{%
  \Doi{10.1103/PhysRevD.82.063519}{\bibinfo {journal} {Phys.Rev.}}\ }%
  \textbf{\bibinfo {volume} {D82}},\ \bibinfo {pages} {063519} (\bibinfo {year}
  {2010}),\ \Eprint{http://arxiv.org/abs/1005.3735}{arXiv:1005.3735
  [astro-ph.CO]}%
  \bibAnnoteFile{NoStop}{brax:2010gi}%
%%CITATION = ARXIV:1005.3735;%%
\bibitem{brax:2011ja}%
  \BibitemOpen
  \bibfield{author}{%
  \bibinfo {author} {\bibfnamefont{P.}~\bibnamefont{Brax}}, \bibinfo {author}
  {\bibfnamefont{C.}~\bibnamefont{van~de Bruck}}, \bibinfo {author}
  {\bibfnamefont{A.-C.}\ \bibnamefont{Davis}}, \bibinfo {author}
  {\bibfnamefont{B.}~\bibnamefont{Li}},\ and\ \bibinfo {author}
  {\bibfnamefont{D.~J.}\ \bibnamefont{Shaw}},\ }%
  \bibfield{journal}{%
  \Doi{10.1103/PhysRevD.83.104026}{\bibinfo {journal} {Phys.Rev.}}\ }%
  \textbf{\bibinfo {volume} {D83}},\ \bibinfo {pages} {104026} (\bibinfo {year}
  {2011}),\ \Eprint{http://arxiv.org/abs/1102.3692}{arXiv:1102.3692
  [astro-ph.CO]}%
  \bibAnnoteFile{NoStop}{brax:2011ja}%
%%CITATION = ARXIV:1102.3692;%%
\bibitem{Vainshtein1972393}%
  \BibitemOpen
  \bibfield{author}{%
  \bibinfo {author} {\bibfnamefont{A.}~\bibnamefont{Vainshtein}},\ }%
  \bibfield{journal}{%
  \Doi{10.1016/0370-2693(72)90147-5}{\bibinfo {journal} {Phys.~Lett.~B}}\ }%
  \textbf{\bibinfo {volume} {39}},\ \bibinfo {pages} {393 } (\bibinfo {year}
  {1972}),\ ISSN \bibinfo {issn} {0370-2693}%
  \bibAnnoteFile{NoStop}{Vainshtein1972393}%
\bibitem{Babichev:2013usa}%
  \BibitemOpen
  \bibfield{author}{%
  \bibinfo {author} {\bibfnamefont{E.}~\bibnamefont{Babichev}}\ and\ \bibinfo
  {author} {\bibfnamefont{C.}~\bibnamefont{Deffayet}},\ }%
  \bibfield{journal}{%
  \Doi{10.1088/0264-9381/30/18/184001}{\bibinfo {journal} {Class.Quant.Grav.}}\
  }%
  \textbf{\bibinfo {volume} {30}},\ \bibinfo {pages} {184001} (\bibinfo {year}
  {2013}),\ \Eprint{http://arxiv.org/abs/1304.7240}{arXiv:1304.7240 [gr-qc]}%
  \bibAnnoteFile{NoStop}{Babichev:2013usa}%
%%CITATION = ARXIV:1304.7240;%%
\bibitem{Koyama:2013paa}%
  \BibitemOpen
  \bibfield{author}{%
  \bibinfo {author} {\bibfnamefont{K.}~\bibnamefont{Koyama}}, \bibinfo {author}
  {\bibfnamefont{G.}~\bibnamefont{Niz}},\ and\ \bibinfo {author}
  {\bibfnamefont{G.}~\bibnamefont{Tasinato}},\ }%
  \bibfield{journal}{%
  \Doi{10.1103/PhysRevD.88.021502}{\bibinfo {journal} {Phys.Rev.}}\ }%
  \textbf{\bibinfo {volume} {D88}},\ \bibinfo {pages} {021502} (\bibinfo {year}
  {2013}),\ \Eprint{http://arxiv.org/abs/1305.0279}{arXiv:1305.0279 [hep-th]}%
  \bibAnnoteFile{NoStop}{Koyama:2013paa}%
%%CITATION = ARXIV:1305.0279;%%
\bibitem{Babichev:2009ee}%
  \BibitemOpen
  \bibfield{author}{%
  \bibinfo {author} {\bibfnamefont{E.}~\bibnamefont{Babichev}}, \bibinfo
  {author} {\bibfnamefont{C.}~\bibnamefont{Deffayet}},\ and\ \bibinfo {author}
  {\bibfnamefont{R.}~\bibnamefont{Ziour}},\ }%
  \bibfield{journal}{%
  \Doi{10.1142/S0218271809016107}{\bibinfo {journal} {Int.J.Mod.Phys.}}\ }%
  \textbf{\bibinfo {volume} {D18}},\ \bibinfo {pages} {2147} (\bibinfo {year}
  {2009}),\ \Eprint{http://arxiv.org/abs/0905.2943}{arXiv:0905.2943 [hep-th]}%
  \bibAnnoteFile{NoStop}{Babichev:2009ee}%
%%CITATION = ARXIV:0905.2943;%%
\bibitem{brax:2014gra}%
  \BibitemOpen
  \bibfield{author}{%
  \bibinfo {author} {\bibfnamefont{P.}~\bibnamefont{Brax}}\ and\ \bibinfo
  {author} {\bibfnamefont{P.}~\bibnamefont{Valageas}},\ }%
  \bibfield{journal}{%
  \Doi{10.1103/PhysRevD.90.123521}{\bibinfo {journal} {Phys.Rev.}}\ }%
  \textbf{\bibinfo {volume} {D90}},\ \bibinfo {pages} {123521} (\bibinfo {year}
  {2014}),\ \Eprint{http://arxiv.org/abs/1408.0969}{arXiv:1408.0969
  [astro-ph.CO]}%
  \bibAnnoteFile{NoStop}{brax:2014gra}%
%%CITATION = ARXIV:1408.0969;%%
\bibitem{2016MNRAS.455.3367G}%
  \BibitemOpen
  \bibfield{author}{%
  \bibinfo {author} {\bibfnamefont{D.}~\bibnamefont{{Gruen}}}, \bibinfo
  {author} {\bibfnamefont{O.}~\bibnamefont{{Friedrich}}},\ and\ \bibinfo
  {author} {\bibnamefont{et~al}},\ }%
  \bibfield{journal}{%
  \Doi{10.1093/mnras/stv2506}{\bibinfo {journal} {MNRAS}}\ }%
  \textbf{\bibinfo {volume} {455}},\ \bibinfo {pages} {3367} (\bibinfo {month}
  {Jan.}\ \bibinfo {year} {2016}),\
  \Eprint{http://arxiv.org/abs/1507.05090}{arXiv:1507.05090}%
  \bibAnnoteFile{NoStop}{2016MNRAS.455.3367G}%
\bibitem{melchior:2013gxd}%
  \BibitemOpen
  \bibfield{author}{%
  \bibinfo {author} {\bibfnamefont{P.}~\bibnamefont{Melchior}}, \bibinfo
  {author} {\bibfnamefont{P.}~\bibnamefont{Sutter}}, \bibinfo {author}
  {\bibfnamefont{E.~S.}\ \bibnamefont{Sheldon}}, \bibinfo {author}
  {\bibfnamefont{E.}~\bibnamefont{Krause}},\ and\ \bibinfo {author}
  {\bibfnamefont{B.~D.}\ \bibnamefont{Wandelt}},\ }%
  \bibfield{journal}{%
  \Doi{10.1093/mnras/stu456}{\bibinfo {journal} {Mon.Not.Roy.Astron.Soc.}},\
  \bibinfo {pages} {2922}}%
   (\bibinfo {year} {2014}),\
  \Eprint{http://arxiv.org/abs/1309.2045}{arXiv:1309.2045 [astro-ph.CO]}%
  \bibAnnoteFile{NoStop}{melchior:2013gxd}%
%%CITATION = ARXIV:1309.2045;%%
\bibitem{clampitt:2014gpa}%
  \BibitemOpen
  \bibfield{author}{%
  \bibinfo {author} {\bibfnamefont{J.}~\bibnamefont{Clampitt}}\ and\ \bibinfo
  {author} {\bibfnamefont{B.}~\bibnamefont{Jain}}}%
   (\bibinfo {year} {2014}),\
  \Eprint{http://arxiv.org/abs/1404.1834}{arXiv:1404.1834 [astro-ph.CO]}%
  \bibAnnoteFile{NoStop}{clampitt:2014gpa}%
%%CITATION = ARXIV:1404.1834;%%
\bibitem{2016arXiv160503982S}%
  \BibitemOpen
  \bibfield{author}{%
  \bibinfo {author} {\bibfnamefont{C.}~\bibnamefont{{S{\'a}nchez}}}\ and\
  \bibinfo {author} {\bibnamefont{et~al}},\ }%
  \bibfield{journal}{%
  \bibinfo {journal} {ArXiv e-prints}}%
   (\bibinfo {month} {May}\ \bibinfo {year} {2016}),\
  \Eprint{http://arxiv.org/abs/1605.03982}{arXiv:1605.03982}%
  \bibAnnoteFile{NoStop}{2016arXiv160503982S}%
\bibitem{2015JCAP...08..028B}%
  \BibitemOpen
  \bibfield{author}{%
  \bibinfo {author} {\bibfnamefont{A.}~\bibnamefont{{Barreira}}}, \bibinfo
  {author} {\bibfnamefont{M.}~\bibnamefont{{Cautun}}}, \bibinfo {author}
  {\bibfnamefont{B.}~\bibnamefont{{Li}}}, \bibinfo {author}
  {\bibfnamefont{C.~M.}\ \bibnamefont{{Baugh}}},\ and\ \bibinfo {author}
  {\bibfnamefont{S.}~\bibnamefont{{Pascoli}}},\ }%
  \bibfield{journal}{%
  \Doi{10.1088/1475-7516/2015/08/028}{\bibinfo {journal} {JCAP}}\ }%
  \textbf{\bibinfo {volume} {8}},\ \bibinfo {eid} {028} (\bibinfo {month}
  {Aug.}\ \bibinfo {year} {2015}),\
  \Eprint{http://arxiv.org/abs/1505.05809}{arXiv:1505.05809}%
  \bibAnnoteFile{NoStop}{2015JCAP...08..028B}%
\bibitem{2000ApJ...530..547J}%
  \BibitemOpen
  \bibfield{author}{%
  \bibinfo {author} {\bibfnamefont{B.}~\bibnamefont{{Jain}}}, \bibinfo {author}
  {\bibfnamefont{U.}~\bibnamefont{{Seljak}}},\ and\ \bibinfo {author}
  {\bibfnamefont{S.}~\bibnamefont{{White}}},\ }%
  \bibfield{journal}{%
  \Doi{10.1086/308384}{\bibinfo {journal} {APJ}}\ }%
  \textbf{\bibinfo {volume} {530}},\ \bibinfo {pages} {547} (\bibinfo {month}
  {Feb.}\ \bibinfo {year} {2000}),\
  \Eprint{http://arxiv.org/abs/astro-ph/9901191}{astro-ph/9901191}%
  \bibAnnoteFile{NoStop}{2000ApJ...530..547J}%
\bibitem{2003ApJ...592..699V}%
  \BibitemOpen
  \bibfield{author}{%
  \bibinfo {author} {\bibfnamefont{C.}~\bibnamefont{{Vale}}}\ and\ \bibinfo
  {author} {\bibfnamefont{M.}~\bibnamefont{{White}}},\ }%
  \bibfield{journal}{%
  \Doi{10.1086/375867}{\bibinfo {journal} {APJ}}\ }%
  \textbf{\bibinfo {volume} {592}},\ \bibinfo {pages} {699} (\bibinfo {month}
  {Aug.}\ \bibinfo {year} {2003}),\
  \Eprint{http://arxiv.org/abs/astro-ph/0303555}{astro-ph/0303555}%
  \bibAnnoteFile{NoStop}{2003ApJ...592..699V}%
\bibitem{Hilbert:2008kb}%
  \BibitemOpen
  \bibfield{author}{%
  \bibinfo {author} {\bibfnamefont{S.}~\bibnamefont{Hilbert}}, \bibinfo
  {author} {\bibfnamefont{J.}~\bibnamefont{Hartlap}}, \bibinfo {author}
  {\bibfnamefont{S.~D.~M.}\ \bibnamefont{White}},\ and\ \bibinfo {author}
  {\bibfnamefont{P.}~\bibnamefont{Schneider}},\ }%
  \bibfield{journal}{%
  \Doi{10.1051/0004-6361/200811054}{\bibinfo {journal} {Astron. Astrophys.}}\
  }%
  \textbf{\bibinfo {volume} {499}},\ \bibinfo {pages} {31} (\bibinfo {year}
  {2009}),\ \Eprint{http://arxiv.org/abs/0809.5035}{arXiv:0809.5035
  [astro-ph]}%
  \bibAnnoteFile{NoStop}{Hilbert:2008kb}%
%%CITATION = ARXIV:0809.5035;%%
\bibitem{2015arXiv151108211G}%
  \BibitemOpen
  \bibfield{author}{%
  \bibinfo {author} {\bibfnamefont{C.}~\bibnamefont{{Giocoli}}}, \bibinfo
  {author} {\bibfnamefont{E.}~\bibnamefont{{Jullo}}}, \bibinfo {author}
  {\bibfnamefont{R.~B.}\ \bibnamefont{{Metcalf}}}, \bibinfo {author}
  {\bibfnamefont{S.}~\bibnamefont{{de la Torre}}}, \bibinfo {author}
  {\bibfnamefont{G.}~\bibnamefont{{Yepes}}}, \bibinfo {author}
  {\bibfnamefont{F.}~\bibnamefont{{Prada}}}, \bibinfo {author}
  {\bibfnamefont{J.}~\bibnamefont{{Comparat}}}, \bibinfo {author}
  {\bibfnamefont{S.}~\bibnamefont{{Goettlober}}}, \bibinfo {author}
  {\bibfnamefont{A.}~\bibnamefont{{Kyplin}}}, \bibinfo {author}
  {\bibfnamefont{J.-P.}\ \bibnamefont{{Kneib}}}, \bibinfo {author}
  {\bibfnamefont{M.}~\bibnamefont{{Petkova}}}, \bibinfo {author}
  {\bibfnamefont{H.}~\bibnamefont{{Shan}}},\ and\ \bibinfo {author}
  {\bibfnamefont{N.}~\bibnamefont{{Tessore}}},\ }%
  \bibfield{journal}{%
  \bibinfo {journal} {ArXiv e-prints}}%
   (\bibinfo {month} {Nov.}\ \bibinfo {year} {2015}),\
  \Eprint{http://arxiv.org/abs/1511.08211}{arXiv:1511.08211}%
  \bibAnnoteFile{NoStop}{2015arXiv151108211G}%
\bibitem{2011PhRvL.106t1102W}%
  \BibitemOpen
  \bibfield{author}{%
  \bibinfo {author} {\bibfnamefont{M.}~\bibnamefont{{Wyman}}},\ }%
  \bibfield{journal}{%
  \Doi{10.1103/PhysRevLett.106.201102}{\bibinfo {journal} {Physical Review
  Letters}}\ }%
  \textbf{\bibinfo {volume} {106}},\ \bibinfo {eid} {201102} (\bibinfo {month}
  {May}\ \bibinfo {year} {2011}),\
  \Eprint{http://arxiv.org/abs/1101.1295}{arXiv:1101.1295 [astro-ph.CO]}%
  \bibAnnoteFile{NoStop}{2011PhRvL.106t1102W}%
\bibitem{2015PhRvD..91f4012P}%
  \BibitemOpen
  \bibfield{author}{%
  \bibinfo {author} {\bibfnamefont{Y.}~\bibnamefont{{Park}}}\ and\ \bibinfo
  {author} {\bibfnamefont{M.}~\bibnamefont{{Wyman}}},\ }%
  \bibfield{journal}{%
  \Doi{10.1103/PhysRevD.91.064012}{\bibinfo {journal} {PRD}}\ }%
  \textbf{\bibinfo {volume} {91}},\ \bibinfo {eid} {064012} (\bibinfo {month}
  {Mar.}\ \bibinfo {year} {2015}),\
  \Eprint{http://arxiv.org/abs/1408.4773}{arXiv:1408.4773}%
  \bibAnnoteFile{NoStop}{2015PhRvD..91f4012P}%
\bibitem{2015MNRAS.454.4085B}%
  \BibitemOpen
  \bibfield{author}{%
  \bibinfo {author} {\bibfnamefont{A.}~\bibnamefont{{Barreira}}}, \bibinfo
  {author} {\bibfnamefont{B.}~\bibnamefont{{Li}}}, \bibinfo {author}
  {\bibfnamefont{E.}~\bibnamefont{{Jennings}}}, \bibinfo {author}
  {\bibfnamefont{J.}~\bibnamefont{{Merten}}}, \bibinfo {author}
  {\bibfnamefont{L.}~\bibnamefont{{King}}}, \bibinfo {author}
  {\bibfnamefont{C.~M.}\ \bibnamefont{{Baugh}}},\ and\ \bibinfo {author}
  {\bibfnamefont{S.}~\bibnamefont{{Pascoli}}},\ }%
  \bibfield{journal}{%
  \Doi{10.1093/mnras/stv2211}{\bibinfo {journal} {MNRAS}}\ }%
  \textbf{\bibinfo {volume} {454}},\ \bibinfo {pages} {4085} (\bibinfo {month}
  {Dec.}\ \bibinfo {year} {2015}),\
  \Eprint{http://arxiv.org/abs/1505.03468}{arXiv:1505.03468}%
  \bibAnnoteFile{NoStop}{2015MNRAS.454.4085B}%
\bibitem{2015MNRAS.451.1036C}%
  \BibitemOpen
  \bibfield{author}{%
  \bibinfo {author} {\bibfnamefont{Y.-C.}\ \bibnamefont{{Cai}}}, \bibinfo
  {author} {\bibfnamefont{N.}~\bibnamefont{{Padilla}}},\ and\ \bibinfo {author}
  {\bibfnamefont{B.}~\bibnamefont{{Li}}},\ }%
  \bibfield{journal}{%
  \Doi{10.1093/mnras/stv777}{\bibinfo {journal} {MNRAS}}\ }%
  \textbf{\bibinfo {volume} {451}},\ \bibinfo {pages} {1036} (\bibinfo {month}
  {Jul.}\ \bibinfo {year} {2015}),\
  \Eprint{http://arxiv.org/abs/1410.1510}{arXiv:1410.1510}%
  \bibAnnoteFile{NoStop}{2015MNRAS.451.1036C}%
\bibitem{2015JCAP...10..036T}%
  \BibitemOpen
  \bibfield{author}{%
  \bibinfo {author} {\bibfnamefont{N.}~\bibnamefont{{Tessore}}}, \bibinfo
  {author} {\bibfnamefont{H.~A.}\ \bibnamefont{{Winther}}}, \bibinfo {author}
  {\bibfnamefont{R.~B.}\ \bibnamefont{{Metcalf}}}, \bibinfo {author}
  {\bibfnamefont{P.~G.}\ \bibnamefont{{Ferreira}}},\ and\ \bibinfo {author}
  {\bibfnamefont{C.}~\bibnamefont{{Giocoli}}},\ }%
  \bibfield{journal}{%
  \Doi{10.1088/1475-7516/2015/10/036}{\bibinfo {journal} {JCAP}}\ }%
  \textbf{\bibinfo {volume} {10}},\ \bibinfo {eid} {036} (\bibinfo {month}
  {Oct.}\ \bibinfo {year} {2015}),\
  \Eprint{http://arxiv.org/abs/1508.04011}{arXiv:1508.04011}%
  \bibAnnoteFile{NoStop}{2015JCAP...10..036T}%
\bibitem{2016MNRAS.459.2762H}%
  \BibitemOpen
  \bibfield{author}{%
  \bibinfo {author} {\bibfnamefont{Y.}~\bibnamefont{{Higuchi}}}\ and\ \bibinfo
  {author} {\bibfnamefont{M.}~\bibnamefont{{Shirasaki}}},\ }%
  \bibfield{journal}{%
  \Doi{10.1093/mnras/stw814}{\bibinfo {journal} {MNRAS}}\ }%
  \textbf{\bibinfo {volume} {459}},\ \bibinfo {pages} {2762} (\bibinfo {month}
  {Jul.}\ \bibinfo {year} {2016}),\
  \Eprint{http://arxiv.org/abs/1603.01325}{arXiv:1603.01325}%
  \bibAnnoteFile{NoStop}{2016MNRAS.459.2762H}%
\bibitem{2016arXiv160102012B}%
  \BibitemOpen
  \bibfield{author}{%
  \bibinfo {author} {\bibfnamefont{A.}~\bibnamefont{{Barreira}}}, \bibinfo
  {author} {\bibfnamefont{C.}~\bibnamefont{{Llinares}}}, \bibinfo {author}
  {\bibfnamefont{S.}~\bibnamefont{{Bose}}},\ and\ \bibinfo {author}
  {\bibfnamefont{B.}~\bibnamefont{{Li}}},\ }%
  \bibfield{journal}{%
  \bibinfo {journal} {ArXiv e-prints}}%
   (\bibinfo {month} {Jan.}\ \bibinfo {year} {2016}),\
  \Eprint{http://arxiv.org/abs/1601.02012}{arXiv:1601.02012}%
  \bibAnnoteFile{NoStop}{2016arXiv160102012B}%
\bibitem{whitehu2000}%
  \BibitemOpen
  \bibfield{author}{%
  \bibinfo {author} {\bibfnamefont{M.}~\bibnamefont{{White}}}\ and\ \bibinfo
  {author} {\bibfnamefont{W.}~\bibnamefont{{Hu}}},\ }%
  \bibfield{journal}{%
  \Doi{10.1086/309009}{\bibinfo {journal} {APJ}}\ }%
  \textbf{\bibinfo {volume} {537}},\ \bibinfo {pages} {1} (\bibinfo {month}
  {Jul.}\ \bibinfo {year} {2000}),\
  \Eprint{http://arxiv.org/abs/astro-ph/9909165}{astro-ph/9909165}%
  \bibAnnoteFile{NoStop}{whitehu2000}%
\bibitem{li2001}%
  \BibitemOpen
  \bibfield{author}{%
  \bibinfo {author} {\bibfnamefont{B.}~\bibnamefont{{Li}}}, \bibinfo {author}
  {\bibfnamefont{L.~J.}\ \bibnamefont{{King}}}, \bibinfo {author}
  {\bibfnamefont{G.-B.}\ \bibnamefont{{Zhao}}},\ and\ \bibinfo {author}
  {\bibfnamefont{H.}~\bibnamefont{{Zhao}}},\ }%
  \bibfield{journal}{%
  \Doi{10.1111/j.1365-2966.2011.18754.x}{\bibinfo {journal} {MNRAS}}\ }%
  \textbf{\bibinfo {volume} {415}},\ \bibinfo {pages} {881} (\bibinfo {month}
  {Jul.}\ \bibinfo {year} {2011}),\
  \Eprint{http://arxiv.org/abs/1012.1625}{arXiv:1012.1625}%
  \bibAnnoteFile{NoStop}{li2001}%
\bibitem{2002A&A...385..337T}%
  \BibitemOpen
  \bibfield{author}{%
  \bibinfo {author} {\bibfnamefont{R.}~\bibnamefont{{Teyssier}}},\ }%
  \bibfield{journal}{%
  \Doi{10.1051/0004-6361:20011817}{\bibinfo {journal} {AAP}}\ }%
  \textbf{\bibinfo {volume} {385}},\ \bibinfo {pages} {337} (\bibinfo {month}
  {Apr.}\ \bibinfo {year} {2002}),\
  \Eprint{http://arxiv.org/abs/astro-ph/0111367}{astro-ph/0111367}%
  \bibAnnoteFile{NoStop}{2002A&A...385..337T}%
\bibitem{2012JCAP...01..051L}%
  \BibitemOpen
  \bibfield{author}{%
  \bibinfo {author} {\bibfnamefont{B.}~\bibnamefont{{Li}}}, \bibinfo {author}
  {\bibfnamefont{G.-B.}\ \bibnamefont{{Zhao}}}, \bibinfo {author}
  {\bibfnamefont{R.}~\bibnamefont{{Teyssier}}},\ and\ \bibinfo {author}
  {\bibfnamefont{K.}~\bibnamefont{{Koyama}}},\ }%
  \bibfield{journal}{%
  \Doi{10.1088/1475-7516/2012/01/051}{\bibinfo {journal} {JCAP}}\ }%
  \textbf{\bibinfo {volume} {1}},\ \bibinfo {eid} {051} (\bibinfo {month}
  {Jan.}\ \bibinfo {year} {2012}),\
  \Eprint{http://arxiv.org/abs/1110.1379}{arXiv:1110.1379 [astro-ph.CO]}%
  \bibAnnoteFile{NoStop}{2012JCAP...01..051L}%
\bibitem{baojiudgp}%
  \BibitemOpen
  \bibfield{author}{%
  \bibinfo {author} {\bibfnamefont{B.}~\bibnamefont{{Li}}}, \bibinfo {author}
  {\bibfnamefont{G.-B.}\ \bibnamefont{{Zhao}}},\ and\ \bibinfo {author}
  {\bibfnamefont{K.}~\bibnamefont{{Koyama}}},\ }%
  \bibfield{journal}{%
  \Doi{10.1088/1475-7516/2013/05/023}{\bibinfo {journal} {JCAP}}\ }%
  \textbf{\bibinfo {volume} {5}},\ \bibinfo {eid} {023} (\bibinfo {month}
  {May}\ \bibinfo {year} {2013}),\
  \Eprint{http://arxiv.org/abs/1303.0008}{arXiv:1303.0008 [astro-ph.CO]}%
  \bibAnnoteFile{NoStop}{baojiudgp}%
\bibitem{Dvali:2000hr}%
  \BibitemOpen
  \bibfield{author}{%
  \bibinfo {author} {\bibfnamefont{G.}~\bibnamefont{Dvali}}, \bibinfo {author}
  {\bibfnamefont{G.}~\bibnamefont{Gabadadze}},\ and\ \bibinfo {author}
  {\bibfnamefont{M.}~\bibnamefont{Porrati}},\ }%
  \bibfield{journal}{%
  \Doi{10.1016/S0370-2693(00)00669-9}{\bibinfo {journal} {Phys.Lett.}}\ }%
  \textbf{\bibinfo {volume} {B485}},\ \bibinfo {pages} {208} (\bibinfo {year}
  {2000}),\ \Eprint{http://arxiv.org/abs/hep-th/0005016}{arXiv:hep-th/0005016
  [hep-th]}%
  \bibAnnoteFile{NoStop}{Dvali:2000hr}%
%%CITATION = HEP-TH/0005016;%%
\bibitem{DeffayetEtal02}%
  \BibitemOpen
  \bibfield{author}{%
  \bibinfo {author} {\bibfnamefont{C.}~\bibnamefont{{Deffayet}}}, \bibinfo
  {author} {\bibfnamefont{G.}~\bibnamefont{{Dvali}}},\ and\ \bibinfo {author}
  {\bibfnamefont{G.}~\bibnamefont{{Gabadadze}}},\ }%
  \bibfield{journal}{%
  \bibinfo {journal} {\prd}\ }%
  \textbf{\bibinfo {volume} {65}},\ \bibinfo {pages} {044023} (\bibinfo {month}
  {Feb.}\ \bibinfo {year} {2002}),\
  \Eprint{http://arxiv.org/abs/arXiv:astro-ph/0105068}{arXiv:astro-ph/0105068}%
  \bibAnnoteFile{NoStop}{DeffayetEtal02}%
\bibitem{2003JHEP...09..029L}%
  \BibitemOpen
  \bibfield{author}{%
  \bibinfo {author} {\bibfnamefont{M.~A.}\ \bibnamefont{{Luty}}}, \bibinfo
  {author} {\bibfnamefont{M.}~\bibnamefont{{Porrati}}},\ and\ \bibinfo {author}
  {\bibfnamefont{R.}~\bibnamefont{{Rattazzi}}},\ }%
  \bibfield{journal}{%
  \Doi{10.1088/1126-6708/2003/09/029}{\bibinfo {journal} {Journal of High
  Energy Physics}}\ }%
  \textbf{\bibinfo {volume} {9}},\ \bibinfo {eid} {029} (\bibinfo {month}
  {Sep.}\ \bibinfo {year} {2003}),\
  \Eprint{http://arxiv.org/abs/hep-th/0303116}{hep-th/0303116}%
  \bibAnnoteFile{NoStop}{2003JHEP...09..029L}%
\bibitem{SahniShtanov}%
  \BibitemOpen
  \bibfield{author}{%
  \bibinfo {author} {\bibfnamefont{V.}~\bibnamefont{{Sahni}}}\ and\ \bibinfo
  {author} {\bibfnamefont{Y.}~\bibnamefont{{Shtanov}}},\ }%
  \bibfield{journal}{%
  \Doi{10.1088/1475-7516/2003/11/014}{\bibinfo {journal} {Journal of Cosmology
  and Astro-Particle Physics}}\ }%
  \textbf{\bibinfo {volume} {11}},\ \bibinfo {pages} {14} (\bibinfo {month}
  {Nov.}\ \bibinfo {year} {2003}),\
  \Eprint{http://arxiv.org/abs/arXiv:astro-ph/0202346}{arXiv:astro-ph/0202346}%
  \bibAnnoteFile{NoStop}{SahniShtanov}%
\bibitem{2004JHEP...06..059N}%
  \BibitemOpen
  \bibfield{author}{%
  \bibinfo {author} {\bibfnamefont{A.}~\bibnamefont{{Nicolis}}}\ and\ \bibinfo
  {author} {\bibfnamefont{R.}~\bibnamefont{{Rattazzi}}},\ }%
  \bibfield{journal}{%
  \Doi{10.1088/1126-6708/2004/06/059}{\bibinfo {journal} {Journal of High
  Energy Physics}}\ }%
  \textbf{\bibinfo {volume} {6}},\ \bibinfo {eid} {059} (\bibinfo {month}
  {Jun.}\ \bibinfo {year} {2004}),\
  \Eprint{http://arxiv.org/abs/hep-th/0404159}{hep-th/0404159}%
  \bibAnnoteFile{NoStop}{2004JHEP...06..059N}%
\bibitem{2007CQGra..24R.231K}%
  \BibitemOpen
  \bibfield{author}{%
  \bibinfo {author} {\bibfnamefont{K.}~\bibnamefont{{Koyama}}},\ }%
  \bibfield{journal}{%
  \Doi{10.1088/0264-9381/24/24/R01}{\bibinfo {journal} {Classical and Quantum
  Gravity}}\ }%
  \textbf{\bibinfo {volume} {24}},\ \bibinfo {pages} {231} (\bibinfo {month}
  {Dec.}\ \bibinfo {year} {2007}),\
  \Eprint{http://arxiv.org/abs/0709.2399}{arXiv:0709.2399 [hep-th]}%
  \bibAnnoteFile{NoStop}{2007CQGra..24R.231K}%
\bibitem{2006PhLB..642..432F}%
  \BibitemOpen
  \bibfield{author}{%
  \bibinfo {author} {\bibfnamefont{M.}~\bibnamefont{{Fairbairn}}}\ and\
  \bibinfo {author} {\bibfnamefont{A.}~\bibnamefont{{Goobar}}},\ }%
  \bibfield{journal}{%
  \Doi{10.1016/j.physletb.2006.07.048}{\bibinfo {journal} {Physics Letters B}}\
  }%
  \textbf{\bibinfo {volume} {642}},\ \bibinfo {pages} {432} (\bibinfo {month}
  {Nov.}\ \bibinfo {year} {2006}),\
  \Eprint{http://arxiv.org/abs/astro-ph/0511029}{astro-ph/0511029}%
  \bibAnnoteFile{NoStop}{2006PhLB..642..432F}%
\bibitem{2006PhRvD..74b3004M}%
  \BibitemOpen
  \bibfield{author}{%
  \bibinfo {author} {\bibfnamefont{R.}~\bibnamefont{{Maartens}}}\ and\ \bibinfo
  {author} {\bibfnamefont{E.}~\bibnamefont{{Majerotto}}},\ }%
  \bibfield{journal}{%
  \Doi{10.1103/PhysRevD.74.023004}{\bibinfo {journal} {PRD}}\ }%
  \textbf{\bibinfo {volume} {74}},\ \bibinfo {eid} {023004} (\bibinfo {month}
  {Jul.}\ \bibinfo {year} {2006}),\
  \Eprint{http://arxiv.org/abs/astro-ph/0603353}{astro-ph/0603353}%
  \bibAnnoteFile{NoStop}{2006PhRvD..74b3004M}%
\bibitem{2008PhRvD..78j3509F}%
  \BibitemOpen
  \bibfield{author}{%
  \bibinfo {author} {\bibfnamefont{W.}~\bibnamefont{{Fang}}}, \bibinfo {author}
  {\bibfnamefont{S.}~\bibnamefont{{Wang}}}, \bibinfo {author}
  {\bibfnamefont{W.}~\bibnamefont{{Hu}}}, \bibinfo {author}
  {\bibfnamefont{Z.}~\bibnamefont{{Haiman}}}, \bibinfo {author}
  {\bibfnamefont{L.}~\bibnamefont{{Hui}}},\ and\ \bibinfo {author}
  {\bibfnamefont{M.}~\bibnamefont{{May}}},\ }%
  \bibfield{journal}{%
  \Doi{10.1103/PhysRevD.78.103509}{\bibinfo {journal} {PRD}}\ }%
  \textbf{\bibinfo {volume} {78}},\ \bibinfo {eid} {103509} (\bibinfo {month}
  {Nov.}\ \bibinfo {year} {2008}),\
  \Eprint{http://arxiv.org/abs/0808.2208}{arXiv:0808.2208}%
  \bibAnnoteFile{NoStop}{2008PhRvD..78j3509F}%
\bibitem{2009PhRvD..80f3536L}%
  \BibitemOpen
  \bibfield{author}{%
  \bibinfo {author} {\bibfnamefont{L.}~\bibnamefont{{Lombriser}}}, \bibinfo
  {author} {\bibfnamefont{W.}~\bibnamefont{{Hu}}}, \bibinfo {author}
  {\bibfnamefont{W.}~\bibnamefont{{Fang}}},\ and\ \bibinfo {author}
  {\bibfnamefont{U.}~\bibnamefont{{Seljak}}},\ }%
  \bibfield{journal}{%
  \Doi{10.1103/PhysRevD.80.063536}{\bibinfo {journal} {PRD}}\ }%
  \textbf{\bibinfo {volume} {80}},\ \bibinfo {eid} {063536} (\bibinfo {month}
  {Sep.}\ \bibinfo {year} {2009}),\
  \Eprint{http://arxiv.org/abs/0905.1112}{arXiv:0905.1112 [astro-ph.CO]}%
  \bibAnnoteFile{NoStop}{2009PhRvD..80f3536L}%
\bibitem{2010PhRvD..82d4032W}%
  \BibitemOpen
  \bibfield{author}{%
  \bibinfo {author} {\bibfnamefont{M.}~\bibnamefont{{Wyman}}}\ and\ \bibinfo
  {author} {\bibfnamefont{J.}~\bibnamefont{{Khoury}}},\ }%
  \bibfield{journal}{%
  \Doi{10.1103/PhysRevD.82.044032}{\bibinfo {journal} {PRD}}\ }%
  \textbf{\bibinfo {volume} {82}},\ \bibinfo {eid} {044032} (\bibinfo {month}
  {Aug.}\ \bibinfo {year} {2010}),\
  \Eprint{http://arxiv.org/abs/1004.2046}{arXiv:1004.2046}%
  \bibAnnoteFile{NoStop}{2010PhRvD..82d4032W}%
\bibitem{2013MNRAS.436...89R}%
  \BibitemOpen
  \bibfield{author}{%
  \bibinfo {author} {\bibfnamefont{A.}~\bibnamefont{{Raccanelli}}}, \bibinfo
  {author} {\bibfnamefont{D.}~\bibnamefont{{Bertacca}}}, \bibinfo {author}
  {\bibfnamefont{D.}~\bibnamefont{{Pietrobon}}}, \bibinfo {author}
  {\bibfnamefont{F.}~\bibnamefont{{Schmidt}}}, \bibinfo {author}
  {\bibfnamefont{L.}~\bibnamefont{{Samushia}}}, \bibinfo {author}
  {\bibfnamefont{N.}~\bibnamefont{{Bartolo}}}, \bibinfo {author}
  {\bibfnamefont{O.}~\bibnamefont{{Dor{\'e}}}}, \bibinfo {author}
  {\bibfnamefont{S.}~\bibnamefont{{Matarrese}}},\ and\ \bibinfo {author}
  {\bibfnamefont{W.~J.}\ \bibnamefont{{Percival}}},\ }%
  \bibfield{journal}{%
  \Doi{10.1093/mnras/stt1517}{\bibinfo {journal} {MNRAS}}\ }%
  \textbf{\bibinfo {volume} {436}},\ \bibinfo {pages} {89} (\bibinfo {month}
  {Nov.}\ \bibinfo {year} {2013}),\
  \Eprint{http://arxiv.org/abs/1207.0500}{arXiv:1207.0500}%
  \bibAnnoteFile{NoStop}{2013MNRAS.436...89R}%
\bibitem{2014JCAP...02..048X}%
  \BibitemOpen
  \bibfield{author}{%
  \bibinfo {author} {\bibfnamefont{L.}~\bibnamefont{{Xu}}},\ }%
  \bibfield{journal}{%
  \Doi{10.1088/1475-7516/2014/02/048}{\bibinfo {journal} {JCAP}}\ }%
  \textbf{\bibinfo {volume} {2}},\ \bibinfo {eid} {048} (\bibinfo {month}
  {Feb.}\ \bibinfo {year} {2014}),\
  \Eprint{http://arxiv.org/abs/1312.4679}{arXiv:1312.4679 [astro-ph.CO]}%
  \bibAnnoteFile{NoStop}{2014JCAP...02..048X}%
\bibitem{2009PhRvD..80d3001S}%
  \BibitemOpen
  \bibfield{author}{%
  \bibinfo {author} {\bibfnamefont{F.}~\bibnamefont{{Schmidt}}},\ }%
  \bibfield{journal}{%
  \Doi{10.1103/PhysRevD.80.043001}{\bibinfo {journal} {PRD}}\ }%
  \textbf{\bibinfo {volume} {80}},\ \bibinfo {eid} {043001} (\bibinfo {month}
  {Aug.}\ \bibinfo {year} {2009}),\
  \Eprint{http://arxiv.org/abs/0905.0858}{arXiv:0905.0858 [astro-ph.CO]}%
  \bibAnnoteFile{NoStop}{2009PhRvD..80d3001S}%
\bibitem{2009PhRvD..80l3003S}%
  \BibitemOpen
  \bibfield{author}{%
  \bibinfo {author} {\bibfnamefont{F.}~\bibnamefont{{Schmidt}}},\ }%
  \bibfield{journal}{%
  \Doi{10.1103/PhysRevD.80.123003}{\bibinfo {journal} {PRD}}\ }%
  \textbf{\bibinfo {volume} {80}},\ \bibinfo {eid} {123003} (\bibinfo {month}
  {Dec.}\ \bibinfo {year} {2009}),\
  \Eprint{http://arxiv.org/abs/0910.0235}{arXiv:0910.0235 [astro-ph.CO]}%
  \bibAnnoteFile{NoStop}{2009PhRvD..80l3003S}%
\bibitem{2012PhRvL.109e1301L}%
  \BibitemOpen
  \bibfield{author}{%
  \bibinfo {author} {\bibfnamefont{T.~Y.}\ \bibnamefont{{Lam}}}, \bibinfo
  {author} {\bibfnamefont{T.}~\bibnamefont{{Nishimichi}}}, \bibinfo {author}
  {\bibfnamefont{F.}~\bibnamefont{{Schmidt}}},\ and\ \bibinfo {author}
  {\bibfnamefont{M.}~\bibnamefont{{Takada}}},\ }%
  \bibfield{journal}{%
  \Doi{10.1103/PhysRevLett.109.051301}{\bibinfo {journal} {Physical Review
  Letters}}\ }%
  \textbf{\bibinfo {volume} {109}},\ \bibinfo {eid} {051301} (\bibinfo {month}
  {Aug.}\ \bibinfo {year} {2012}),\
  \Eprint{http://arxiv.org/abs/1202.4501}{arXiv:1202.4501 [astro-ph.CO]}%
  \bibAnnoteFile{NoStop}{2012PhRvL.109e1301L}%
\bibitem{2014MNRAS.445.1885Z}%
  \BibitemOpen
  \bibfield{author}{%
  \bibinfo {author} {\bibfnamefont{Y.}~\bibnamefont{{Zu}}}, \bibinfo {author}
  {\bibfnamefont{D.~H.}\ \bibnamefont{{Weinberg}}}, \bibinfo {author}
  {\bibfnamefont{E.}~\bibnamefont{{Jennings}}}, \bibinfo {author}
  {\bibfnamefont{B.}~\bibnamefont{{Li}}},\ and\ \bibinfo {author}
  {\bibfnamefont{M.}~\bibnamefont{{Wyman}}},\ }%
  \bibfield{journal}{%
  \Doi{10.1093/mnras/stu1739}{\bibinfo {journal} {MNRAS}}\ }%
  \textbf{\bibinfo {volume} {445}},\ \bibinfo {pages} {1885} (\bibinfo {month}
  {Dec.}\ \bibinfo {year} {2014}),\
  \Eprint{http://arxiv.org/abs/1310.6768}{arXiv:1310.6768}%
  \bibAnnoteFile{NoStop}{2014MNRAS.445.1885Z}%
\bibitem{2014JCAP...07..058F}%
  \BibitemOpen
  \bibfield{author}{%
  \bibinfo {author} {\bibfnamefont{B.}~\bibnamefont{{Falck}}}, \bibinfo
  {author} {\bibfnamefont{K.}~\bibnamefont{{Koyama}}}, \bibinfo {author}
  {\bibfnamefont{G.-b.}\ \bibnamefont{{Zhao}}},\ and\ \bibinfo {author}
  {\bibfnamefont{B.}~\bibnamefont{{Li}}},\ }%
  \bibfield{journal}{%
  \Doi{10.1088/1475-7516/2014/07/058}{\bibinfo {journal} {JCAP}}\ }%
  \textbf{\bibinfo {volume} {7}},\ \bibinfo {eid} {058} (\bibinfo {month}
  {Jul.}\ \bibinfo {year} {2014}),\
  \Eprint{http://arxiv.org/abs/1404.2206}{arXiv:1404.2206}%
  \bibAnnoteFile{NoStop}{2014JCAP...07..058F}%
\bibitem{2015JCAP...07..049F}%
  \BibitemOpen
  \bibfield{author}{%
  \bibinfo {author} {\bibfnamefont{B.}~\bibnamefont{{Falck}}}, \bibinfo
  {author} {\bibfnamefont{K.}~\bibnamefont{{Koyama}}},\ and\ \bibinfo {author}
  {\bibfnamefont{G.-B.}\ \bibnamefont{{Zhao}}},\ }%
  \bibfield{journal}{%
  \Doi{10.1088/1475-7516/2015/07/049}{\bibinfo {journal} {JCAP}}\ }%
  \textbf{\bibinfo {volume} {7}},\ \bibinfo {eid} {049} (\bibinfo {month}
  {Jul.}\ \bibinfo {year} {2015}),\
  \Eprint{http://arxiv.org/abs/1503.06673}{arXiv:1503.06673}%
  \bibAnnoteFile{NoStop}{2015JCAP...07..049F}%
\bibitem{codecomp}%
  \BibitemOpen
  \bibfield{author}{%
  \bibinfo {author} {\bibfnamefont{H.~A.}\ \bibnamefont{{Winther}}}, \bibinfo
  {author} {\bibfnamefont{F.}~\bibnamefont{{Schmidt}}}, \bibinfo {author}
  {\bibfnamefont{A.}~\bibnamefont{{Barreira}}}, \bibinfo {author}
  {\bibfnamefont{C.}~\bibnamefont{{Arnold}}}, \bibinfo {author}
  {\bibfnamefont{S.}~\bibnamefont{{Bose}}}, \bibinfo {author}
  {\bibfnamefont{C.}~\bibnamefont{{Llinares}}}, \bibinfo {author}
  {\bibfnamefont{M.}~\bibnamefont{{Baldi}}}, \bibinfo {author}
  {\bibfnamefont{B.}~\bibnamefont{{Falck}}}, \bibinfo {author}
  {\bibfnamefont{W.~A.}\ \bibnamefont{{Hellwing}}}, \bibinfo {author}
  {\bibfnamefont{K.}~\bibnamefont{{Koyama}}}, \bibinfo {author}
  {\bibfnamefont{B.}~\bibnamefont{{Li}}}, \bibinfo {author}
  {\bibfnamefont{D.~F.}\ \bibnamefont{{Mota}}}, \bibinfo {author}
  {\bibfnamefont{E.}~\bibnamefont{{Puchwein}}}, \bibinfo {author}
  {\bibfnamefont{R.~E.}\ \bibnamefont{{Smith}}},\ and\ \bibinfo {author}
  {\bibfnamefont{G.-B.}\ \bibnamefont{{Zhao}}},\ }%
  \bibfield{journal}{%
  \Doi{10.1093/mnras/stv2253}{\bibinfo {journal} {MNRAS}}\ }%
  \textbf{\bibinfo {volume} {454}},\ \bibinfo {pages} {4208} (\bibinfo {month}
  {Dec.}\ \bibinfo {year} {2015}),\
  \Eprint{http://arxiv.org/abs/1506.06384}{arXiv:1506.06384}%
  \bibAnnoteFile{NoStop}{codecomp}%
\bibitem{2015JCAP...12..059B}%
  \BibitemOpen
  \bibfield{author}{%
  \bibinfo {author} {\bibfnamefont{A.}~\bibnamefont{{Barreira}}}, \bibinfo
  {author} {\bibfnamefont{S.}~\bibnamefont{{Bose}}},\ and\ \bibinfo {author}
  {\bibfnamefont{B.}~\bibnamefont{{Li}}},\ }%
  \bibfield{journal}{%
  \Doi{10.1088/1475-7516/2015/12/059}{\bibinfo {journal} {JCAP}}\ }%
  \textbf{\bibinfo {volume} {12}},\ \bibinfo {eid} {059} (\bibinfo {month}
  {Dec.}\ \bibinfo {year} {2015}),\
  \Eprint{http://arxiv.org/abs/1511.08200}{arXiv:1511.08200}%
  \bibAnnoteFile{NoStop}{2015JCAP...12..059B}%
\bibitem{2016arXiv160503965B}%
  \BibitemOpen
  \bibfield{author}{%
  \bibinfo {author} {\bibfnamefont{A.}~\bibnamefont{{Barreira}}}, \bibinfo
  {author} {\bibfnamefont{A.~G.}\ \bibnamefont{{S{\'a}nchez}}},\ and\ \bibinfo
  {author} {\bibfnamefont{F.}~\bibnamefont{{Schmidt}}},\ }%
  \enquote{\bibinfo {title} {{Validating estimates of the growth rate of
  structure with modified gravity simulations}},}\  (\bibinfo {month} {May}\
  \bibinfo {year} {2016}),\
  \Eprint{http://arxiv.org/abs/1605.03965}{arXiv:1605.03965}%
  \bibAnnoteFile{NoStop}{2016arXiv160503965B}%
\bibitem{2007PhRvD..75h4040K}%
  \BibitemOpen
  \bibfield{author}{%
  \bibinfo {author} {\bibfnamefont{K.}~\bibnamefont{{Koyama}}}\ and\ \bibinfo
  {author} {\bibfnamefont{F.~P.}\ \bibnamefont{{Silva}}},\ }%
  \bibfield{journal}{%
  \Doi{10.1103/PhysRevD.75.084040}{\bibinfo {journal} {PRD}}\ }%
  \textbf{\bibinfo {volume} {75}},\ \bibinfo {eid} {084040} (\bibinfo {month}
  {Apr.}\ \bibinfo {year} {2007}),\
  \Eprint{http://arxiv.org/abs/hep-th/0702169}{hep-th/0702169}%
  \bibAnnoteFile{NoStop}{2007PhRvD..75h4040K}%
\bibitem{2015arXiv150503539W}%
  \BibitemOpen
  \bibfield{author}{%
  \bibinfo {author} {\bibfnamefont{H.~A.}\ \bibnamefont{{Winther}}}\ and\
  \bibinfo {author} {\bibfnamefont{P.~G.}\ \bibnamefont{{Ferreira}}},\ }%
  \bibfield{journal}{%
  \bibinfo {journal} {ArXiv e-prints}}%
   (\bibinfo {month} {May}\ \bibinfo {year} {2015}),\
  \Eprint{http://arxiv.org/abs/1505.03539}{arXiv:1505.03539 [gr-qc]}%
  \bibAnnoteFile{NoStop}{2015arXiv150503539W}%
\bibitem{PhysRevD.79.064036}%
  \BibitemOpen
  \bibfield{author}{%
  \bibinfo {author} {\bibfnamefont{A.}~\bibnamefont{Nicolis}}, \bibinfo
  {author} {\bibfnamefont{R.}~\bibnamefont{Rattazzi}},\ and\ \bibinfo {author}
  {\bibfnamefont{E.}~\bibnamefont{Trincherini}},\ }%
  \bibfield{journal}{%
  \Doi{10.1103/PhysRevD.79.064036}{\bibinfo {journal} {Phys. Rev. D}}\ }%
  \textbf{\bibinfo {volume} {79}},\ \bibinfo {pages} {064036} (\bibinfo {year}
  {2009})%
  \bibAnnoteFile{NoStop}{PhysRevD.79.064036}%
\bibitem{PhysRevD.79.084003}%
  \BibitemOpen
  \bibfield{author}{%
  \bibinfo {author} {\bibfnamefont{C.}~\bibnamefont{Deffayet}}, \bibinfo
  {author} {\bibfnamefont{G.}~\bibnamefont{Esposito-Far\`ese}},\ and\ \bibinfo
  {author} {\bibfnamefont{A.}~\bibnamefont{Vikman}},\ }%
  \bibfield{journal}{%
  \Doi{10.1103/PhysRevD.79.084003}{\bibinfo {journal} {Phys. Rev. D}}\ }%
  \textbf{\bibinfo {volume} {79}},\ \bibinfo {pages} {084003} (\bibinfo {year}
  {2009})%
  \bibAnnoteFile{NoStop}{PhysRevD.79.084003}%
\bibitem{Deffayet:2009mn}%
  \BibitemOpen
  \bibfield{author}{%
  \bibinfo {author} {\bibfnamefont{C.}~\bibnamefont{Deffayet}}, \bibinfo
  {author} {\bibfnamefont{S.}~\bibnamefont{Deser}},\ and\ \bibinfo {author}
  {\bibfnamefont{G.}~\bibnamefont{Esposito-Farese}},\ }%
  \bibfield{journal}{%
  \Doi{10.1103/PhysRevD.80.064015}{\bibinfo {journal} {Phys.Rev.}}\ }%
  \textbf{\bibinfo {volume} {D80}},\ \bibinfo {pages} {064015} (\bibinfo {year}
  {2009}),\ \Eprint{http://arxiv.org/abs/0906.1967}{arXiv:0906.1967 [gr-qc]}%
  \bibAnnoteFile{NoStop}{Deffayet:2009mn}%
%%CITATION = ARXIV:0906.1967;%%
\bibitem{2012PhRvD..86l4016B}%
  \BibitemOpen
  \bibfield{author}{%
  \bibinfo {author} {\bibfnamefont{A.}~\bibnamefont{{Barreira}}}, \bibinfo
  {author} {\bibfnamefont{B.}~\bibnamefont{{Li}}}, \bibinfo {author}
  {\bibfnamefont{C.~M.}\ \bibnamefont{{Baugh}}},\ and\ \bibinfo {author}
  {\bibfnamefont{S.}~\bibnamefont{{Pascoli}}},\ }%
  \bibfield{journal}{%
  \Doi{10.1103/PhysRevD.86.124016}{\bibinfo {journal} {PRD}}\ }%
  \textbf{\bibinfo {volume} {86}},\ \bibinfo {eid} {124016} (\bibinfo {month}
  {Dec.}\ \bibinfo {year} {2012}),\
  \Eprint{http://arxiv.org/abs/1208.0600}{arXiv:1208.0600 [astro-ph.CO]}%
  \bibAnnoteFile{NoStop}{2012PhRvD..86l4016B}%
\bibitem{Barreira:2014jha}%
  \BibitemOpen
  \bibfield{author}{%
  \bibinfo {author} {\bibfnamefont{A.}~\bibnamefont{Barreira}}, \bibinfo
  {author} {\bibfnamefont{B.}~\bibnamefont{Li}}, \bibinfo {author}
  {\bibfnamefont{C.}~\bibnamefont{Baugh}},\ and\ \bibinfo {author}
  {\bibfnamefont{S.}~\bibnamefont{Pascoli}},\ }%
  \bibfield{journal}{%
  \Doi{10.1088/1475-7516/2014/08/059}{\bibinfo {journal} {JCAP}}\ }%
  \textbf{\bibinfo {volume} {1408}},\ \bibinfo {pages} {059} (\bibinfo {year}
  {2014}),\ \Eprint{http://arxiv.org/abs/1406.0485}{arXiv:1406.0485
  [astro-ph.CO]}%
  \bibAnnoteFile{NoStop}{Barreira:2014jha}%
%%CITATION = ARXIV:1406.0485;%%
\bibitem{2016arXiv160403487R}%
  \BibitemOpen
  \bibfield{author}{%
  \bibinfo {author} {\bibfnamefont{J.}~\bibnamefont{{Renk}}}, \bibinfo {author}
  {\bibfnamefont{M.}~\bibnamefont{{Zumalacarregui}}},\ and\ \bibinfo {author}
  {\bibfnamefont{F.}~\bibnamefont{{Montanari}}},\ }%
  \bibfield{journal}{%
  \bibinfo {journal} {ArXiv e-prints}}%
   (\bibinfo {month} {Apr.}\ \bibinfo {year} {2016}),\
  \Eprint{http://arxiv.org/abs/1604.03487}{arXiv:1604.03487}%
  \bibAnnoteFile{NoStop}{2016arXiv160403487R}%
\bibitem{2015JCAP...10..064T}%
  \BibitemOpen
  \bibfield{author}{%
  \bibinfo {author} {\bibfnamefont{A.}~\bibnamefont{{Terukina}}}, \bibinfo
  {author} {\bibfnamefont{K.}~\bibnamefont{{Yamamoto}}}, \bibinfo {author}
  {\bibfnamefont{N.}~\bibnamefont{{Okabe}}}, \bibinfo {author}
  {\bibfnamefont{K.}~\bibnamefont{{Matsushita}}},\ and\ \bibinfo {author}
  {\bibfnamefont{T.}~\bibnamefont{{Sasaki}}},\ }%
  \bibfield{journal}{%
  \Doi{10.1088/1475-7516/2015/10/064}{\bibinfo {journal} {JCAP}}\ }%
  \textbf{\bibinfo {volume} {10}},\ \bibinfo {eid} {064} (\bibinfo {month}
  {Oct.}\ \bibinfo {year} {2015}),\
  \Eprint{http://arxiv.org/abs/1505.03692}{arXiv:1505.03692}%
  \bibAnnoteFile{NoStop}{2015JCAP...10..064T}%
\bibitem{Deser:2007jk}%
  \BibitemOpen
  \bibfield{author}{%
  \bibinfo {author} {\bibfnamefont{S.}~\bibnamefont{Deser}}\ and\ \bibinfo
  {author} {\bibfnamefont{R.}~\bibnamefont{Woodard}},\ }%
  \bibfield{journal}{%
  \Doi{10.1103/PhysRevLett.99.111301}{\bibinfo {journal} {Phys.Rev.Lett.}}\ }%
  \textbf{\bibinfo {volume} {99}},\ \bibinfo {pages} {111301} (\bibinfo {year}
  {2007}),\ \Eprint{http://arxiv.org/abs/0706.2151}{arXiv:0706.2151
  [astro-ph]}%
  \bibAnnoteFile{NoStop}{Deser:2007jk}%
%%CITATION = ARXIV:0706.2151;%%
\bibitem{Deser:2013uya}%
  \BibitemOpen
  \bibfield{author}{%
  \bibinfo {author} {\bibfnamefont{S.}~\bibnamefont{Deser}}\ and\ \bibinfo
  {author} {\bibfnamefont{R.}~\bibnamefont{Woodard}},\ }%
  \bibfield{journal}{%
  \Doi{10.1088/1475-7516/2013/11/036}{\bibinfo {journal} {JCAP}}\ }%
  \textbf{\bibinfo {volume} {1311}},\ \bibinfo {pages} {036} (\bibinfo {year}
  {2013}),\ \Eprint{http://arxiv.org/abs/1307.6639}{arXiv:1307.6639
  [astro-ph.CO]}%
  \bibAnnoteFile{NoStop}{Deser:2013uya}%
%%CITATION = ARXIV:1307.6639;%%
\bibitem{Maggiore:2014sia}%
  \BibitemOpen
  \bibfield{author}{%
  \bibinfo {author} {\bibfnamefont{M.}~\bibnamefont{Maggiore}}\ and\ \bibinfo
  {author} {\bibfnamefont{M.}~\bibnamefont{Mancarella}},\ }%
  \bibfield{journal}{%
  \Doi{10.1103/PhysRevD.90.023005}{\bibinfo {journal} {Phys.Rev.}}\ }%
  \textbf{\bibinfo {volume} {D90}},\ \bibinfo {pages} {023005} (\bibinfo {year}
  {2014}),\ \Eprint{http://arxiv.org/abs/1402.0448}{arXiv:1402.0448 [hep-th]}%
  \bibAnnoteFile{NoStop}{Maggiore:2014sia}%
%%CITATION = ARXIV:1402.0448;%%
\bibitem{Barreira:2014kra}%
  \BibitemOpen
  \bibfield{author}{%
  \bibinfo {author} {\bibfnamefont{A.}~\bibnamefont{Barreira}}, \bibinfo
  {author} {\bibfnamefont{B.}~\bibnamefont{Li}}, \bibinfo {author}
  {\bibfnamefont{W.~A.}\ \bibnamefont{Hellwing}}, \bibinfo {author}
  {\bibfnamefont{C.~M.}\ \bibnamefont{Baugh}},\ and\ \bibinfo {author}
  {\bibfnamefont{S.}~\bibnamefont{Pascoli}},\ }%
  \bibfield{journal}{%
  \Doi{10.1088/1475-7516/2014/09/031}{\bibinfo {journal} {JCAP}}\ }%
  \textbf{\bibinfo {volume} {1409}},\ \bibinfo {pages} {031} (\bibinfo {year}
  {2014}),\ \Eprint{http://arxiv.org/abs/1408.1084}{arXiv:1408.1084
  [astro-ph.CO]}%
  \bibAnnoteFile{NoStop}{Barreira:2014kra}%
%%CITATION = ARXIV:1408.1084;%%
\bibitem{2016arXiv160203558D}%
  \BibitemOpen
  \bibfield{author}{%
  \bibinfo {author} {\bibfnamefont{Y.}~\bibnamefont{{Dirian}}}, \bibinfo
  {author} {\bibfnamefont{S.}~\bibnamefont{{Foffa}}}, \bibinfo {author}
  {\bibfnamefont{M.}~\bibnamefont{{Kunz}}}, \bibinfo {author}
  {\bibfnamefont{M.}~\bibnamefont{{Maggiore}}},\ and\ \bibinfo {author}
  {\bibfnamefont{V.}~\bibnamefont{{Pettorino}}},\ }%
  \bibfield{journal}{%
  \bibinfo {journal} {ArXiv e-prints}}%
   (\bibinfo {month} {Feb.}\ \bibinfo {year} {2016}),\
  \Eprint{http://arxiv.org/abs/1602.03558}{arXiv:1602.03558}%
  \bibAnnoteFile{NoStop}{2016arXiv160203558D}%
\bibitem{Brax:2014wla}%
  \BibitemOpen
  \bibfield{author}{%
  \bibinfo {author} {\bibfnamefont{P.}~\bibnamefont{Brax}}\ and\ \bibinfo
  {author} {\bibfnamefont{P.}~\bibnamefont{Valageas}},\ }%
  \bibfield{journal}{%
  \Doi{10.1103/PhysRevD.90.023507}{\bibinfo {journal} {Phys.Rev.}}\ }%
  \textbf{\bibinfo {volume} {D90}},\ \bibinfo {pages} {023507} (\bibinfo {year}
  {2014}),\ \Eprint{http://arxiv.org/abs/1403.5420}{arXiv:1403.5420
  [astro-ph.CO]}%
  \bibAnnoteFile{NoStop}{Brax:2014wla}%
%%CITATION = ARXIV:1403.5420;%%
\bibitem{Brax:2014yla}%
  \BibitemOpen
  \bibfield{author}{%
  \bibinfo {author} {\bibfnamefont{P.}~\bibnamefont{Brax}}\ and\ \bibinfo
  {author} {\bibfnamefont{P.}~\bibnamefont{Valageas}},\ }%
  \bibfield{journal}{%
  \Doi{10.1103/PhysRevD.90.023508}{\bibinfo {journal} {Phys.Rev.}}\ }%
  \textbf{\bibinfo {volume} {D90}},\ \bibinfo {pages} {023508} (\bibinfo {year}
  {2014}),\ \Eprint{http://arxiv.org/abs/1403.5424}{arXiv:1403.5424
  [astro-ph.CO]}%
  \bibAnnoteFile{NoStop}{Brax:2014yla}%
%%CITATION = ARXIV:1403.5424;%%
\bibitem{2015PhRvD..91f3528B}%
  \BibitemOpen
  \bibfield{author}{%
  \bibinfo {author} {\bibfnamefont{A.}~\bibnamefont{{Barreira}}}, \bibinfo
  {author} {\bibfnamefont{P.}~\bibnamefont{{Brax}}}, \bibinfo {author}
  {\bibfnamefont{S.}~\bibnamefont{{Clesse}}}, \bibinfo {author}
  {\bibfnamefont{B.}~\bibnamefont{{Li}}},\ and\ \bibinfo {author}
  {\bibfnamefont{P.}~\bibnamefont{{Valageas}}},\ }%
  \bibfield{journal}{%
  \Doi{10.1103/PhysRevD.91.063528}{\bibinfo {journal} {PRD}}\ }%
  \textbf{\bibinfo {volume} {91}},\ \bibinfo {eid} {063528} (\bibinfo {month}
  {Mar.}\ \bibinfo {year} {2015}),\
  \Eprint{http://arxiv.org/abs/1411.5965}{arXiv:1411.5965}%
  \bibAnnoteFile{NoStop}{2015PhRvD..91f3528B}%
\bibitem{Horndeski:1974wa}%
  \BibitemOpen
  \bibfield{author}{%
  \bibinfo {author} {\bibfnamefont{G.~W.}\ \bibnamefont{Horndeski}},\ }%
  \bibfield{journal}{%
  \Doi{10.1007/BF01807638}{\bibinfo {journal} {Int.J.Theor.Phys.}}\ }%
  \textbf{\bibinfo {volume} {10}},\ \bibinfo {pages} {363} (\bibinfo {year}
  {1974})%
  \bibAnnoteFile{NoStop}{Horndeski:1974wa}%
%%CITATION = IJTPB,10,363;%%
\bibitem{2014PhRvD..89f4046Z}%
  \BibitemOpen
  \bibfield{author}{%
  \bibinfo {author} {\bibfnamefont{M.}~\bibnamefont{{Zumalac{\'a}rregui}}}\
  and\ \bibinfo {author}
  {\bibfnamefont{J.}~\bibnamefont{{Garc{\'{\i}}a-Bellido}}},\ }%
  \bibfield{journal}{%
  \Doi{10.1103/PhysRevD.89.064046}{\bibinfo {journal} {PRD}}\ }%
  \textbf{\bibinfo {volume} {89}},\ \bibinfo {eid} {064046} (\bibinfo {month}
  {Mar.}\ \bibinfo {year} {2014}),\
  \Eprint{http://arxiv.org/abs/1308.4685}{arXiv:1308.4685 [gr-qc]}%
  \bibAnnoteFile{NoStop}{2014PhRvD..89f4046Z}%
\bibitem{2015PhRvL.114u1101G}%
  \BibitemOpen
  \bibfield{author}{%
  \bibinfo {author} {\bibfnamefont{J.}~\bibnamefont{{Gleyzes}}}, \bibinfo
  {author} {\bibfnamefont{D.}~\bibnamefont{{Langlois}}}, \bibinfo {author}
  {\bibfnamefont{F.}~\bibnamefont{{Piazza}}},\ and\ \bibinfo {author}
  {\bibfnamefont{F.}~\bibnamefont{{Vernizzi}}},\ }%
  \bibfield{journal}{%
  \Doi{10.1103/PhysRevLett.114.211101}{\bibinfo {journal} {Physical Review
  Letters}}\ }%
  \textbf{\bibinfo {volume} {114}},\ \bibinfo {eid} {211101} (\bibinfo {month}
  {May}\ \bibinfo {year} {2015}),\
  \Eprint{http://arxiv.org/abs/1404.6495}{arXiv:1404.6495 [hep-th]}%
  \bibAnnoteFile{NoStop}{2015PhRvL.114u1101G}%
\bibitem{2015arXiv151005964S}%
  \BibitemOpen
  \bibfield{author}{%
  \bibinfo {author} {\bibfnamefont{J.}~\bibnamefont{{Sakstein}}},\ }%
  \bibfield{journal}{%
  \bibinfo {journal} {ArXiv e-prints}}%
   (\bibinfo {month} {Oct.}\ \bibinfo {year} {2015}),\
  \Eprint{http://arxiv.org/abs/1510.05964}{arXiv:1510.05964}%
  \bibAnnoteFile{NoStop}{2015arXiv151005964S}%
\bibitem{2016arXiv160306368S}%
  \BibitemOpen
  \bibfield{author}{%
  \bibinfo {author} {\bibfnamefont{J.}~\bibnamefont{{Sakstein}}}, \bibinfo
  {author} {\bibfnamefont{H.}~\bibnamefont{{Wilcox}}}, \bibinfo {author}
  {\bibfnamefont{D.}~\bibnamefont{{Bacon}}}, \bibinfo {author}
  {\bibfnamefont{K.}~\bibnamefont{{Koyama}}},\ and\ \bibinfo {author}
  {\bibfnamefont{R.~C.}\ \bibnamefont{{Nichol}}},\ }%
  \bibfield{journal}{%
  \bibinfo {journal} {ArXiv e-prints}}%
   (\bibinfo {month} {Mar.}\ \bibinfo {year} {2016}),\
  \Eprint{http://arxiv.org/abs/1603.06368}{arXiv:1603.06368}%
  \bibAnnoteFile{NoStop}{2016arXiv160306368S}%
\bibitem{2010RvMP...82..451S}%
  \BibitemOpen
  \bibfield{author}{%
  \bibinfo {author} {\bibfnamefont{T.~P.}\ \bibnamefont{{Sotiriou}}}\ and\
  \bibinfo {author} {\bibfnamefont{V.}~\bibnamefont{{Faraoni}}},\ }%
  \bibfield{journal}{%
  \Doi{10.1103/RevModPhys.82.451}{\bibinfo {journal} {Reviews of Modern
  Physics}}\ }%
  \textbf{\bibinfo {volume} {82}},\ \bibinfo {pages} {451} (\bibinfo {month}
  {Jan.}\ \bibinfo {year} {2010}),\
  \Eprint{http://arxiv.org/abs/0805.1726}{arXiv:0805.1726 [gr-qc]}%
  \bibAnnoteFile{NoStop}{2010RvMP...82..451S}%
\bibitem{2008ApJS..178..179P}%
  \BibitemOpen
  \bibfield{author}{%
  \bibinfo {author} {\bibfnamefont{S.}~\bibnamefont{{Prunet}}}, \bibinfo
  {author} {\bibfnamefont{C.}~\bibnamefont{{Pichon}}}, \bibinfo {author}
  {\bibfnamefont{D.}~\bibnamefont{{Aubert}}}, \bibinfo {author}
  {\bibfnamefont{D.}~\bibnamefont{{Pogosyan}}}, \bibinfo {author}
  {\bibfnamefont{R.}~\bibnamefont{{Teyssier}}},\ and\ \bibinfo {author}
  {\bibfnamefont{S.}~\bibnamefont{{Gottloeber}}},\ }%
  \bibfield{journal}{%
  \Doi{10.1086/590370}{\bibinfo {journal} {APJS}}\ }%
  \textbf{\bibinfo {volume} {178}},\ \bibinfo {eid} {179-188} (\bibinfo {month}
  {Oct.}\ \bibinfo {year} {2008}),\
  \Eprint{http://arxiv.org/abs/0804.3536}{arXiv:0804.3536}%
  \bibAnnoteFile{NoStop}{2008ApJS..178..179P}%
\bibitem{2003MNRAS.341.1311S}%
  \BibitemOpen
  \bibfield{author}{%
  \bibinfo {author} {\bibfnamefont{R.~E.}\ \bibnamefont{{Smith}}}, \bibinfo
  {author} {\bibfnamefont{J.~A.}\ \bibnamefont{{Peacock}}}, \bibinfo {author}
  {\bibfnamefont{A.}~\bibnamefont{{Jenkins}}}, \bibinfo {author}
  {\bibfnamefont{S.~D.~M.}\ \bibnamefont{{White}}}, \bibinfo {author}
  {\bibfnamefont{C.~S.}\ \bibnamefont{{Frenk}}}, \bibinfo {author}
  {\bibfnamefont{F.~R.}\ \bibnamefont{{Pearce}}}, \bibinfo {author}
  {\bibfnamefont{P.~A.}\ \bibnamefont{{Thomas}}}, \bibinfo {author}
  {\bibfnamefont{G.}~\bibnamefont{{Efstathiou}}},\ and\ \bibinfo {author}
  {\bibfnamefont{H.~M.~P.}\ \bibnamefont{{Couchman}}},\ }%
  \bibfield{journal}{%
  \Doi{10.1046/j.1365-8711.2003.06503.x}{\bibinfo {journal} {MNRAS}}\ }%
  \textbf{\bibinfo {volume} {341}},\ \bibinfo {pages} {1311} (\bibinfo {month}
  {Jun.}\ \bibinfo {year} {2003}),\
  \Eprint{http://arxiv.org/abs/astro-ph/0207664}{astro-ph/0207664}%
  \bibAnnoteFile{NoStop}{2003MNRAS.341.1311S}%
\bibitem{2012ApJ...761..152T}%
  \BibitemOpen
  \bibfield{author}{%
  \bibinfo {author} {\bibfnamefont{R.}~\bibnamefont{{Takahashi}}}, \bibinfo
  {author} {\bibfnamefont{M.}~\bibnamefont{{Sato}}}, \bibinfo {author}
  {\bibfnamefont{T.}~\bibnamefont{{Nishimichi}}}, \bibinfo {author}
  {\bibfnamefont{A.}~\bibnamefont{{Taruya}}},\ and\ \bibinfo {author}
  {\bibfnamefont{M.}~\bibnamefont{{Oguri}}},\ }%
  \bibfield{journal}{%
  \Doi{10.1088/0004-637X/761/2/152}{\bibinfo {journal} {APJ}}\ }%
  \textbf{\bibinfo {volume} {761}},\ \bibinfo {eid} {152} (\bibinfo {month}
  {Dec.}\ \bibinfo {year} {2012}),\
  \Eprint{http://arxiv.org/abs/1208.2701}{arXiv:1208.2701}%
  \bibAnnoteFile{NoStop}{2012ApJ...761..152T}%
\bibitem{2008MNRAS.391..435F}%
  \BibitemOpen
  \bibfield{author}{%
  \bibinfo {author} {\bibfnamefont{P.}~\bibnamefont{{Fosalba}}}, \bibinfo
  {author} {\bibfnamefont{E.}~\bibnamefont{{Gazta{\~n}aga}}}, \bibinfo {author}
  {\bibfnamefont{F.~J.}\ \bibnamefont{{Castander}}},\ and\ \bibinfo {author}
  {\bibfnamefont{M.}~\bibnamefont{{Manera}}},\ }%
  \bibfield{journal}{%
  \Doi{10.1111/j.1365-2966.2008.13910.x}{\bibinfo {journal} {MNRAS}}\ }%
  \textbf{\bibinfo {volume} {391}},\ \bibinfo {pages} {435} (\bibinfo {month}
  {Nov.}\ \bibinfo {year} {2008}),\
  \Eprint{http://arxiv.org/abs/0711.1540}{arXiv:0711.1540}%
  \bibAnnoteFile{NoStop}{2008MNRAS.391..435F}%
\bibitem{2013ApJ...762..109B}%
  \BibitemOpen
  \bibfield{author}{%
  \bibinfo {author} {\bibfnamefont{P.~S.}\ \bibnamefont{{Behroozi}}}, \bibinfo
  {author} {\bibfnamefont{R.~H.}\ \bibnamefont{{Wechsler}}},\ and\ \bibinfo
  {author} {\bibfnamefont{H.-Y.}\ \bibnamefont{{Wu}}},\ }%
  \bibfield{journal}{%
  \Doi{10.1088/0004-637X/762/2/109}{\bibinfo {journal} {APJ}}\ }%
  \textbf{\bibinfo {volume} {762}},\ \bibinfo {eid} {109} (\bibinfo {month}
  {Jan.}\ \bibinfo {year} {2013}),\
  \Eprint{http://arxiv.org/abs/1110.4372}{arXiv:1110.4372 [astro-ph.CO]}%
  \bibAnnoteFile{NoStop}{2013ApJ...762..109B}%
\end{thebibliography}%

\end{document}